\documentclass[letterpaper]{aa}
	\usepackage{txfonts}
	\usepackage{natbib}
	\usepackage{graphicx}
	\usepackage{amssymb}
	\usepackage{epsfig}
	\usepackage{hyperref}
	\begin{document}
	\def\sizex{16.0 cm}
	\def\bigx{10.0 cm}
	\def\smallerxsize{7.0 cm}
	\def\smallxsize{10.0 cm}
	\def\smallysize{12.0 cm}
        \def\citeNP#1{\citeauthor{#1} \citeyear{#1}}

	\title {Photometric redshifts for the CFHTLS T0004 Deep and Wide fields\thanks{Based on
	    observations obtained with MegaPrime/MegaCam, a joint project of
	    CFHT and CEA/DAPNIA, at the Canada-France-Hawaii Telescope (CFHT)
	    which is operated by the National Research Council (NRC) of
	    Canada, the Institut National des Sciences de l'Univers of the
	    Centre National de la Recherche Scientifique (CNRS) of France, and
	    the University of Hawaii. This work is based in part on data
	    products produced at {\sc Terapix} and the Canadian Astronomy Data
	    Centre as part of the Canada-France-Hawaii Telescope Legacy
	    Survey, a collaborative project of NRC and CNRS.}}
	\offprints {J.\ Coupon} \date{\fbox{\sc Draft Version: \today}}
	\titlerunning{Photometric redshifts for the CFHTLS T0004 Deep and
          Wide fields}
	\authorrunning{Coupon et al.}

        \author{
          J. Coupon\inst{1}\fnmsep\thanks{\email{coupon@iap.fr}},
          O. Ilbert\inst{2},
          M. Kilbinger\inst{1},
          H. J. McCracken\inst{1},
          Y. Mellier\inst{1},
          S. Arnouts\inst{3},
          E. Bertin\inst{1},
          P. Hudelot\inst{1},
          M. Schultheis\inst{4},
          O. Le F\`evre\inst{5},
          V. Le Brun\inst{5},
          L. Guzzo\inst{6},
          S. Bardelli\inst{7},
          E. Zucca\inst{7},
          M. Bolzonella\inst{7},
          B. Garilli\inst{6},
          G. Zamorani\inst{7}
          and A. Zanichelli\inst{8}
        }
        
	\institute{Institut d'Astrophysique de Paris, UMR7095 CNRS,
	  Universit\'e Pierre et Marie Curie, 98 bis Boulevard Arago, 75014
	  Paris, France \and Institute for Astronomy, 2680 Woodlawn
	  Dr.,University of Hawaii, Honolulu, Hawaii, 96822  \and
          Canada-France-Hawaii telescope, 65-1238 Mamalahoa Highway, Kamuela,
          HI 9674 \and Observatoire des Sciences de l'Univers de
          Besan\c con, UMR6213 CNRS, 41 bis, avenue de l'Observatoire, 
          25010 Besan\c con, France \and Laboratoire d'Astrophysique de
          Marseille, BP 8, Traverse du Siphon, 13376 Marseille Cedex 12,
          France \and INAF-Osservatorio Astronomico di Brera, via Bianchi 46, 
          I-23807 Merate (LC), Italy \and Osservatorio Astronomicoq
          di Bologna, Via Ranzani 1, 40127 Bologna, Italy \and IRA-INAF -
          via Gobetti, 101, 40129, Bologna, Italy}

	\abstract {}
{We compute photometric redshifts in the fourth public release
of the Canada-France-Hawaii Telescope Legacy Survey. This
unique multi-colour catalogue comprises $u^*,g',r',i',z'$
photometry in four deep fields of $1\deg^2$ each and
$35~\deg^2$ distributed over three Wide fields.}
{Our photometric redshifts, computed using a
  template-fitting method,  are calibrated with a
large catalogue of 16,983 high-quality spectroscopic redshifts
from several spectroscopic surveys.}
 {From the comparison with
   spectroscopic redshifts, we find a photometric redshift
dispersion of 0.028 and an outlier rate of 3.5\% in the Deep
field at $i'_{AB} < 24$. In the Wide fields, we find a
dispersion of 0.036 and outlier rate of 2.8\% at 
$i'_{AB} < 22.5$. Beyond $i'_{AB} = 22.5$ in the Wide fields the number of
outliers rises from 5\% to 10\% at $i'_{AB}<23$ and $i'_{AB}<24$
respectively.  For the Wide sample, we find the systematic redshift bias 
  keeps below 1\% to  $i'_{AB} < 22.5$, whereas we find no significant bias in the Deep 
  field.
    We investigated the effect of tile-to-tile
photometric variations and demonstrate that the accuracy of our
photometric redshifts is reduced by at most 21\%.
 We separate stars from galaxies using both
the size and colour information. Applying our star-galaxy classifier
we reduce the contamination by stars in our catalogues from 50\% to
8\% at $i'_{AB} < 22.5$ in our field with the highest stellar density
while keeping a complete galaxy sample. Our CFHTLS T0004 photometric
redshifts are distributed to the community. Our
release include $592,891$ ($i'_{AB}<22.5$) and $244,701$
($i'_{AB}<24$) reliable galaxy photometric redshifts in the
Wide and Deep fields, respectively.} 
{}


	\maketitle

\section{Introduction}

It is now evident that the exploration of large scale
structure and the high-redshift Universe with the Canada
France Hawaii Telescope Legacy Survey (CFHTLS) requires
precise magnitudes and redshifts for millions of sources
(\citeNP{2007MNRAS.381..702B}, \citeNP{2007ApJ...669...21P},
\citeNP{2008A&A...479..321M}, \citeNP{2008A&A...479....9F},
\citeNP{2008MNRAS.385..695B},
\citeNP{2008arXiv0810.5129K},\citeNP{2008arXiv0810.0555T} ).

To date, only ``photometric redshift'' techniques can provide
(with a comparatively modest expenditure of telescope time)
redshifts of enormous numbers of galaxies with sufficient
precision ($\sim$ 1-5\%) to the faintest limiting magnitudes
of the CFHTLS cosmological surveys.  The construction of well
defined, accurate and reliable photometric redshift catalogues
is therefore an indispensable task following photometric
catalogue production.

Several photometric redshift codes are now publicly available and have been
applied with reasonable success to many photometric catalogues of
galaxies (see \cite{2008A&A...480..703H}, and references
therein).  Some recent
photometric redshift studies, like the COMBO-17 survey
\citep{2003A&A...401...73W},
 CFHTLS (\citeNP{2006A&A...457..841I}, hereafter I06), 
SWIRE \citep{2008MNRAS.386..697R}, or COSMOS
(\citeNP{2007ApJS..172..117M}, \citeNP{2008arXiv0809.2101I}), 
contain up to 1,000,000
galaxies as faint as $i\sim 25$.  In particular, the ``{\it Le Phare}''
photometric redshift code (\citeNP{1999MNRAS.310..540A},
\citeNP{2002MNRAS.329..355A}, I06)
 has shown to be well adapted
for joint photometric and spectroscopic surveys like the CFHTLS.
I06 used {\it Le Phare} with the CFHTLS-Deep photometric catalogues and
VVDS spectroscopic redshifts \citep{2005A&A...439..845L} 
to calculate photometric redshifts with
an accuracy of $\sim 3\%$ at $i<24$.
The redshift distribution of sources has been used to calibrate the
absolute gravitational shear signal presented in \cite{2008A&A...479....9F}.

By combining photometric and
spectroscopic measurements they were able to derive a set of optimised
Spectral Energy Distributions (SEDs). By computing the mean difference
between magnitudes in each filter of objects with known redshifts
and those derived from these optimised SEDs one can ``tune'' the CFHTLS
photometry. This method improves photometric redshifts for all
galaxies, even those for which there are no spectroscopic
measurements.

More spectra and deeper data can lead to a better
calibration. With the advent of new CFHTLS photometric data and much
larger spectroscopic catalogues we can now extend the application of
 {\it Le Phare} to more complex surveys composed of many MegaCam fields
like the CFHTLS-Wide.

Most photometric redshift studies either explore shallow very wide fields
covering thousand of deg$^2$ (see \citeNP{2008ApJ...683...12B} 
for the SDSS and references therein) where visible photometric 
data are sufficient to sample the whole redshift range of galaxies, 
or those which focus on deep beams of few deg$^2$
 (like the CFHTLS Deep or COSMOS) comprising
both visible and near infrared photometric data.  To date, the only
moderately deep visible survey currently available for photometric redshift studies
covering a large area is the CFHTLS Wide.  

The aim of this work is the calibration and production of a flux
limited photometric redshift catalogues based on the CFHTLS Deep and Wide
surveys.  When completed, the CFHTLS Wide will cover $170~\deg^2$ in
five optical filters spread over four separate regions of the sky
which also contain subsets of several deep spectroscopic surveys.
Ultimately, a catalogue of more than ten million galaxies down to
$i'_{AB} = 24.5$, with reliable five-band photometry will be available
for photometric redshift measurement.  In this paper, we use {\it Le Phare} to
compute photometric redshift for the ``T0004'' CFHTLS release.  In addition to the
four CFHTLS Deep fields that were already analysed in previous works,
T0004 includes a new large Wide catalogue, covering $35~\deg^2$ in five
bands $u^*,g',r',i',z'$ in three independent fields.

To construct our photometric redshift catalogues we follow the
method described in I06, but with some important changes.  To
calibrate the photometric redshifts new spectroscopic data are
added to the $VVDS$ ``deep'' sample
\citep{2005A&A...439..845L}.  It includes the $DEEP2$ redshift
survey (\citeNP{2003SPIE.4834..161D} and
  \citeNP{2007ApJ...660L...1D}), the new $VVDS$ ``wide''
  sample \citep{2008A&A...486..683G} and the ``zCOSMOS''
  sample \citep{2007ApJS..172...70L}.  We first compute the
  new Deep field photometric redshift are then computed first
  and compared to the previous I06 Deep study. Once validated,
  we extend the analysis further to Wide data.  We then
  challenge the photometric redshift catalogues against the
  new spectroscopic redshift samples to assess the robustness
  of the calibration, to derive a detailed error budget for
  each photometric redshift and to estimate the sensitivity of
  the method to parameters like redshift or magnitude.
Finally, we derive the redshift distribution and its field-to-field
variance (cosmic variance).

The paper is organised as follows. In Section \ref{photoData}
 we describe the data used in this analysis.
 We show how the ``T0004'' photometry is used
and we explain how the spectroscopic redshifts are evaluated in the
\emph{VVDS F02} and \emph{F22} fields and the \emph{DEEP2} survey.
  The next Section describes the principles of {\it Le Phare}, and discusses
  recent improvements for the photometric redshift calibration using new
  spectroscopic redshift samples. Sections \ref{SecStargal} and
  \ref{SecAnalysis} focus on the photometric redshifts of the Deep and
  Wide CFHTLS sample.  We then discuss the photometric redshift
  accuracy and their robustness and reliability.  The redshift
  distribution and its variance are discussed in the last section.

Throughout the paper, we use a flat lambda cosmology
($\Omega_m=0.3$, $\Omega_\Lambda=0.7$) and we define
$h=H_{\rm0}/100$, with $H_{\rm0}=70$~km~s$^{-1}$~Mpc$^{-1}$.
 Magnitudes are given in the
AB system. Photometric and spectroscopic redshifts are denoted by $z_p$
and $z_s$, respectively, and $\Delta z$ represents $z_p-z_s$.


 \section{Data}
 \label{SecData}

 \subsection{Photometric data}
 \label{photoData}

 \begin{figure*}
   \centering
   \includegraphics[width=9cm]{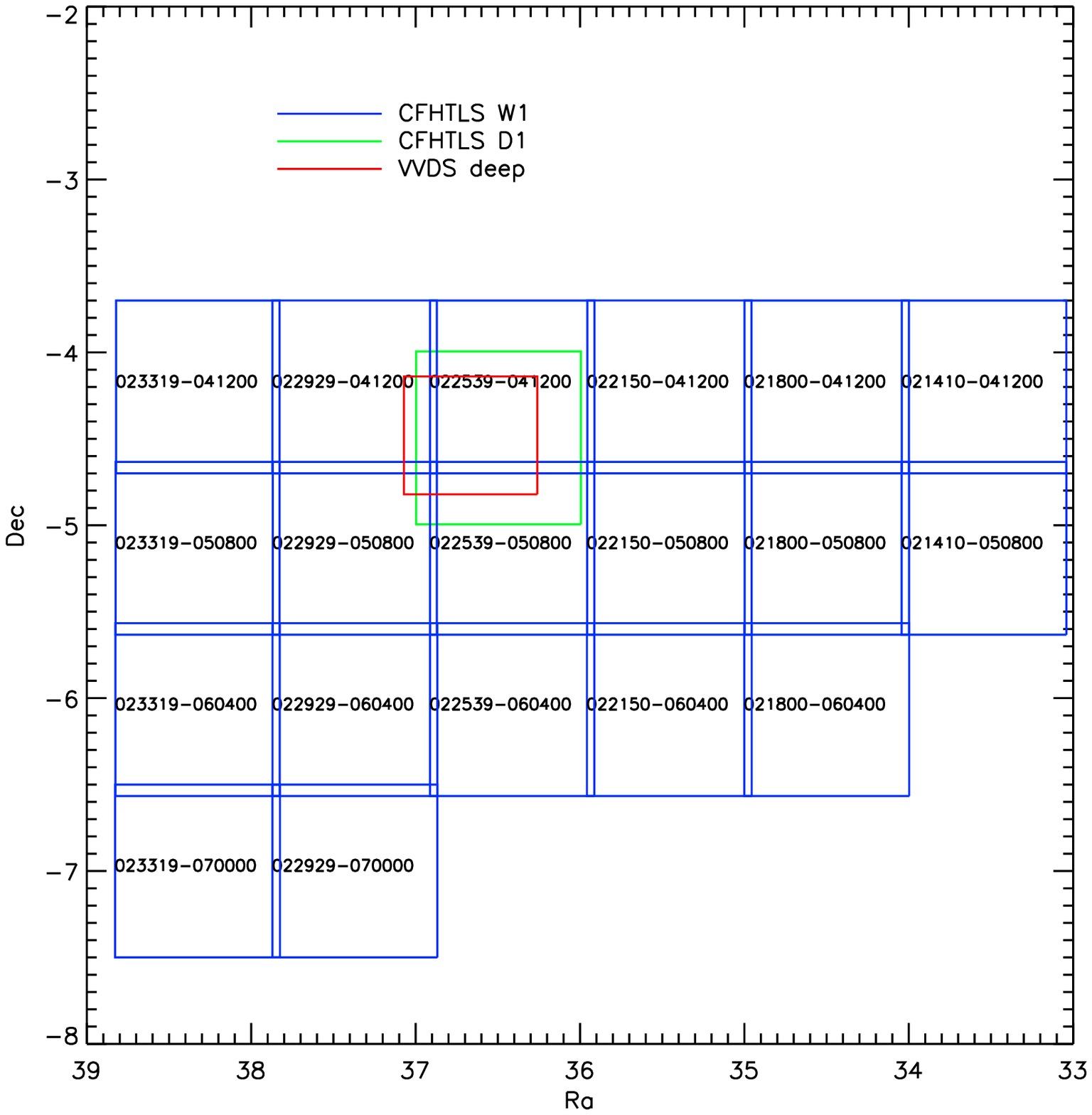}
   \includegraphics[width=9cm]{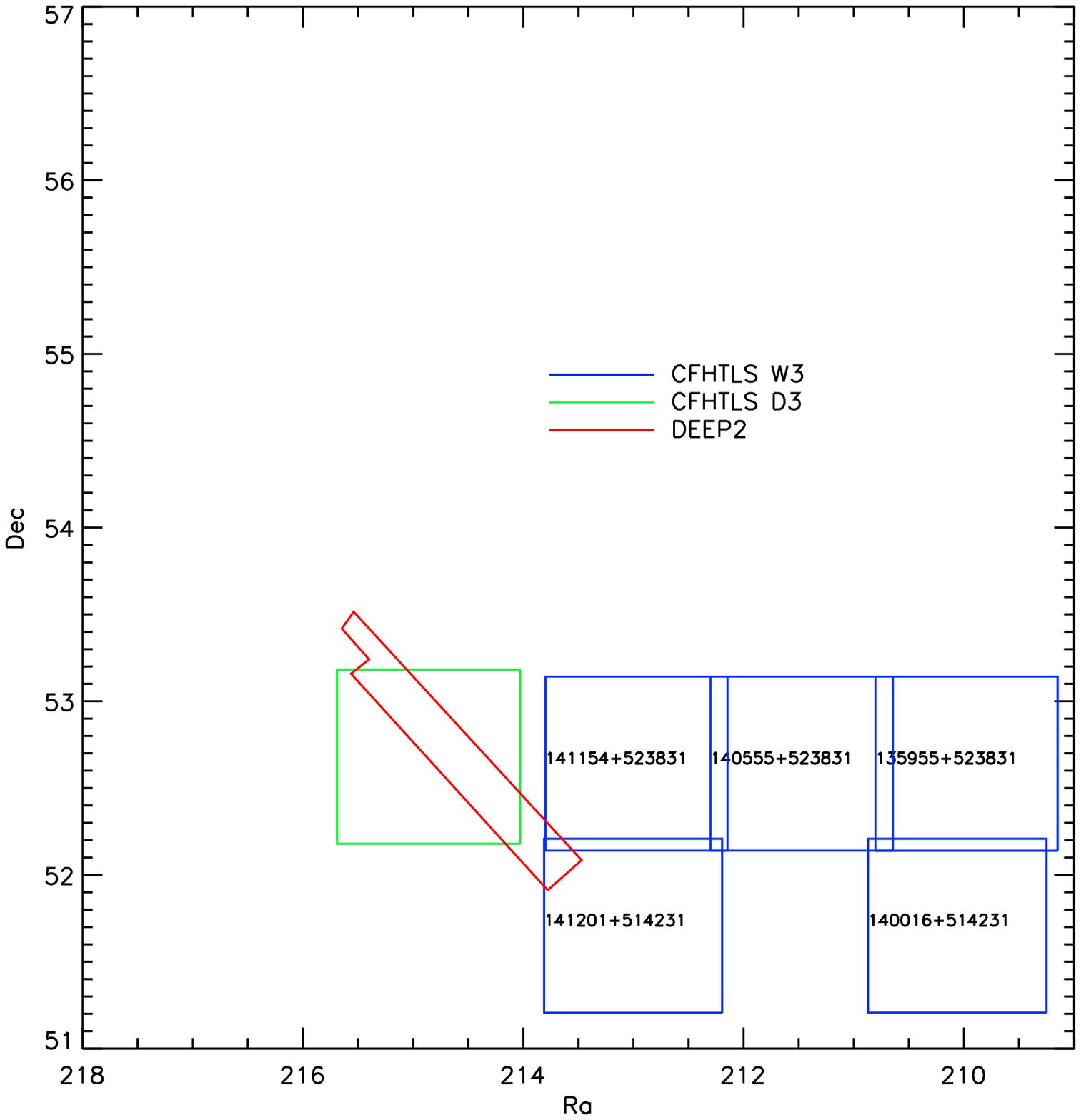}
   \includegraphics[width=9cm]{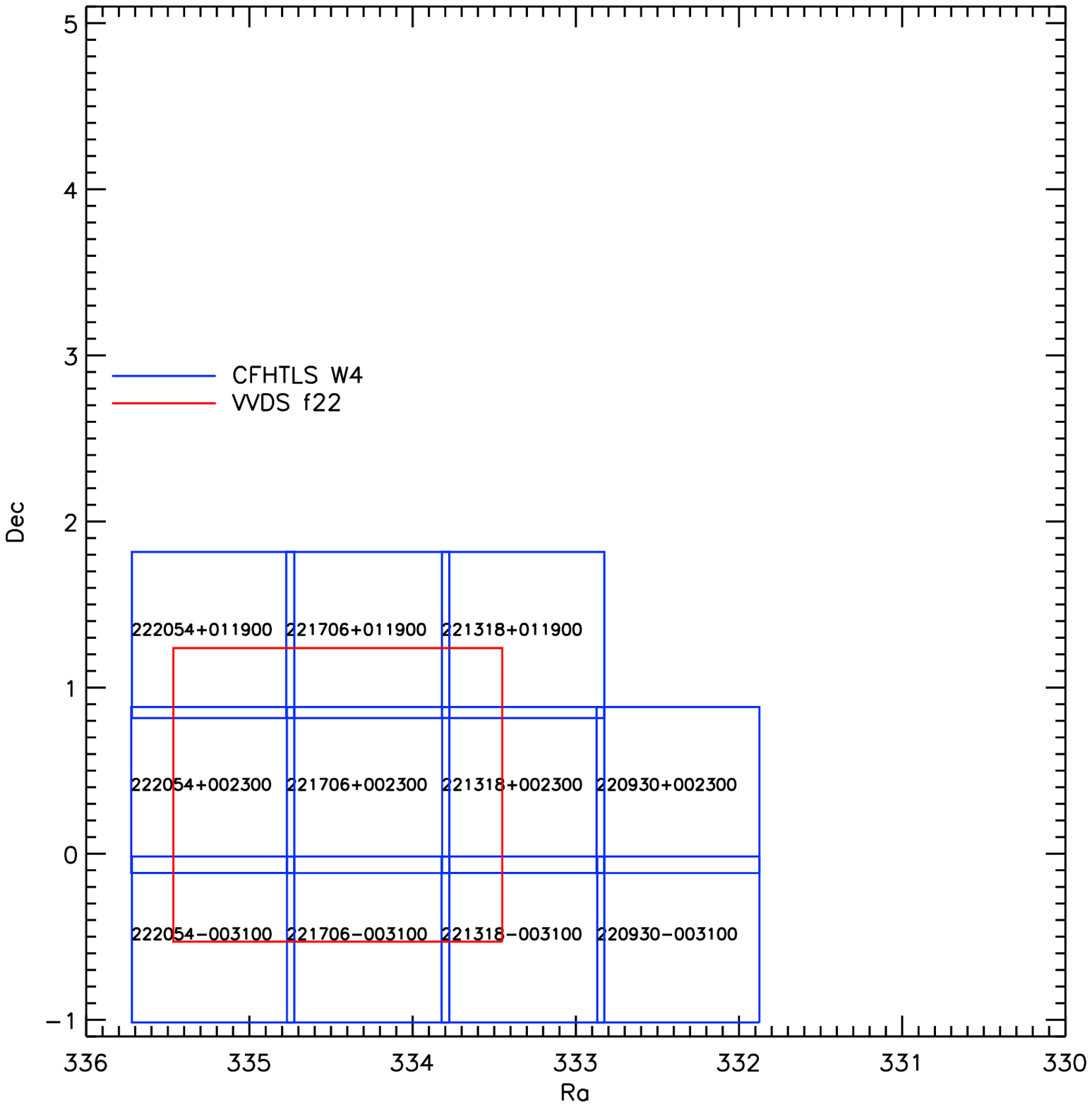}
   \includegraphics[width=9cm]{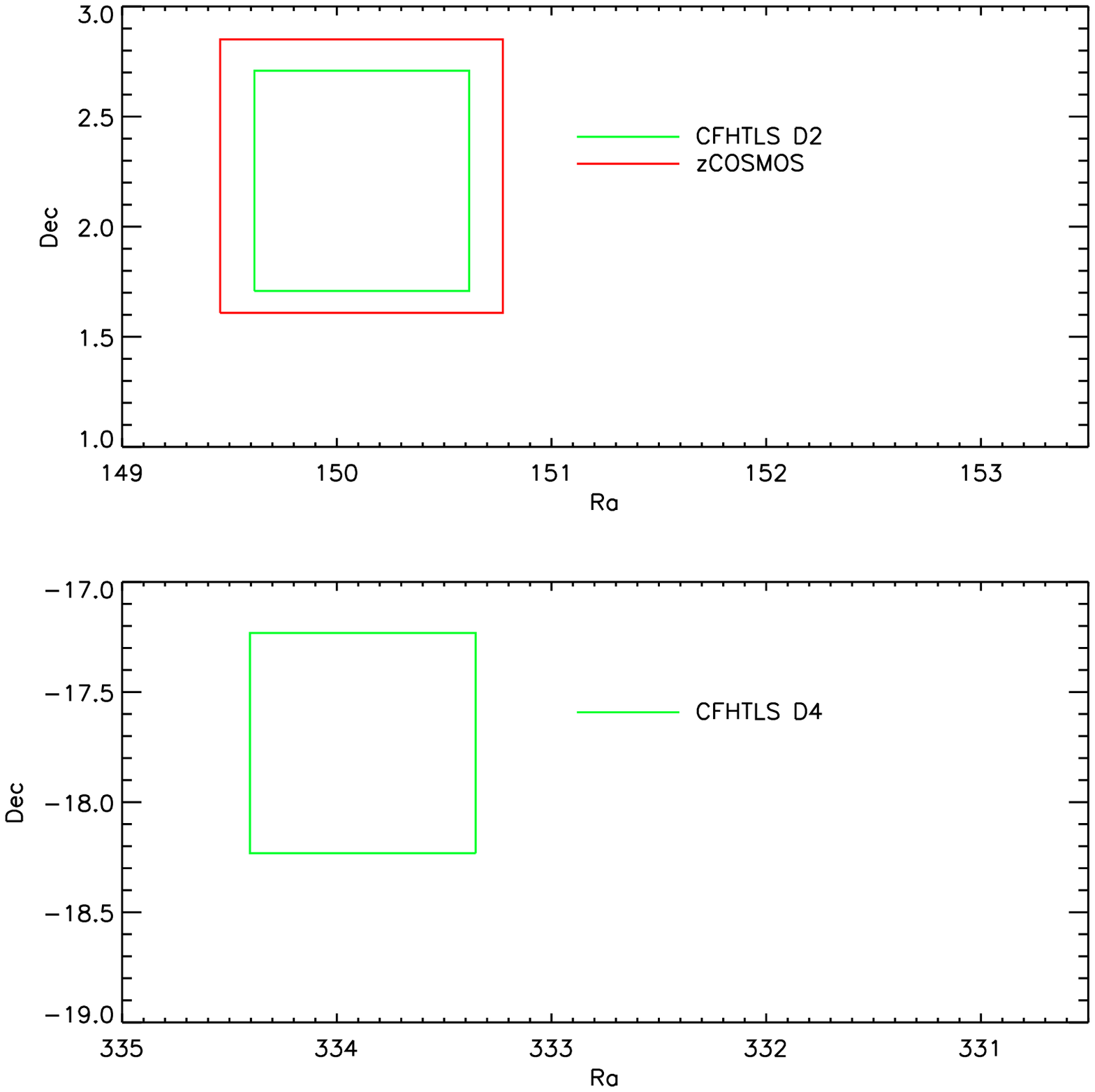}
   \caption{Sky coverage of the CFHTLS Deep fields (D1,D2,D3,D4) and Wide 
 fields (W1, W3 and W4) overlapped 
 with spectroscopic surveys: VVDS F02, F22 and DEEP2.}
   \label{CFHTLS_maps}
 \end{figure*}

 \subsubsection{The CFHTLS}

 The CFHTLS, a joint Canadian and French survey, is composed of three
 imaging surveys of different depths, shape and sky coverage: the
 CFHTLS Deep, the CFHTLS Wide and the CFHTLS Very Wide. The survey
 started in 2003 and will last for 450 nights until finishing in
 January 2009. When completed, the CFHTLS Deep will cover 4$\times$1
 ~deg$^2$ at a limiting magnitude of $i'_{AB}=27.5$ (point source,
 S/N=5, 1.15" aperture, seeing $0.8\arcsec$) and the Wide will cover
 $170~\deg^2$ over 4 fields with a limiting magnitude of $i'_{AB}=25.5$.
  
 The CFHTLS is conducted at the 3.6m Canada-France-Hawaii Telescope,
 equipped with the new MegaPrime prime-focus assembly and the MegaCam
 wide field camera \citep{2000SPIE.4008..657B}.  MegaCam is composed
 of 36 2080$\times$4644 pixel EEV CCDs. MegaCam has a pixel scale of
 $0.186\arcsec$/pixel and therefore covers the entire 0.96$\times$0.98
 deg$^2$ unvingnetted MegaPrime focal plane. The CCD assembly leaves
 small gaps between each detector and two blind lanes of $62\arcsec$
 width along the X-axis (E-W direction).  For this reason each field is
 observed in a dithering pattern to fill the gaps between the CCDs
 and help in removing systematic CCD features.

 MegaPrime is equipped of $u^*,g',r',i',z'$ broad band filters that
 provide continuous coverage over the whole spectral range
 $3500\rm{\AA} < \lambda< 9400\rm{\AA}$.  Only the CFHTLS Deep and Wide
 surveys which are observed in the five MegaCam $u^*,g',r',i',z'$ broad
 band filters are suitable for photometric redshift computation. 

  The positions of the four deep fields and the three wide fields 
   used in this work are listed in table \ref{datadiff}.
 Each Wide field has a different geometry and  sky coverage:  W1
 ($8^\circ \times 9^\circ$ ``tiles'' of $1^\circ \times 1^\circ$ each),
  W3 ($7^\circ \times 7^\circ$), W4  (a $7^\circ \times 3.5^\circ$ strip along the 
   South-East/North-West direction), and
 is composed of a contiguous mosaic of MegaCam fields. The
 identification name and position of all fields can be found on the
 {\sc Terapix} web
 pages\footnote{{\tt http://terapix.iap.fr/cplt/oldSite/\\
 Descart/summarycfhtlswide.html}}.

 W1 and W3 overlap with D1 and D3, respectively,  while W4  
 located a few degrees apart D4.  Each MegaCam ``tile''
 (which is an elementary $1^\circ \times 1^\circ$ MegaCam area paving the sky)
 composing a Wide field overlaps with its four neighbouring fields.
 The width of overlap regions is around $2\arcmin$ in both directions. 
 The overlap between pointings can be used for the astrometric and
 photometric calibrations.

 \subsubsection{The T0004 release}

 The data used in this work are part of the CFHTLS T0004 release
 produced at the {\sc Terapix} data center. The release consists of all
 CFHTLS Deep and Wide images observed from June 1st, 2003 to October
 24th, 2006.

 For the photometric redshift study, we consider only fields with observations
 in all five photometric bands.  Our parent sample is therefore
 reduced to $u^*,g',r',i',z'$ images, namely
 \begin{itemize}
 \item   the 4$\times$1~deg$^2$ Deep fields with longer integration times
   than the previous T0003 release ($\sim 0.2 mag$ deeper) used in I06, and
 \item the $35~\deg^2$ of the Wide field 
 that are completed in all filters and distributed as follows: 
   $19~\deg^2$ in W1, $5~\deg^2$ in W3 and $11~\deg^2$ in W4.
 \end{itemize}
 The  sky coverage of the complete parent data set is is shown in 
  figure \ref{CFHTLS_maps}.

 For each deep field, {\sc Terapix} produces two stacks per filter. One
 comprises the 25\% best seeing images and the other consists of the
 85\% best seeing images.  Since both types of deep stack have seeing
 better than one arc-second and we are primarily interested in very
 deep galaxy catalogues,we used the deepest 85\% to compute the
 photometric redshifts.

 Table \ref{datadiff} summarises the data used for this study and also
 in \cite{2006A&A...457..841I}. 

 \begin{table*}
   \centering
  \caption{Summary of the data used in this study and in I06.}
  \label{datadiff}

  \begin{tabular}{c c c c c}
      \multicolumn{5}{c}{Ilbert et al (2006). CFHTLS release T0003}\\
       \hline
       \hline
       Field & Center Position & Area (unmasked) & 80\% completeness limit ($i'_{AB}$)& Spectroscopic data\\
       \hline
       Deep - D1& $02^h25^m59^s ,\ -04^\circ 29'40''$ & $0.77~\deg^2$ & 25.1 & VVDS Deep\\
       Deep - D2& $10^h00^m28^s ,\ +02^\circ 12'30''$ & $0.69~\deg^2$ & 24.9 & \emph{no}\\
       Deep - D3& $14^h19^m27^s ,\ +52^\circ 40'56''$ & $0.83~\deg^2$ & 25.7 & \emph{no}\\
       Deep - D4& $22^h15^m31^s ,\ -17^\circ 43'56''$ & $0.82~\deg^2$ & 25.3 & \emph{no}\\
       \hline
    \end{tabular}

    \begin{tabular}{c c c c c}
      \multicolumn{5}{c}{}\\
      \multicolumn{5}{c}{This study. CFHTLS release T0004.}\\
       \hline
       \hline
       Field & Center Position & Area (unmasked) & 80\% completeness limit ($i'_{AB}$)& Spectroscopic data\\
       \hline
       Deep - D1 &  $02^h25^m59^s ,\ -04^\circ 29'40''$ & $0.78~\deg^2$ & 25.3 & VVDS Deep\\
       Deep - D2 & $10^h00^m28^s ,\ +02^\circ 12'30''$ & $0.80~\deg^2$ & 25.1 & zCOSMOS\\
       Deep - D3 & $14^h19^m27^s ,\ +52^\circ 40'56''$ & $0.83~\deg^2$ & 25.9 & DEEP2\\
       Deep - D4 & $22^h15^m31^s ,\ -17^\circ 43'56''$ & $0.82~\deg^2$ & 25.5 & \emph{no}\\
       Wide - W1 & $02^h18^m00^s ,\ -07^\circ 00'00''$ &19 (15.73) deg$^2$& $> 24.3$ & VVDS Deep\\
       Wide - W3 & $14^h17^m54^s ,\ +54^\circ 30'31''$ & 5 (4.05) deg$^2$ & $> 24.4$ & DEEP2\\
       Wide - W4 & $22^h13^m18^s ,\ +01^\circ 19'00''$ &11 (8.87) deg$^2$ & $> 24.4$ & VVDS F22\\
       \hline
    \end{tabular}
 \end{table*}

 \subsubsection{Production of T0004 catalogues}
 \label{SecCats}

 Full details of the processing and content of the T0004 release are
 described in the {\sc Terapix} pages\footnote{{\tt
     http://terapix.iap.fr/rubrique.php?id\_rubrique=241 }}.  Further
 details on calibration and stack and catalogue production can also be
 found also in \cite{2008A&A...479..321M} and in the Mellier et al (2005)
 explanatory document\footnote{{\tt
     http://terapix.iap.fr/cplt/oldSite/Descart/\\
 NewterapixdocT0002.pdf}}.
 In what follows we summarise briefly how the input catalogues used for
 the photometric redshifts have been produced.

\begin{figure*}
   \centering
   \includegraphics[width=15cm]{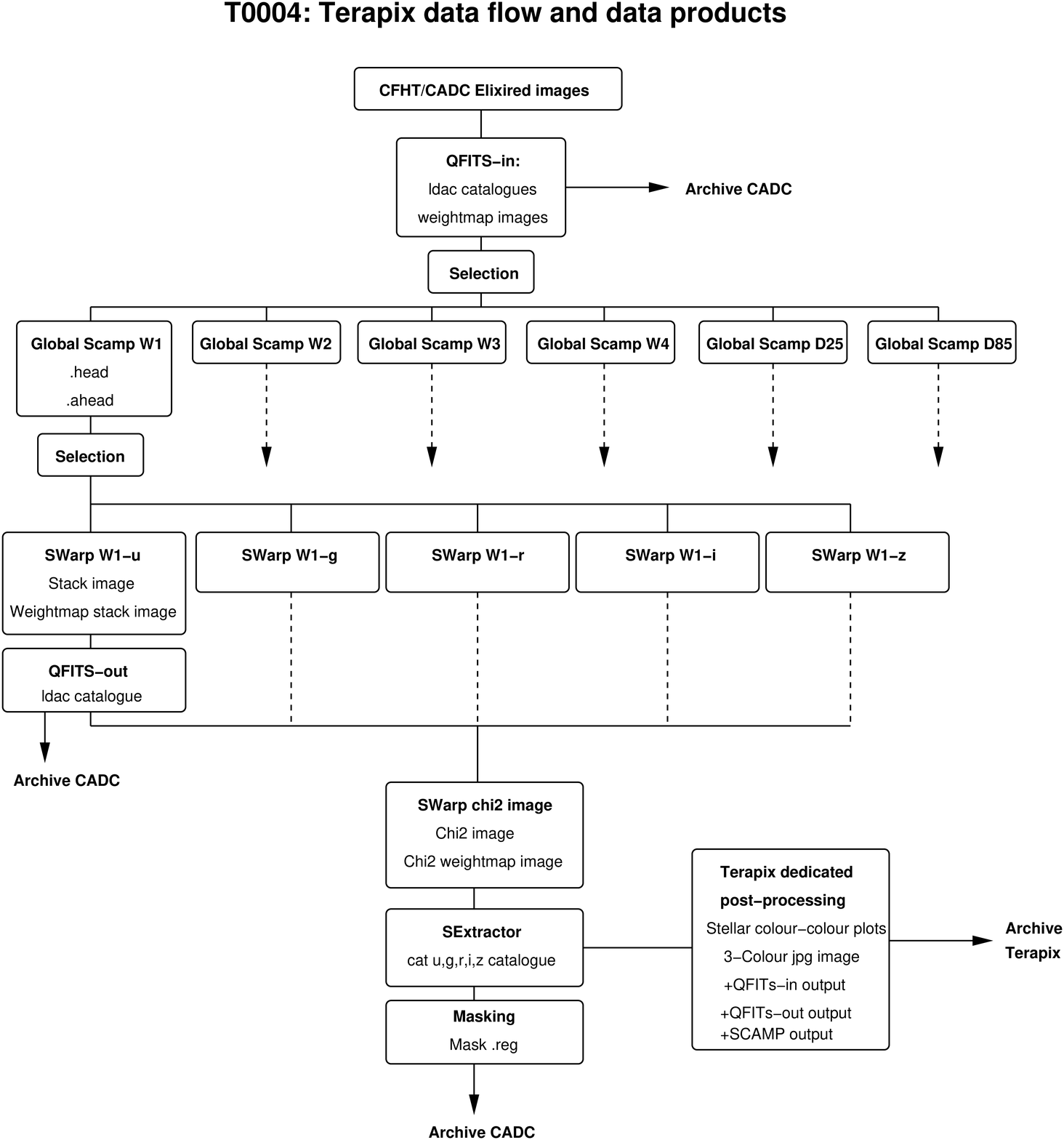}
  \caption{Details of the data processing performed by {\sc Terapix}.}
   \label{dataFlow}
 \end{figure*}

 Pre-processing of raw images (masking bad pixels, removing the
 overscan, subtracting the dark and the bias, flat fielding and
 illumination correction) is performed by the Elixir pipeline at CFHT
 \citep{2004PASP..116..449M}.  The data are then transferred to {\sc
   Terapix} to produce the stacked images and the final
 catalogues. Figure \ref{dataFlow} shows the flowchart of the T0004
 release production.  Each image is first examined
  by the {\sc  Terapix} {\tt QualityFITS} image quality control tool. 
   During the {\tt QualityFITS} step (hereafter QFITS-in) a 
  set of quality assessements is produced,   
 all individual input images are inspected and evaluated and a
 weightmap image as well as an input catalogue are produced.  This
 catalogue will be used later for the astrometric calibration.  It is
 produced using {\tt SExtractor} \citep{1996A&AS..117..393B} with 
 appropriate settings for saturation levels to eliminate any
  spurious objects which could lead to an 
 incorrect computation of the flux re-scaling
 during the astrometric matching process.  A QFITS-in web page
 summarises the inspection and is used as an ID-card of each image.
 All QFITS-ed images are then graded ``A'', ``B'' or ``C'', after a
 visual inspection of each web-page.
  
 With the QFITS-in information available, {\sc Terapix} selected the
 T0004 image sample by applying the following criteria:
 \begin{itemize}
 \item {\sc Terapix} class: A or B grades  (all images within the CFHTLS specifications;
  grade B are images that  may show minor problems or with specifications very close to 
   limits), 
 \item Exposure time higher than 60 s.,
 \item Seeing less than $1.3\arcsec$, except for $u*$ (less than $1.4\arcsec$)
 \item Airmass less than 1.7
 \item Skyprobe value lower than 2.0 magnitudes (security limit)
 \end{itemize}

 Rejected images will no longer be considered. {\sc Terapix} then uses
 the QFITS-in catalogues of the remaining images to derive the
 astrometric and photometric calibrations for the release.

 The astrometric solution is computed using 
 {\tt SCAMP}\footnote{{\tt http://terapix.iap.fr/soft/scamp}}
 \citep{2006ASPC..351..112B}.
 {\tt SCAMP} first examines all image headers and then split the exposures
 into a series of astrometric contexts. Each context separates blocks
 of observing epochs where the instrument focal plane is in a fixed and
 (almost) stable configuration.  The detections and positions of
 astrometric sources on MegaCam images are derived by the
 cross-identification of sources of the QFITS-in catalogue with the
 2MASS astrometric reference catalogue.  For the T0004 release, the
 source matching exploration radius is set to $2\arcsec$ for all Deep
 and Wide fields.  A polynomial distortion model is then derived by
 minimising a weighted quadratic sum of differences in positions
 between the 2MASS and the QFITS-in matched sources, and, internally,
 between different QFITS-in catalogues with overlap regions. Using this
 information {\tt SCAMP} can then compute external and internal errors.

 For the Wide data, the astrometric solution is performed only once for
 each Wide field by using all image catalogues simultaneously,
 regardless the filter and the epoch.  All images of a given Wide field
 are then calibrated globally and in a homogeneous way. For the 85\%
 Deep data, the solution must be computed differently. The number of
 observing runs produces too many astrometric contexts which results in
 a matrix that cannot be inverted with the current {\sc Terapix}
 computing resources.  The Deep field images are then divided into five
 sub-samples, one for each filter.  In order to improve the consistency
 and the robustness of the solutions found for each filter, a common
 set of extra images surrounding the field and shifted by about
 $30\arcmin$ with respect to Deep centre positions is added to each
 sub-sample. The consistency of each solution has been checked
 afterwards.

 In both cases, the Deep and the Wide calibrations worked well.  For
 T0004, the rms internal error of Wide and Deep astrometric solution is
 $0.017''$ and the mean rms external error is $0.21''$ in both
 directions.  After inspection, if acceptable, the astrometric solution
 is then written in the {\tt .head } file attached to each image.

 {\tt SCAMP} is also used to derive a relative photometric 
   calibration by minimizing a weighted, quadratic sum of magnitude 
   differences on the overlap regions between images.
 Images flagged as
 ``photometric'' by CFHT are used as anchor points and their CFHT
 magnitude zero point is written in the {\tt .ahead} file.  For the
 photometric calibration, {\tt SCAMP} minimizes the quadratic sum of
 magnitudes using the overlap region between images.  The flux of
 non-photometric images are then re-scaled accordingly.  Typical
 re-scaling amplitudes in T0004 are $\pm 0.02$ magnitude.  However, for
 some highly non-photometric images it may reach $\pm 0.50$
 magnitudes. For these extreme outliers, the re-scaling itself becomes
 more and more unreliable and the observed error on the rescaling value
 can be significantly higher than the typical $\pm 0.02$ value.  For
 this reason all images with rescaling greater than 0.15 magnitudes
 were removed from the input list.

 After this post-{\tt SCAMP} selection process images are divided into filter
 and tile positions and images are then resampled and co-added with
 {\tt SWarp}\footnote{{\tt http://terapix.iap.fr/soft/swarp}}
 \citep{2002ASPC..281..228B}.
 An ASCII polygon mask is produced at each tile position which
 can be visualised using the ``DS9'' image display software.  Each Wide
 stack is only composed of images centred at the tile position and the
 overlapping pixels of nearest neighbour tiles are not used.  For each 
  stack a 1 deg$\times$1' strip masks the boudaries of fields in order to 
   drop regions with highly elongated PSF and low signal-to-noise.   

 T0004 stacks are produced by a median filter and a Lanczos-3 interpolation
 kernel.  All stacks  have $19354\times19354$ pixels of $0.186\arcsec$
  ({\it i.e.} exactly 1$\times$1 deg$^2$) 
   and have a magnitude zero point set to 30. For the Wide survey, a
 stack is produced at each centre position listed in the {\sc Terapix}
 web page\footnote{{\tt
   http://terapix.iap.fr/cplt/oldSite/Descart/\\
 summarycfhtlswide.html}}.
 A {\tt SExtractor} catalogue is then produced for each stack that is
 used to run {\tt QualityFITS} and produce a QFITS-out ``ID-card'' of
 each stack. QFITS-out  as well as all quality control data of all stacks 
   are publicly available at {\tt http://terapix.iap.fr/article.php?id\_article=724}.
   They have been used to carry out  visual inspections
   and the validation of the T0004 release.

 If $g'-$, $r'-$ and $i'-$ band data are available, {\sc Terapix}
 automatically produces a ``chi2 image'' based on these three
 stacks \citep{1999AJ....117...68S}.
 Then {\tt SExtractor} is run in dual-image mode on both the
 chi2 image and each stacked image ($r'$, $g'$, $i'$ , as well as $u^*$ and
 $z'$, if any).  All catalogues contain parameter values for all
 quantities listed in {\tt
   http://terapix.iap.fr/article.php?id\_article=628}.  
 In this study, we use the merged ($u$, $g'$, $r'$, $i'$, $z'$,)
  catalogues produced by {\sc Terapix}
 that include a limited number of parameters (only MAG\_AUTO
 magnitudes, for example) plus the $E(B-V)$ value at each source
 position derived from dust map images \cite{1998ApJ...500..525S}.

 After removing the masked areas, the effective field-of-view is
 0.78, 0.80, 0.83 and 0.82$~\deg^2$ for the D1, D2, D3 and D4,
 respectively and 15.73, 4.05 and 8.87 for the W1, W3 and W4, 
 respectively.
 Effective areas for each individual tile in the Wide fields 
 are given in the {\tt
   http://terapix.iap.fr/cplt/table\_syn\_T0004.html} table.
  
 All catalogues include flux with uncertainties.  However, in this
 work we follow I06 and we scaled the {\tt SExtractor} flux error
 estimates by a factor of 1.5 to compensate for the slight noise
 correlation introduced by the image re-sampling during image stacking.

 Finally, a series of post-processing analysis is then carried out in
 order to make quality assessments for each stack and, globally, for
 the whole release.  The QFITS-in, {\tt SCAMP}, QFITS-out output files are
 part of the quality assessment data. More specific control files are
 also created using the merged ($u^*$, $g'$, $r'$ , $i'$, $z'$) catalogues,
 like stellar colour-colour plots and comparisons between the
 T0004 stellar photometry and the Sloan Digital Sky Survey
 (SDSS)\footnote{{\tt http://www.sdss.org/data}}. This quality control data
 are available at {\tt
   http://terapix.iap.fr/cplt/table\_syn\_T0004.html}.

 The comparison between CFHTLS T0004 and SDSS photometry is made using
 stars in common in the magnitude range $17<AB<21.1$. The results are
 displayed at {\tt http://terapix.iap.fr/article.php?id\_article=713}.
 The mean offset between CFHTLS and SDSS, $\Delta_{{\rm CFHTLS}-{\rm SDSS}}$,
     varies from stack to stack in the range  
    $-0.03<\Delta_{{\rm CFHTLS}-{\rm SDSS}}<0.03$ 
   magnitude. 
 A calibration problem was noticed on
 $u^*$-band images obtained during the period from March 2006 until
 October 24th, 2006 showing a mean offset of 0.21 mag. with respect to SDSS.
 Following the {\sc Terapix} table all $u^*$-band values obtained during
 this period were corrected for the offset derived from the SDSS.
 In the few CFHTLS fields with no overlap with SDSS, we applied a correction 
 of 0.2 mag.

 \subsubsection{Field-to-field photometric stability}

 The {\sc Terapix} pipeline uses common sources in overlap regions
 between each contiguous MegaCam field to derive a photometric
 magnitude zero-point correction after a field-to-field flux
 rescaling. It is computed by {\tt SCAMP} to produce a Wide survey as uniform
 as possible over all W1, W2, W3 and W4 fields. The recipe relies on
 the CFHT magnitude zero points and photometric flags written in the
 FITS header and in the observing logs, respectively. It is possible
 that some series of marginally photometric images produce a poor flux
 rescaling solution.  We minimised this risk by removing photometric
 outlier images from the T0004 sample during the post-{\tt SCAMP} selection
 process (See Section \ref{SecCats} ). However, some fields may still show
 residuals, in particular at the borders of the survey where the
 field-to-field rescaling cannot be done using the four MegaCam
 boundaries of each stack.  The comparison between CFHTLS and SDSS
 photometry confirms the uniformity is good to within $\pm 0.03$
 mag. over the whole fields, but there are a few fields that are off by
 as much as $\sim$0.10 mag.
  
 This problem can be partially overcome by  {\it Le Phare}. The code
 uses spectroscopic redshifts to correct the magnitude offsets
 (hereafter systematic offsets) in the CFHTLS wide fields that overlap
 with spectroscopic fields. The estimate of the field-to-field
 magnitude variations and how it can affect the photometric redshift
 accuracy is discussed in a forthcoming section.

 \subsection{Spectroscopic data}
 \label{spectro}

 We used spectroscopic redshifts (spectro-$z$'s) from the VVDS survey
 for the D1 and W1 fields.  The VVDS data were obtained with the
 VIsible Multi-Object Spectrograph
 (VIMOS\footnote{{\tt http://www.eso.org/sci/facilities/paranal\\
 /instruments/vimos/overview.html}})
 installed at the ESO-VLT. The deep spectroscopic sample VVDS-0226-04
 was selected in the magnitude range $17.5 \le I_{AB} \le 24.0$ and has
 a median redshift of $\sim 0.67$ \citep{2005A&A...439..877L}. Keeping
 only sources with a confidence level in the redshift measurement
 greater than 97\% (class 3 and 4), our parent VVDS-0226-04
 spectroscopic sample is composed of 3880 galaxies. We matched 3276
 galaxies in the D1 field and 3356 galaxies in the W1 field with
 $i'_{AB} \le 24$.

 We also used the public spectroscopic redshifts from the zCOSMOS
 survey \citep{2007ApJS..172...70L} which overlaps with the D2
 field. The zCOSMOS-bright spectra were obtained with the
 VIMOS spectrograph and were selected at $I_{AB} \le 22.5$. 3915
 spectro-$z$'s with a confidence level greater than 99\% are usable for the
 D2 field.

 In addition, our spectroscopic calibration sample includes 5936
 spectro-$z$'s from the third data release of the DEEP2 survey
 (\citeNP{2003SPIE.4834..161D} and \citeNP{2007ApJ...660L...1D}).  This
 sample overlaps with the D3 field and covers a small area in the W3
 field (see Fig.\ref{CFHTLS_maps}).  Spectra were obtained by the
 spectrograph
 DEIMOS\footnote{{\tt http://loen.ucolick.org//Deimos/deimos.html}} mounted
 on the Keck II telescope. As for the VVDS sample we used the most
 secure spectro-$z$'s, with quality flag 3 or 4, corresponding to
 confidence level greater than 95\%. The DEEP2 data taken in the
 Extended Groth Strip region have been preselected in the range $18.5
 \le R_{AB} \le 24.1$, with a selection based on color and surface
 brightness only aimed at maximizing the number of galaxies over the
 number of stars (in contrast with the rest of the DEEP2 survey where
 selection criteria to target only higher-redshift galaxies are
 used). A smaller subsample of 310 spectro-$z$'s are in the W3 field.

 Finally, the VVDS-Wide survey (VVDS-F22 field,
 \cite{2008A&A...486..683G}) overlaps with the W4 field. The VVDS-Wide
 covers a $4~\deg^2$ contiguous area in the W4 field. The spectro-$z$
 catalogue includes 11\,228 galaxies, 167 type I AGNs, and 6\,748
 stars. Using the most secure spectro-$z$ (confidence level greater
 than 97\%, class 3 or 4) our final sample comprises 3\,854 galaxies in
 the magnitude range $17.5 \le I_{AB} \le 22.5$.

 D4 is the only field without any spectroscopic coverage.


 \section{Photometric redshifts for the CFHTLS-T0004 catalogue}

 \subsection{Photometric redshift computation with {\it Le Phare}}

 Photometric redshifts were computed using a standard $\chi^2$
 template-fitting procedure and calibrated using spectroscopic
 redshifts. We used the code {\it Le Phare}\footnote{{\tt
     http://www.oamp.fr/people/arnouts/LE\_PHARE.html}} described in
 \cite{1999MNRAS.310..540A} and \cite{2002MNRAS.329..355A} with the 
 addition of the optimisation procedure presented in I06. 

 We first selected a set of reference
 SED templates. These were the same CFHTLS-optimised
 templates as I06 used. 

 The original set of templates is composed of four observed galaxy
 spectra from \cite{1980ApJS...43..393C} (hereafter CWW) and two
 starburst galaxy spectra from \cite{1996ApJ...467...38K}. The
 ultraviolet ($\lambda < 2000\AA$) parts of the spectra have been
 linearly extrapolated and the near infrared parts are extrapolated
 using the synthetic models proposed by \cite{2003MNRAS.344.1000B}. It
 is clear that the large variety of galaxy spectra observed in the
 Universe cannot be represented by only a small number of optimised
 templates. However, adding too many templates creates degeneracies
 between observed colors and redshifts. For this reason we use only a
 small number of spectral types. 

 The six templates were ``optimised'' for the CFHTLS using 2867
 spectroscopic redshifts from the VVDS deep survey.  The optimisation
 procedure, described in I06, consists of ``blue-shifting'' the
 observed photometric data to rest-frame using the VVDS spectroscopic
 redshifts. A low resolution optimised version of the template is
 obtained by averaging the rest-frame fluxes of all the galaxies (as
 shown in Fig.5 of I06). Finally, the templates were linearly
 interpolated between spectral types in order to cover the full
 redshift-color space. In total, 62 optimised templates are generated.

 Each SED template was then redshifted onto a grid of interval $\delta
 z=0.04$ and convolved with the filter transmission curves (including
 instrument efficiency). The photometric redshifts were derived by
 determining which SED template provides the best match to the observed
 colours (minimisation of the $\chi^2$ merit function). The galaxy
 internal reddening, $E(B-V)$, was included as a free parameter in the
 template-fitting procedure. The values allowed for $E(B-V)$ were
 derived from the Small Magellanic Cloud extinction law
 \citep{1984A&A...132..389P}, varying from 0 to 0.2 for Scd and later
 types, and no reddening was allowed for earlier types.

 Since no near infrared data were available to break the
 colour-redshift degeneracies between $z<0.5$ and $z>2.5$ (in general
 caused by an inability to distinguish between between the Balmer and
 Lyman breaks), a ``prior'' on the redshift distribution has been
 applied following the Bayesian approach described in
 \cite{2000ApJ...536..571B}. We used the parameterised redshift
 distribution estimated from the VVDS deep spectroscopic survey. In
 I06, the redshift distribution was estimated by spectral type and as a
 function of apparent magnitude . On the basis of the luminosity
 functions measured by \cite{2005A&A...439..863I} we reject 
 unrealistic objects with $M_{g'} < -24$.

 A redshift Probability Distribution Function (PDF$z$) is computed for
 each object using the $\chi^2$ merit function, PDF$z$ $\propto {\rm
   exp}(-\chi^2(z)/2)$. The PDF$z$ is measured every $\delta z=0.04$.
 The best redshift is estimated via a parabolic interpolation of the 
 PDF.
 If a second peak is found in the PDF with a height larger than 
 5\% of the main peak, the corresponding redshift is given as a second
 solution.

 In addition to the best $\chi^2$ derived from the galaxy library
 (hereafter $\chi^2_{gal}$), a best $\chi^2$ computed using the stellar
 library of \cite{1998PASP..110..863P} is derived for each object
 (hereafter $\chi^2_{star}$). Both $\chi^2$'s are use to compute the
 star-galaxy classification as will be explained in a later Section.

 \subsection{Correction of systematic offsets}
 \label{offsets}  

 As demonstrated by \cite{2006ApJ...651..791B}, I06 and
 \cite{2008arXiv0809.2101I}, systematic offsets are often found between
 the best-fit SED templates and the observed apparent magnitudes in a
 given filter. Uncertainties in the zero-point calibration of the
 photometric data as well as imperfect knowledge of galaxy SEDs or of
 filter transmission curves are responsible for these offsets. They may
 produce additional biases in the photometric redshift measurements and must be
 corrected.

 In this work, we recomputed these offsets to account for the changes
 in our spectroscopic and photometric catalogues with respect to those
 used in I06. These changes are:

 \begin{itemize}

 \item The T0004 catalogues have been regenerated after a complete
   re-processing of all data;.

 \item The CFHTLS Wide data were not used by I06. These new stacks
   cover $35~\deg^2$ and combine many MegaCam tiles; 
   
 \item New spectroscopic samples are now available for many CFHTLS
   fields (which were not available in I06). It is now possible to
   compute offsets for many new fields and to derive field-to-field
   systematic offsets.

   \end{itemize}
   
 Following the procedure used in I06, we minimised the SED template
 $\chi^2$'s at a fixed spectroscopic redshift with an additional 
 free offset per band that account for the systematics.
 The offsets are then derived iteratively.  For all spectroscopic surveys, the
 calibration was performed using the most secure spectroscopic
 redshifts. Stars were not used for the offset computation. 

 In this study we limited the spectroscopic samples to $i'_{AB} \leq
 22.5$.  As shown by I06 this correction depends only weakly on the
 spectroscopic limiting magnitude. Each offset correction was derived
 and applied independently for each field covered with spectroscopic
 data (see section \ref{spectro}).  Table \ref{sysoffset} shows the
 systematic offsets derived for the Deep and Wide fields in each band.
 The systematic offsets vary from -0.041 ($g'$) to 0.077 ($u^*$)
 magnitudes and show a small dispersion between the fields, on the
 order of 0.01 mag.  For this reason we added 0.01 magnitudes in
 quadrature to the error estimate to account for the systematic offset
 uncertainties.

 D4 is the only field which does not overlap with any spectroscopic survey. 
 For this field, we used a mean offset correction, computed using a combined
 catalogue of D1/VVDS deep, D2/zCOSMOS and D3/DEEP2.

 \begin{table*}
   \centering
   \caption{Systematic offsets of $i'_{AB} < 22.5$ limited samples
  for the Deep and Wide fields.}
  \label{sysoffset}
 \begin{center}
    \begin{tabular}{c c c c c c c c c}
       \hline
       \hline
       Band & D1/VVDS & D2/zCOSMOS & D3/DEEP2 &  W1/VVDS & W3/DEEP2 & W4/VVDS f22 & mean & dispersion \\
       \hline
       $u*$ & 0.068   &  0.078     &  0.045   & 0.055    &  0.064   &  0.077      & 0.065  &  0.013\\
       $g'$ & -0.055  &  -0.047    &  -0.038  & -0.035   &  -0.030  &  -0.040     & -0.041 &  0.009 \\
       $r'$ & 0.016   &  0.025     &  0.021   & 0.003    &   0.040  &  0.032      & 0.022  &  0.015\\
       $i'$ & 0.003   &  -0.001    &  0.001   & 0.022    &  0.008   &  0.001      & 0.006  &  0.009\\
       $z'$ & 0.001   &  -0.009    &  -0.004  & -0.022   &  -0.037  &  -0.010     & -0.014 &  0.014\\
        \hline
      \end{tabular}
 \end{center}
 \end{table*}
  
 \subsection{Photometric/spectroscopic comparison for the T0004 Deep fields}
 \label{zpzsDeep}
 We evaluate in the two following sections the photometric redshift
 accuracy by comparing photometric redshifts against spectro-$z$'s. We first
 consider the Deep fields; in the following Section we consider
 the Wide fields.

 As in I06, we define two relevant quantities: 
 \begin{itemize}
 \item the {\it photometric redshift dispersion},  $\sigma_{\Delta z/(1+z_s)}$. It is defined using the normalised
 median absolute deviation $\sigma_{\Delta z/(1+z_s)} = 1.48 \times median(|\Delta 
 z|/(1+z_s))$, a robust approximate of the standard deviation,
 \item and the {\it outlier rate}, $\eta$, also called catastrophic 
 errors, defined as the proportion of objects
 with $|\Delta z| \ge 0.15\times (1+z_s)$,
 \end{itemize}
 where $z_p$ is the photometric redshift, $z_s$ the spectroscopic redshift, and 
 $\Delta z = z_p - z_s$.

\begin{figure*}
   \centering
   \includegraphics[width=6cm]{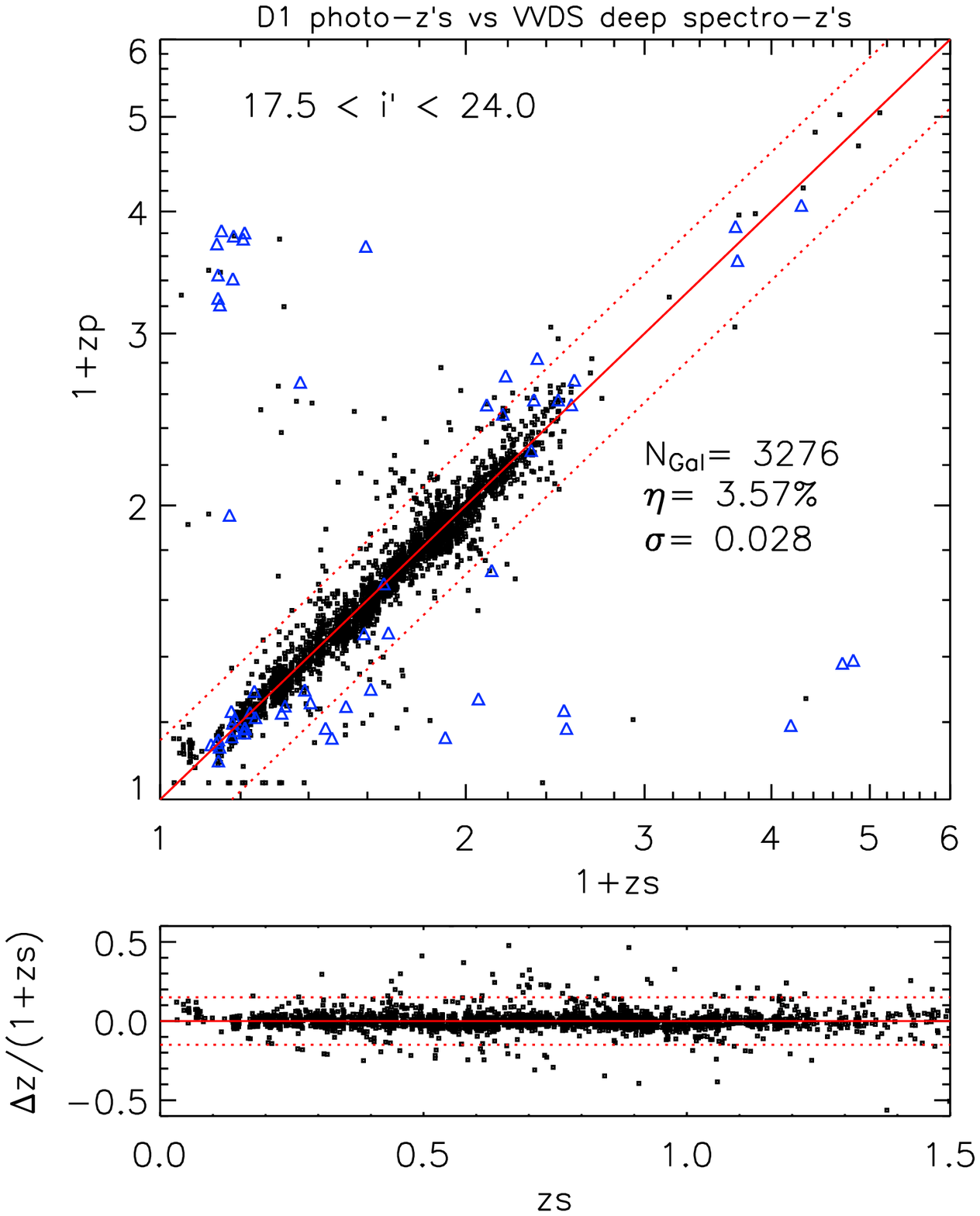}
   \centering
   \includegraphics[width=6cm]{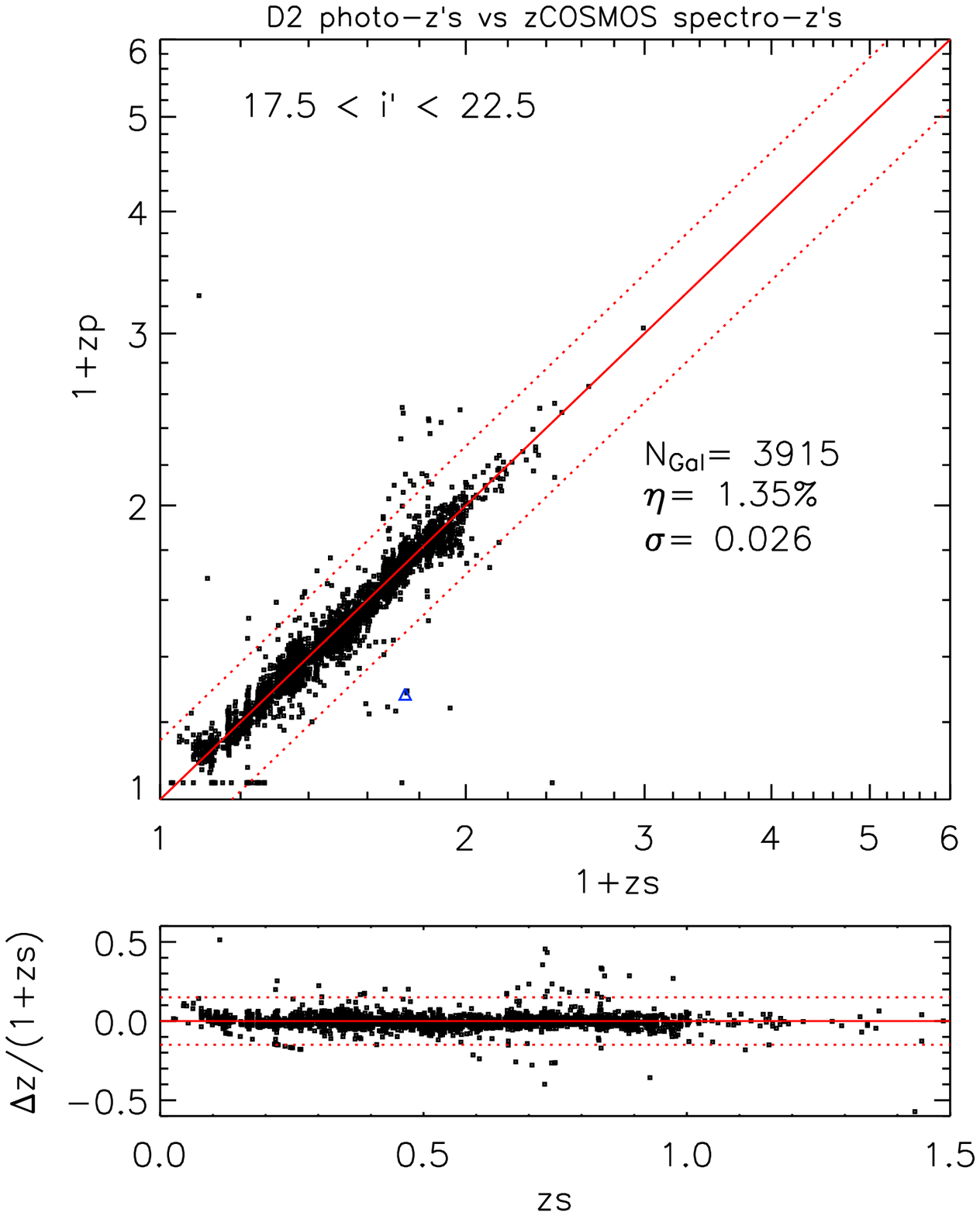}
  \centering
   \includegraphics[width=6cm]{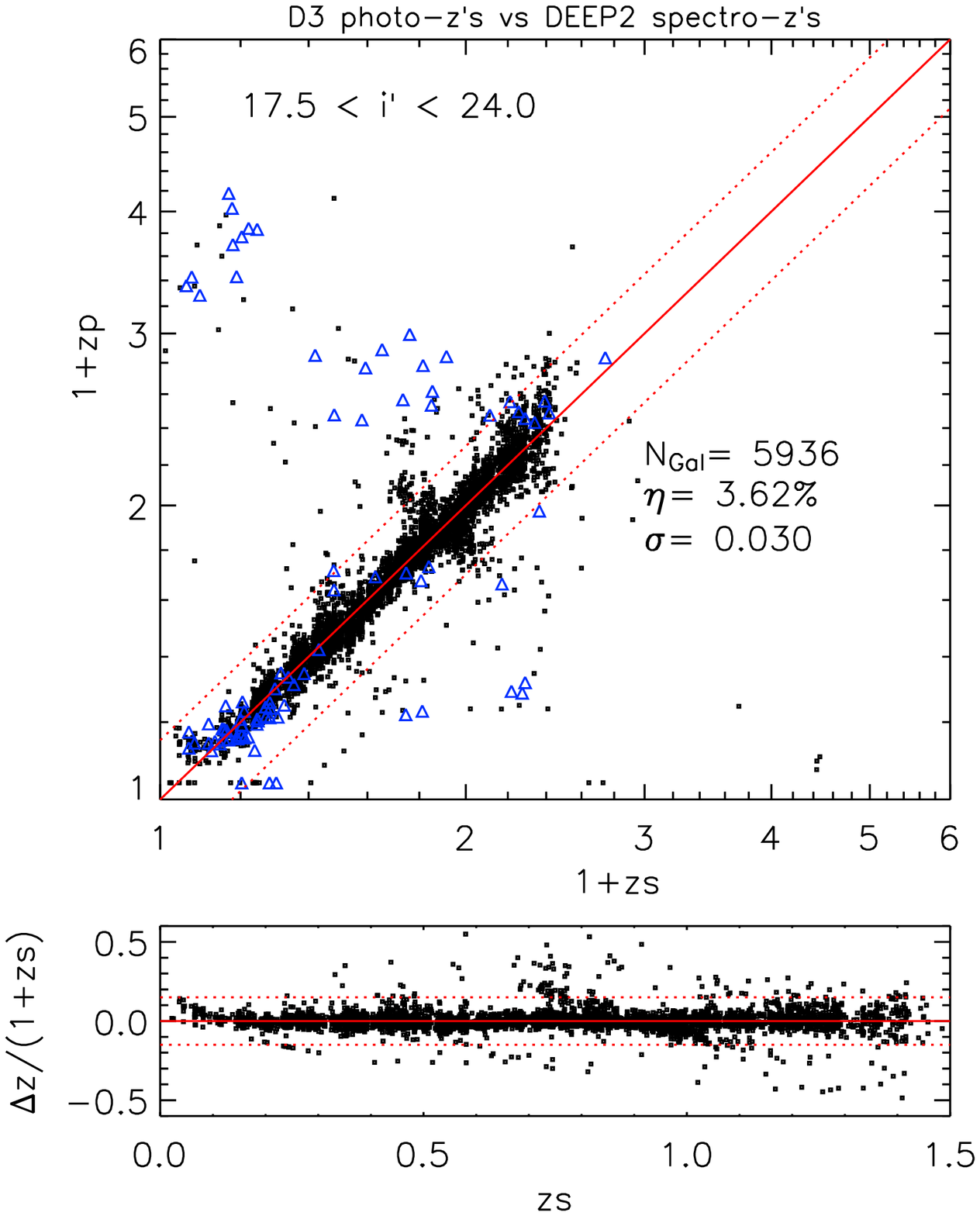}
  \centering
   \includegraphics[width=6cm]{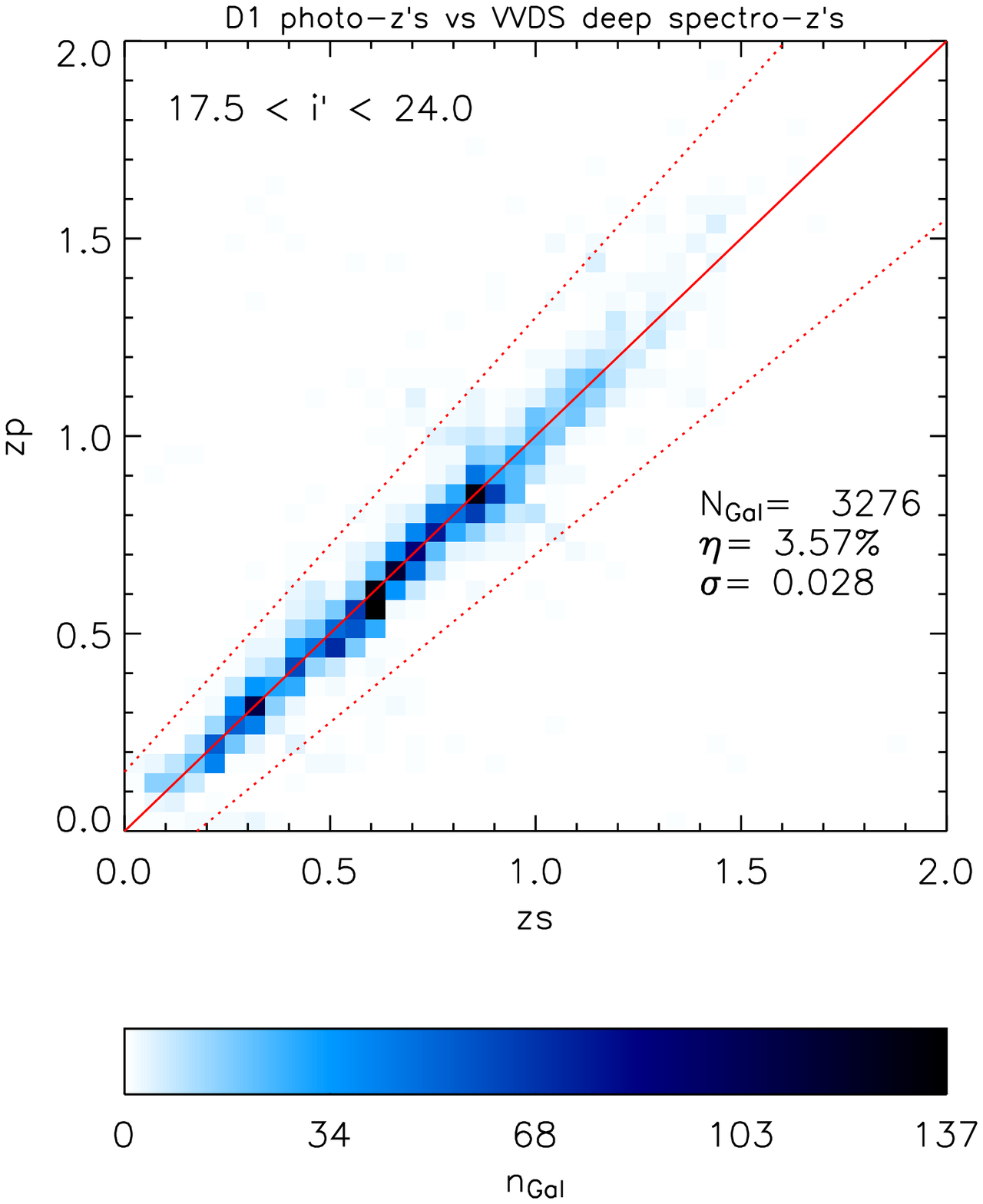}
   \centering
   \includegraphics[width=6cm]{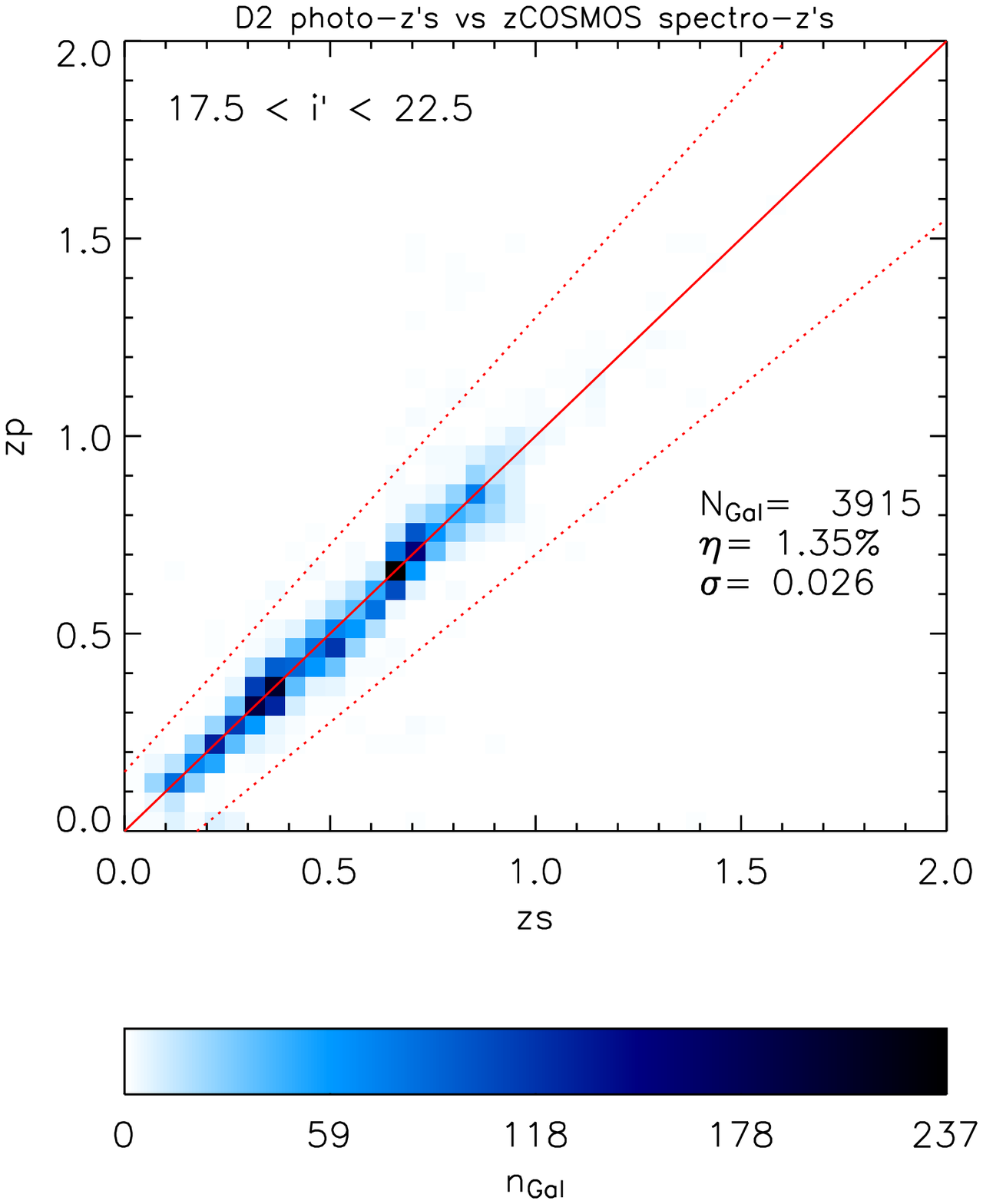}
  \centering
   \includegraphics[width=6cm]{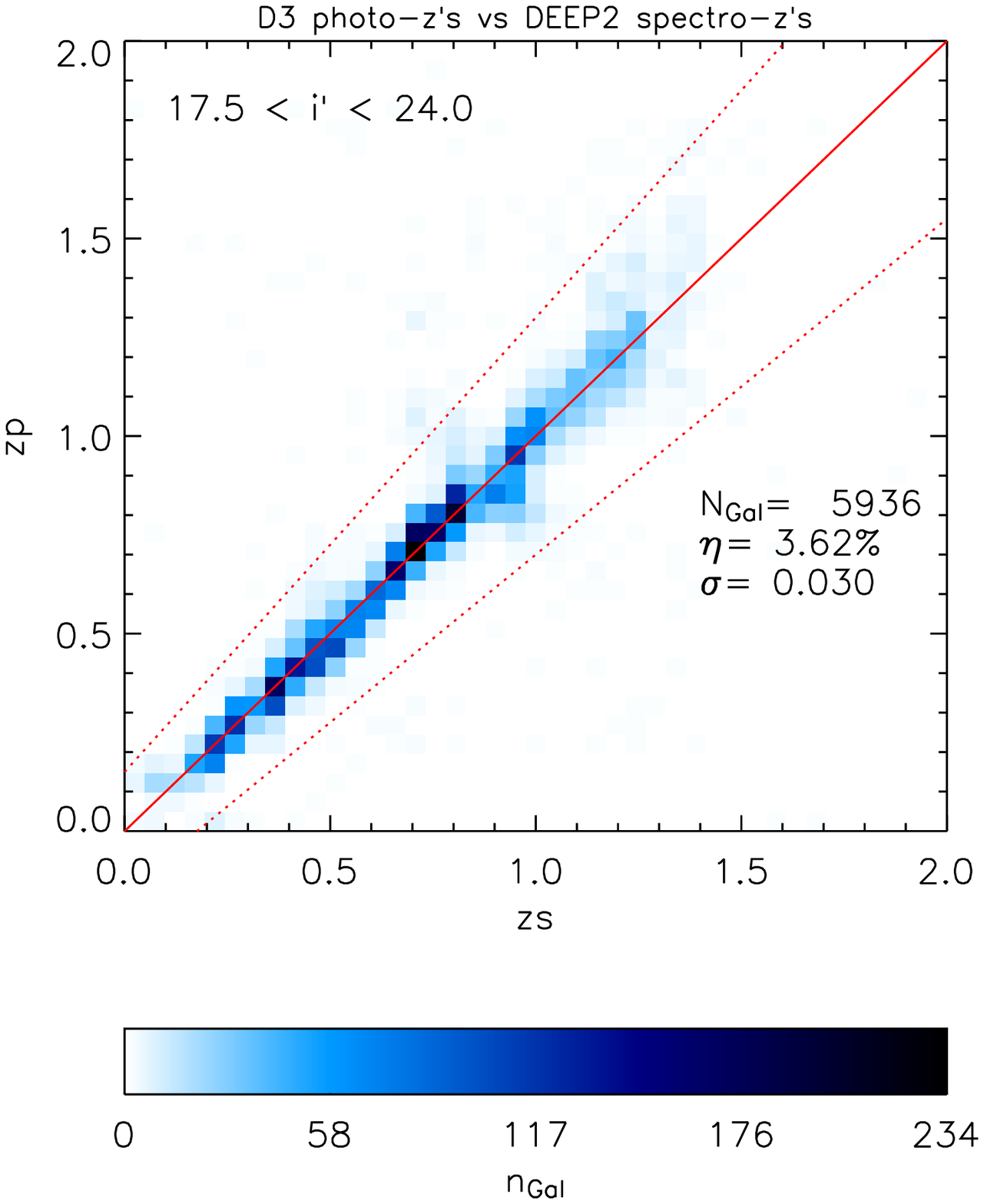}

   \caption{Photometric redshift accuracy in the Deep survey. We
     compare photometric redshifts with spectro-$z$'s for D1 with VVDS deep
     (left), for D2 with zCOSMOS (middle) and for D3 against DEEP2
     (right). The brighter zCOSMOS survey is limited to $i'_{AB}=22.5$.  In
     each panel we show the dispersion, $\sigma_{\Delta z/(1+z_s)}
     = 1.48 \times median(|\Delta z|/(1+z_s))$, and the outlier rate,
     $\eta$, the proportion of objects with $|\Delta z| \ge 0.15\times
     (1+z_s)$.  In the top panel, triangles represent objects
     for which we found a second solution in the PDF.
     In the bottom panel, the object density in the
     zphot-zspec plane is plotted.}

 \label{DeepT04}
 \end{figure*}

 For the VVDS Deep and DEEP2 surveys we can make comparisons to
 $i'_{AB} < 24$.  In the D2 field, for the brighter zCOSMOS survey
 comparisons cannot be made fainter than $i'_{AB} = 22.5$.  Figure
 \ref{DeepT04} shows the comparison between photometric and VVDS
 spectroscopic redshifts for the D1 field; zCOSMOS for the D2 field and
 DEEP2 for the D3 field. The dispersion ($\sigma_{\Delta z/(1+z_s)}$)
 for D1 and D3 in the range $17.5 < i'_{AB} < 24$ is 0.028 and 0.030
 respectively, and the dispersion for D2 in the range $17.5 < i'_{AB} <
 22.5$ is 0.026. The outlier rate ($\eta$) is 3.57\% in D1, 3.62\% in
 D3 and 1.35\% in D2.  These results demonstrate that the photometric
 redshift accuracy is comparable for the three deep fields. Slightly
 better results are found in D2 as this sample has a brighter limiting
 magnitude. We checked this assumption by cutting the limiting
 magnitudes for the D1 and D3 at $17.5 < i'_{AB} < 22.5$.  We found the
 same dispersion for the three fields,
 $\sigma_{\Delta_z/(1+z_s)}=0.026$, as well as slightly different but
 better outlier rates for the three fields: 1.68 for D1, 1.35 for D2
 and 2.36 for D3.

 The consistent results found for these three deep fields are
 reassuring. Although our templates have been optimised using one single
 field (D1) with one set of spectra (VVDS Deep), they can be used with
 confidence on other deep fields without degrading the photometric redshift
 accuracy or increasing the fraction of outliers.




 \subsection{Photometric/spectroscopic comparison for the T0004 Wide fields}

\begin{figure}
   \centering \includegraphics[width=8cm]{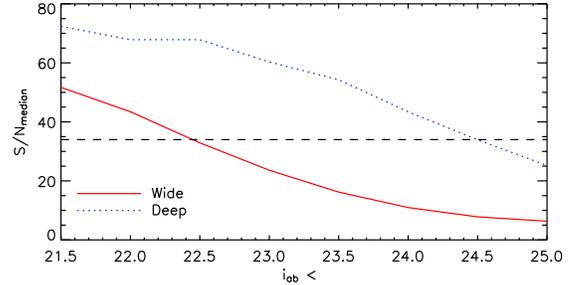}
   \caption{Median signal-to-noise ratio with respect to the limiting
   magnitude in the Wide (W1) and the Deep (D1) field. The $S/N$ is
   $1.0857/\Delta m$, taken in the $i'$ band. The dashed line represents
   the median $S/N=33$, corresponding to the value of the Wide sample 
   cut at $i'_{AB} < 22.5$.}
 \label{SN}
 \end{figure}

 The photometric redshift accuracy is function of the signal-to-noise ratio
 (hereafter $S/N$) of the photometric data (e.g. Margoniner and Wittman
 2007). Since the Wide survey is much shallower than the Deep survey
 the median $S/N$ is lower at a given magnitude in the Wide than in the
 Deep (see Figure \ref{SN} for the D1 and W1 fields). Therefore, we
 expect less accurate photometric redshifts for the wide in comparison with the
 deep fields at the same magnitude. 

 First, we measured the photometric redshift accuracy as a function of decreasing
 $S/N$ (i.e. increasing magnitude) in the wide fields. 

 Noise is estimated using magnitude errors from {\tt SExtractor}'s
 MAG\_AUTO, $\Delta m$, rescaled by a factor 1.5 (See Section 2.1.3). 
 Using the VVDS deep spectroscopic sample, we measured the photometric redshift
 precision to $i'_{AB} = 24$. In Table \ref{widemagvaries}, we show the
 dispersion and the outlier rate for the W1 field as a function of
 limiting magnitude. We found that $\sigma_{\Delta z/(1+z_s)}$ is
 larger at fainter magnitudes. At $i'_{AB} < 24$ the dispersion is
 0.053. However the number of outliers increases dramatically beyond
 $i'_{AB}=23$, reaching 10\% at $i'_{AB} < 24$. We conclude that
 photometric redshifts are reliable down to $i'_{AB} = 22.5$ in the Wide field
 (with a median $S/N=33$). Fainter than $i'_{AB}=23$ the number of
 outliers becomes important. 

\begin{figure*}
   \centering \includegraphics[width=6cm]{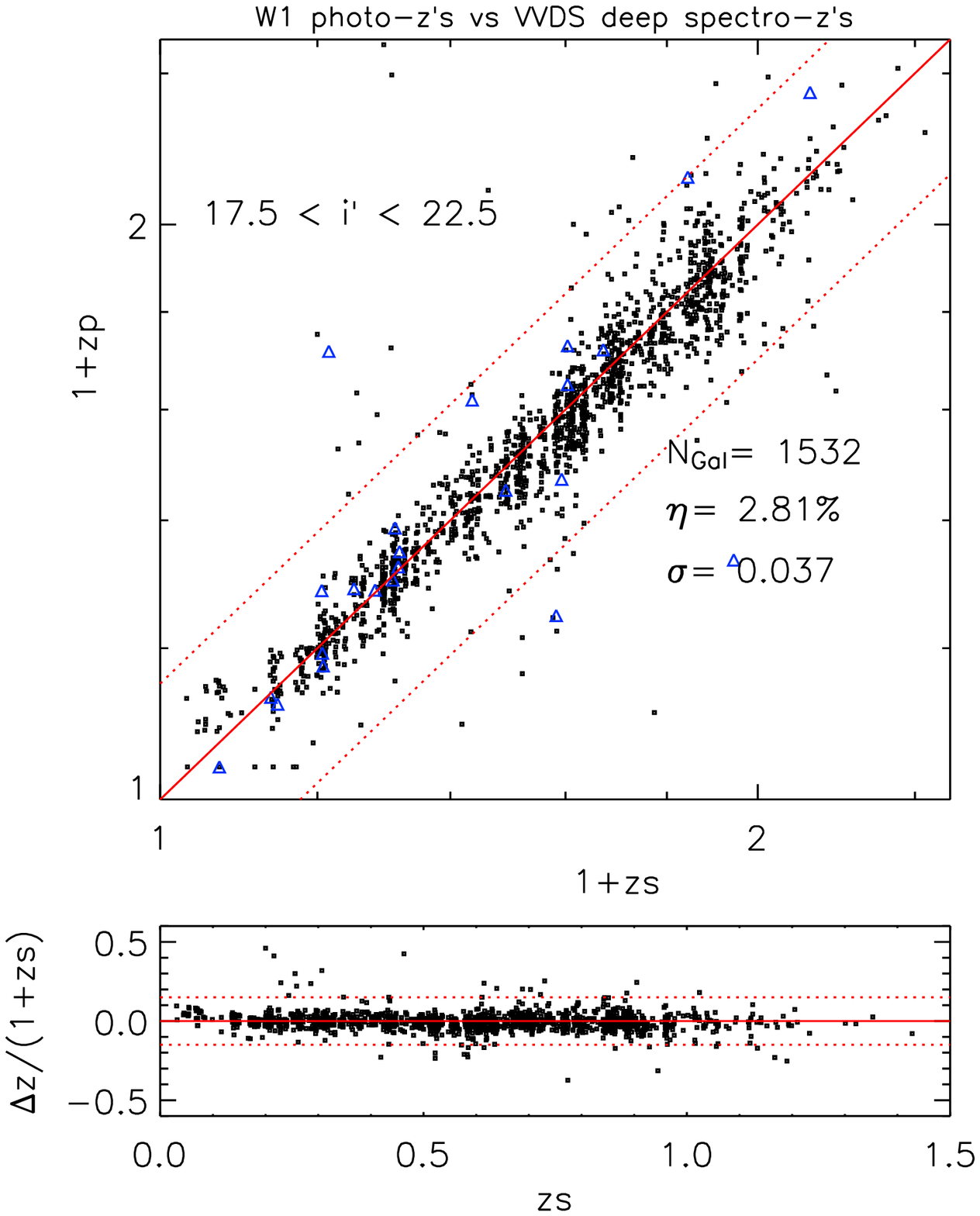}
   \centering \includegraphics[width=6cm]{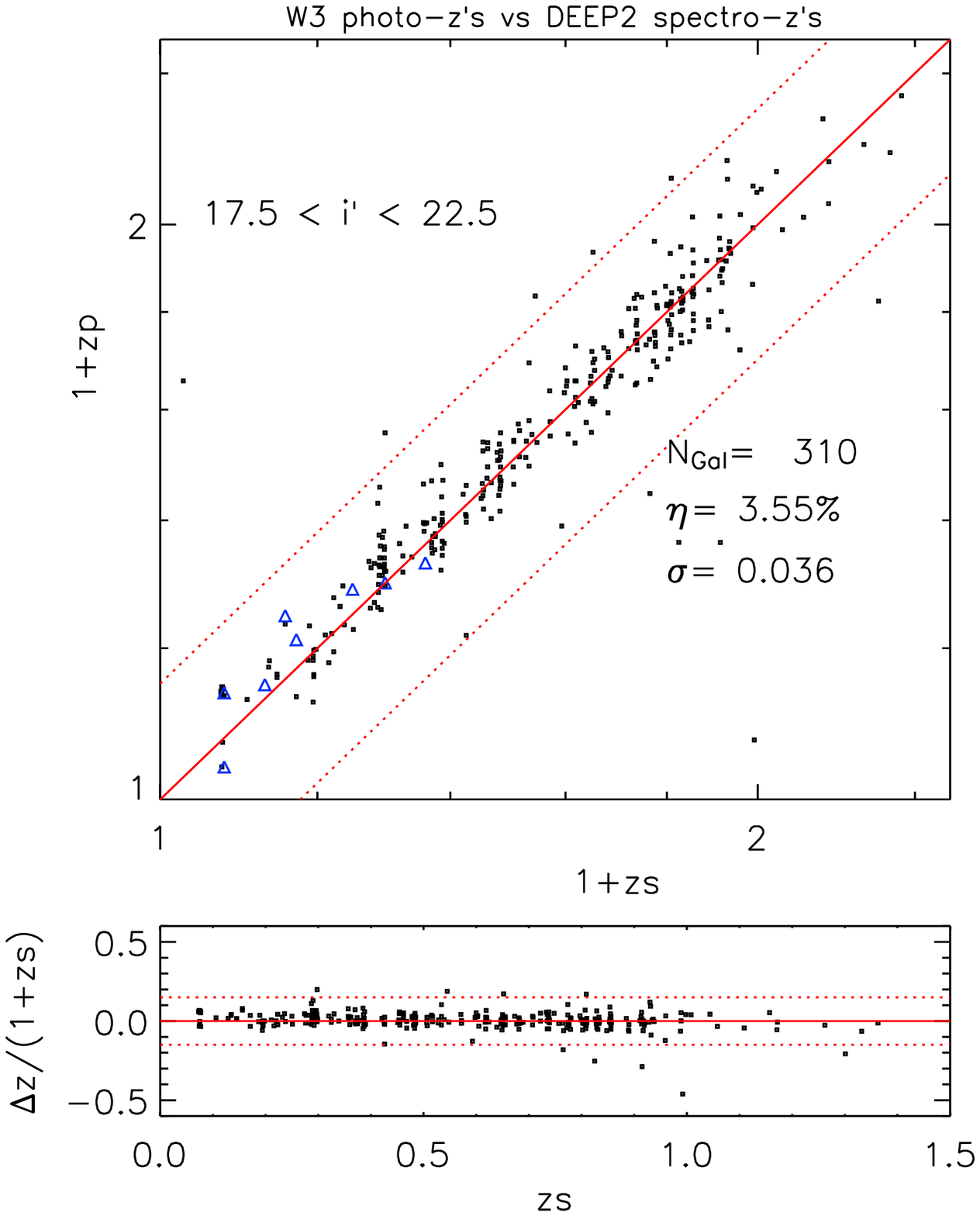}
   \centering \includegraphics[width=6cm]{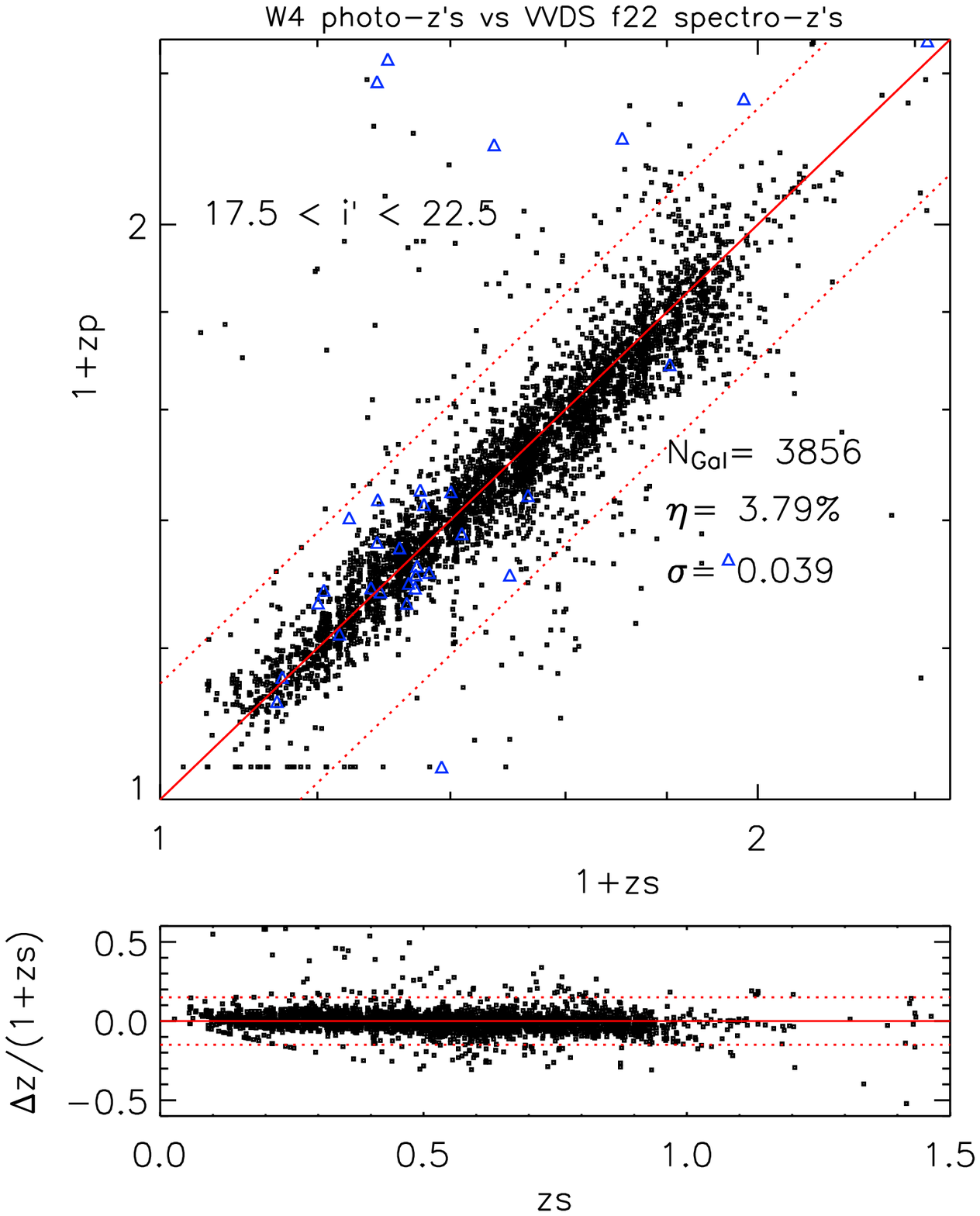}
   \centering \includegraphics[width=6cm]{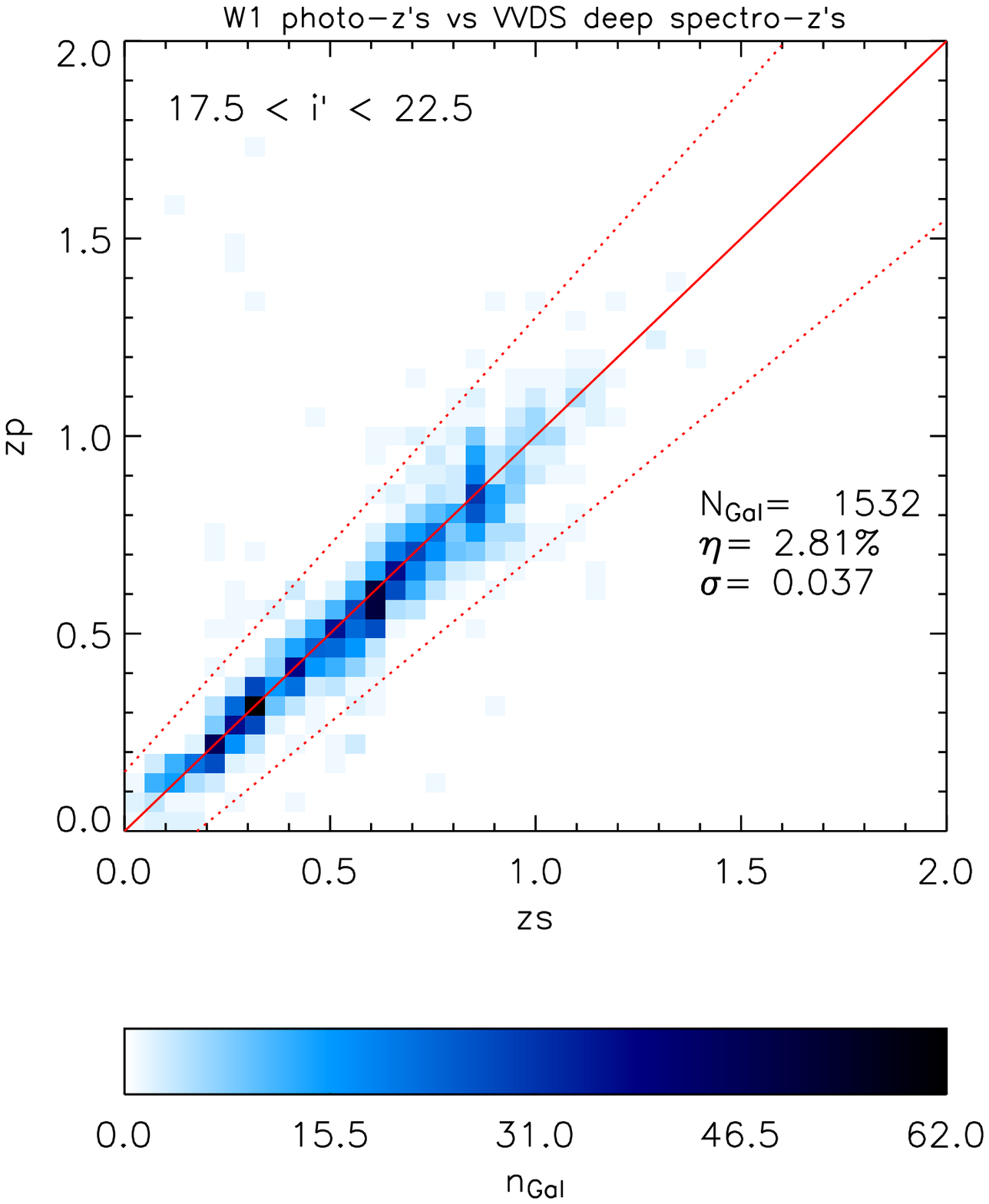}
   \centering \includegraphics[width=6cm]{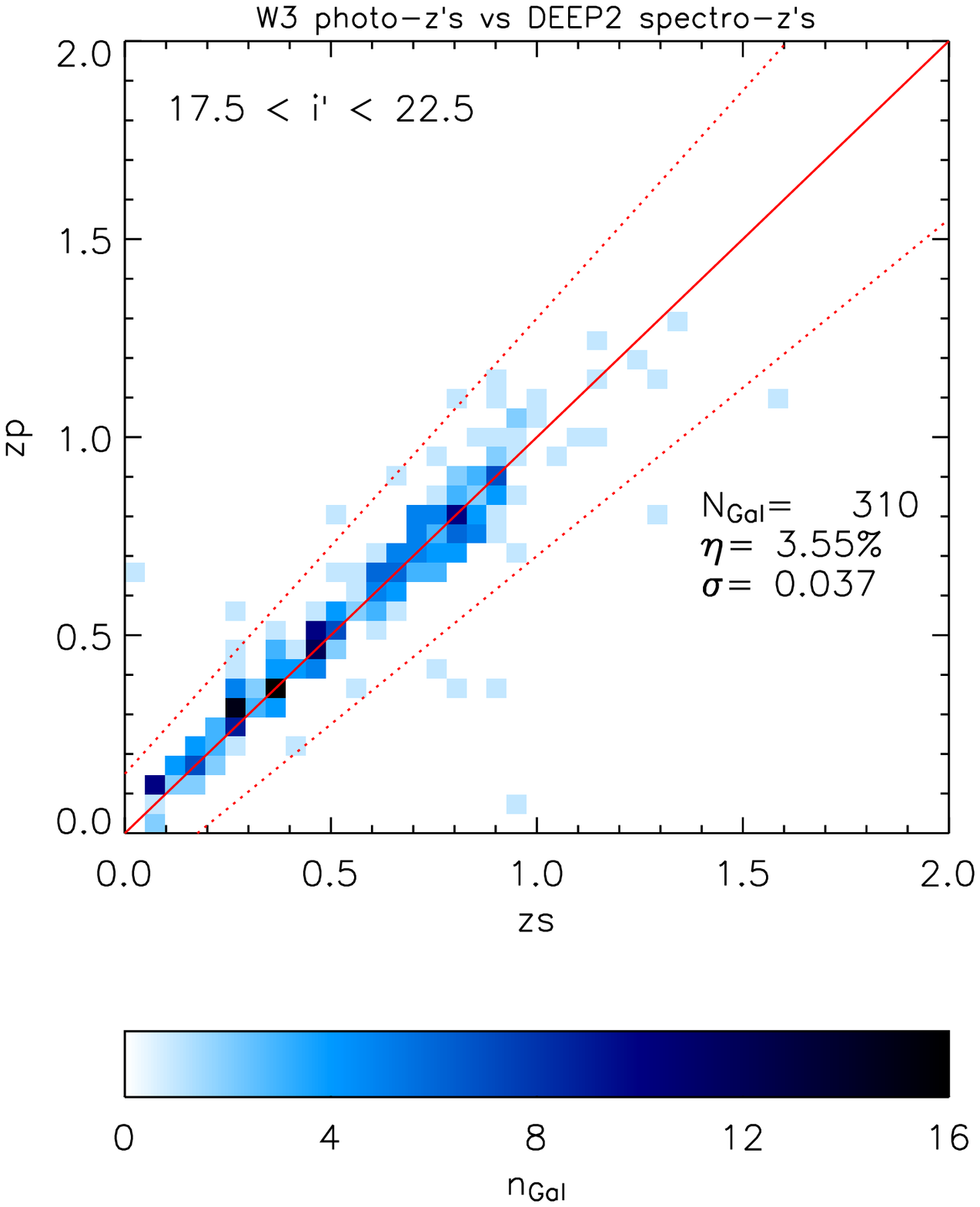}
   \centering \includegraphics[width=6cm]{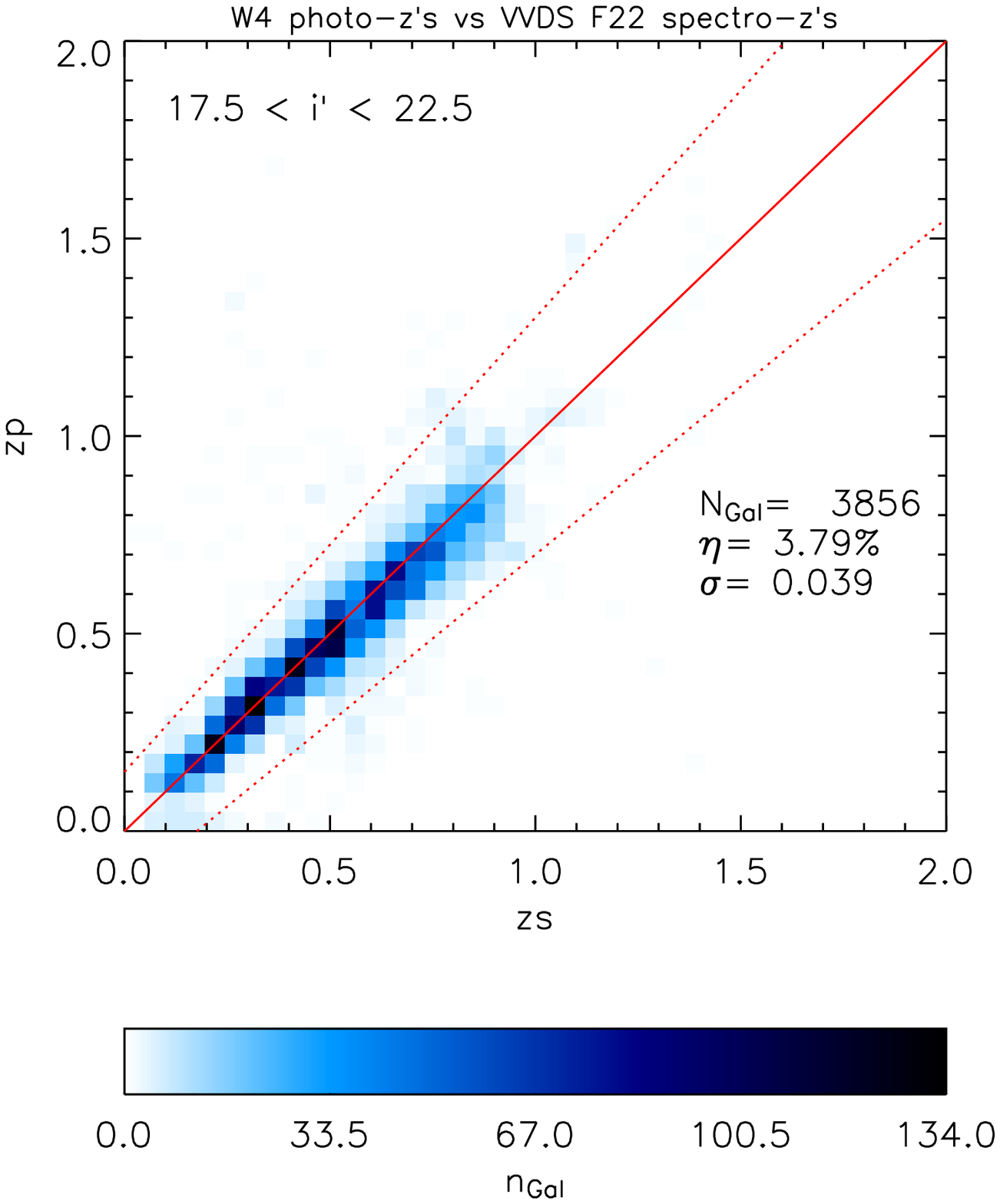}
   \caption{Photometric redshift accuracy in the Wide survey at $i'_{AB} < 22.5$.
     We compared photometric redshifts with spectro-$z$'s for W1 with VVDS deep
     (left), for W3 with DEEP2 (middle) and for W4 with VVDS Wide
     (right).  In each panel we show the dispersion,
     $\sigma_{\Delta z/(1+z_s)} = 1.48 \times median(|\Delta
     z|/(1+z_s))$, and the outlier rate, $\eta$, the proportion of
     objects with $|\Delta z| \ge 0.15\times (1+z_s)$.
     In the top panel, triangles represent objects
     for which we found a second solution in the PDF. Bottom:
     density of objects in the zphot-zspec plane is plotted.}  \label{WideT04}
 \end{figure*}

 Next we compared the photometric redshift accuracy between the different Wide
 fields. We limited the comparison at $i'_{AB}<22.5$ (median $S/N=33$)
 which is the limiting magnitude of the zCOSMOS and VVDS-Wide
 spectrocopic surveys. Figure \ref{WideT04} shows the comparison
 between photometric and spectroscopic  redshifts (W1 against
 VVDS deep, W3 against DEEP2 and W4 against VVDS Wide). The dispersion
 $\sigma_{\Delta z/(1+z_s)}$ is found to be identical for all
 fields, ranging from 0.036 to 0.039 (0.037 for W1, 0.036 for W3 and
 0.039 for W4).  The outlier rate, $\eta$, is 2.81\%, 3.55\% and 3.79\%
 for W1, W3 and W4, respectively.  As for the Deep survey, these
 results show that one can successfully use templates optimised in an
 independent field to reach comparable accuracy. 

 We also noticed that despite being calibrated with a small
 spectroscopic sample, the photometric redshift accuracy and outlier rate in the
 D3 field are comparable to the other fields. For the purpose of
 photometric redshift calibration, it appears that for the same amount
 of time dedicated to a spectroscopic survey it is more important to
 cover a larger area with bright spectroscopic redshifts (which can
 correct for inhomogeneities in the photometric calibration) rather
 than a smaller.

 Finally, in order to perform a fair comparison between the Deep and
 Wide photometric redshift one must consider samples with equal $S/N$. Therefore,
 in order to match the median $S/N=43$ corresponding to $i'_{AB} < 24$
 in the Deep, we limit the magnitude to $i'_{AB}=22$ in the Wide (from
 figure \ref{SN}) and measure a dispersion of 0.032, similar to the
 value of 0.028 found for the D1 photometric redshifts.

 %

 \begin{table}
   \centering
   \caption{Photometric redshift accuracy in the W1 and D1 fields for several
   limiting magnitudes.}
  \label{widemagvaries}
 \begin{center}
    \begin{tabular}{c c c c}
      \multicolumn{4}{c}{}\\
      \multicolumn{4}{c}{W1}\\
       \hline
       \hline
       $i'_{AB} < $ & $S/N$  & $\sigma_{\Delta z/(1+z_s)}$ & $\eta$ (\%)  \\
       \hline
       20.5 & 64 & 0.025 & 1.12 \\
       21.0 & 60 & 0.026 & 1.57 \\
       21.5 & 52 & 0.029 & 1.39 \\
       22.0 & 43 & 0.032 & 2.25 \\
       22.5 & 33 & 0.037 & 2.81 \\
       23.0 & 24 & 0.043 & 4.91 \\
       23.5 & 16 & 0.048 & 7.63 \\
       24.0 & 11 & 0.053 & 10.13 \\
       \hline
    \end{tabular}

  \begin{tabular}{c c c c}
     \multicolumn{4}{c}{}\\
      \multicolumn{4}{c}{D1}\\
       \hline
       \hline
       $i'_{AB} < $ & $S/N$ & $\sigma_{\Delta z/(1+z_s)}$ & $\eta$ (\%)  \\
       \hline
       20.5 & 72 & 0.030 & 0.54 \\
       21.0 & 72 & 0.027 & 1.17 \\
       21.5 & 72 & 0.025 & 1.32 \\
       22.0 & 68 & 0.026 & 1.30 \\
       22.5 & 68 & 0.026 & 1.82 \\
       23.0 & 60 & 0.026 & 2.11 \\
       23.5 & 54 & 0.027 & 2.59 \\
       24.0 & 43 & 0.028 & 3.57 \\
       \hline
    \end{tabular}

 \end{center}
 \end{table}

 \subsection{Effects of the tile-to-tile systematic offset variation}
 \label{sysOffVar}

 From Fig.\ref{CFHTLS_maps} one can see that CFHTLS fields have only
 partial spectroscopic coverage. The offset calibration can only
 be performed in sub-areas where there is spectroscopic coverage, and
 we assume the systematic offsets measured in a sub-area of the field
 can be applied to the whole field.

 This assumption is most valid for the CFHTLS deep fields since each
 spectroscopic sample fits within one tile, but more
 questionable for the CFHTLS wide fields which are ten times larger and
 are composed of many continuous tiles. As it was discussed in Section
 2.1.4, the $\Delta m_{{\rm CFHTLS}-{\rm SDSS}}$ photometric comparison
 performed by {\sc Terapix} shows the differences with SDSS have a mean
 scatter of $\pm0.03$ magnitudes which corresponds to the photometric
 scatter from tile to tile. Table \ref{sysoffset} confirms these trends
 in our calibrations.

 Only 15 of 35 tiles in the Wide fields have spectroscopic data (not
 all centered in the tile). Therefore, the calibration of the
 systematic offsets cannot be performed tile by tile in the wide
 fields. Instead, they are averaged over several tiles (W1 has four
 overlapping tiles over 19 with VVDS deep, W3 has two overlapping tiles
 in five with DEEP2 and W4 has nine overlapping tiles in 11 with VVDS
 wide). Therefore, tile-to-tile photometric calibration variations are
 an additional source of error which degrades the photometric redshift 
 accuracy in the Wide fields.

 We assessed the impact on the photometric redshift accuracy of this additional
 source of error, assuming a tile-to-tile variation of 0.03 magnitudes  (see
 Sect. 2.1.3). We simulated variations of the photometric calibration on fields
 with spectroscopic data as follows:

 \begin{enumerate}
 \item we computed the systematic offsets from a field with
 spectroscopic data (with unchanged photometry);

 \item we modelled the tile-to-tile zero-point variation with a Gaussian
   distribution with zero mean and a dispersion of $0.03$. We applied
   offsets to this field, but on the 5 bands independently ({\it i.e.}
   we drew an independent value for each band);

 \item we computed the photometric redshifts but we kept the offset
   corrections derived at the first step. Since we did not compute the
   systematic offsets again this sequence is equivalent to derive
   systematic offsets from one field and apply them on another field;

 \item finally, we estimated the uncertainties (photometric redshift
   dispersion and outlier rate).

 \end{enumerate}
 We repeated this procedure 1000 times and computed the mean and
 standard deviation for the photometric redshift dispersion
 $\sigma$, and for the outlier rate, $\eta$.

 \begin{figure}
    \centering \includegraphics[width=8cm]{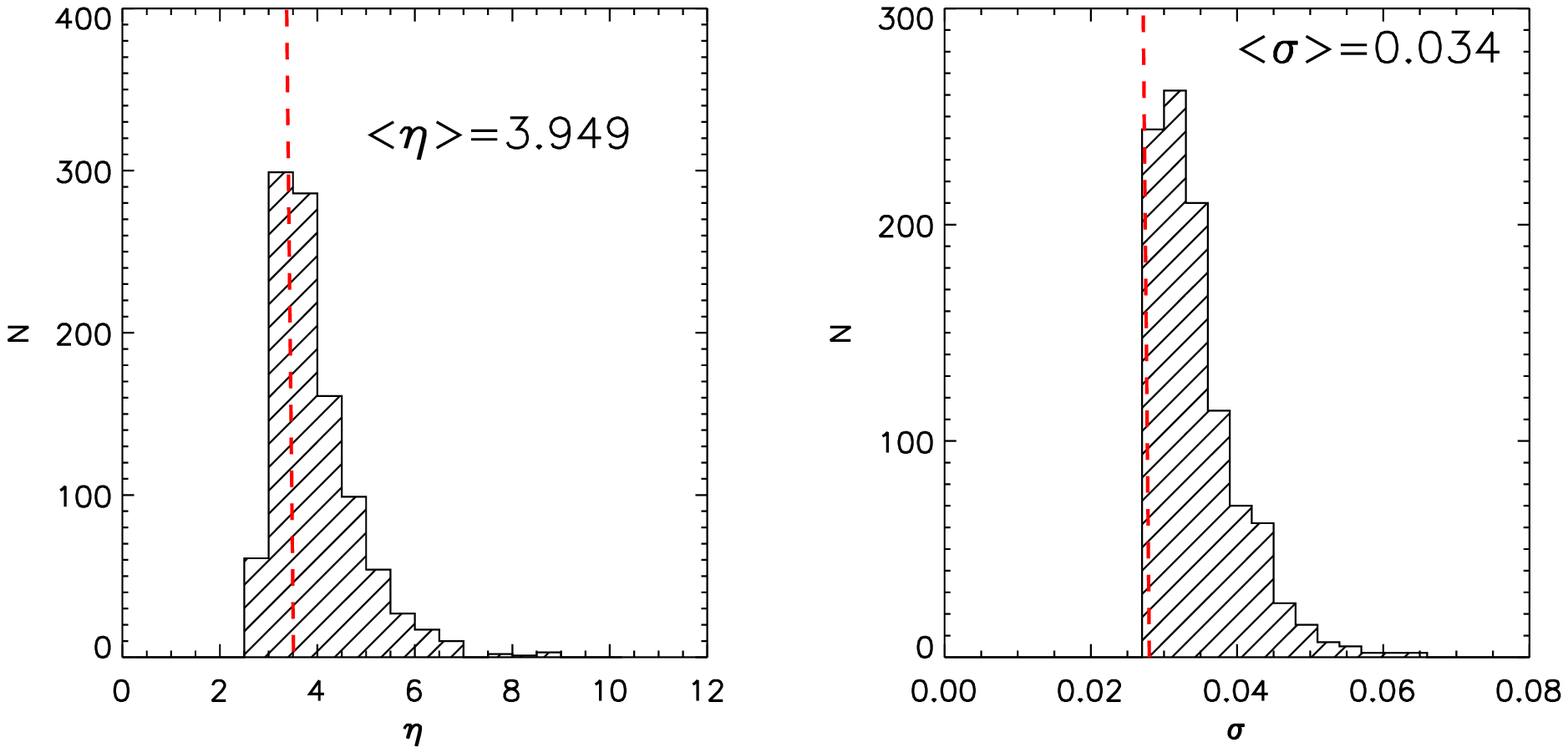}
    \includegraphics[width=8cm]{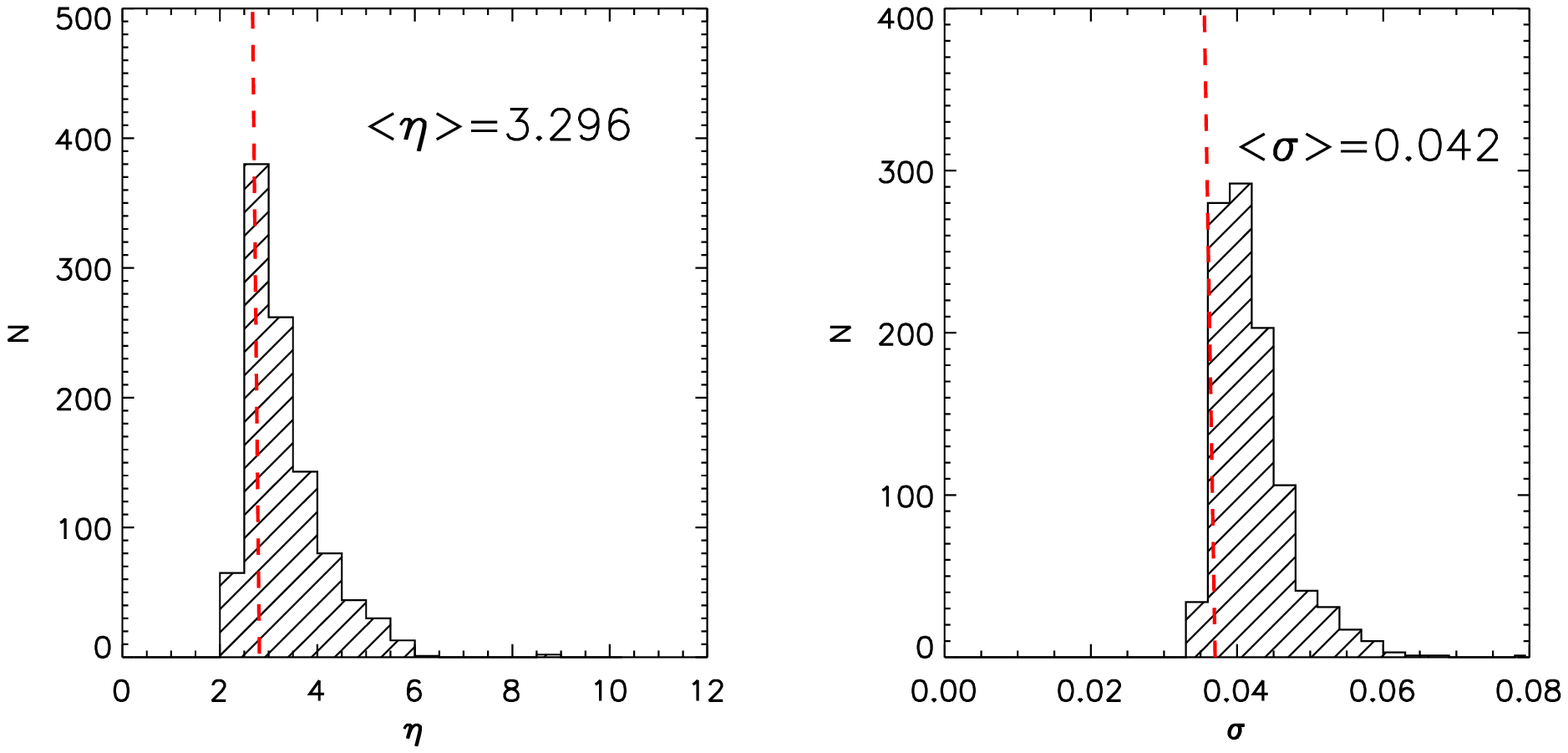}
    \caption{Degradation of the photometric redshift accuracy if the
    offset correction is not computed on the same tile. The results
    show the variation of the outlier rate (left panels) and
    photometric redshift accuracy (right panels) if the photometry varies of the
    order of 0.03 magnitude. Top panels: D1 Deep field at $i'_{AB} <
    24$ (compared with $\sigma_{\Delta z/(1+z_s)}=0.028$ and $\eta=3.57\%$,
    represented by the red dashed 
    lines, with no variation added in the photometry). 
    Lower panels: W1 Wide field at
    $i'_{AB} < 22.5$ (compared with $\sigma_{\Delta z/(1+z_s)}=0.037$ and $\eta=2.81\%$,
     represented by the red dashed lines,
    with no photometric perturbation applied).}\label{photovariations}
 \end{figure}

 Figure \ref{photovariations} shows the results for the Deep and Wide
 fields.  As expected, photometric redshift accuracy degrades. On
 average, $\sigma_{\Delta z/(1+z_s)}$ is $21\%$ worse in the Deep field and $14\%$ worse
 in the Wide field, whereas they were selected to have comparable
 $S/N$. The worse degradation observed for the Deep field is likely due
 to the numerous faint objects, more sensible to photometric
 variations.  The outlier rate shows a similar trend, with a value
 $13\%$ worse in the Deep sample and $17\%$ in the Wide sample.  This
 analysis shows the need for spectroscopic surveys spread over large
 areas.

 \subsection{Wide/Deep comparison}
 \label{WDcomp}

 As described in section \ref{photoData}, Wide and Deep fields 
 have been reduced independently.
 The systematic offsets were also calibrated
 independently.  For this reason, galaxies common to a Deep and a Wide
 field may have two different photometric redshift values.

 To check if results from Wide and Deep surveys are consistent, we
 performed a photometric redshift/photometric redshift comparison using overlapping sources
 in the W1 and D1 fields. We selected 14484 common sources in D1 and W1
 in the range $17.5 < i'_{AB} < 22.5$.

\begin{figure}
   \centering \includegraphics[width=8cm]{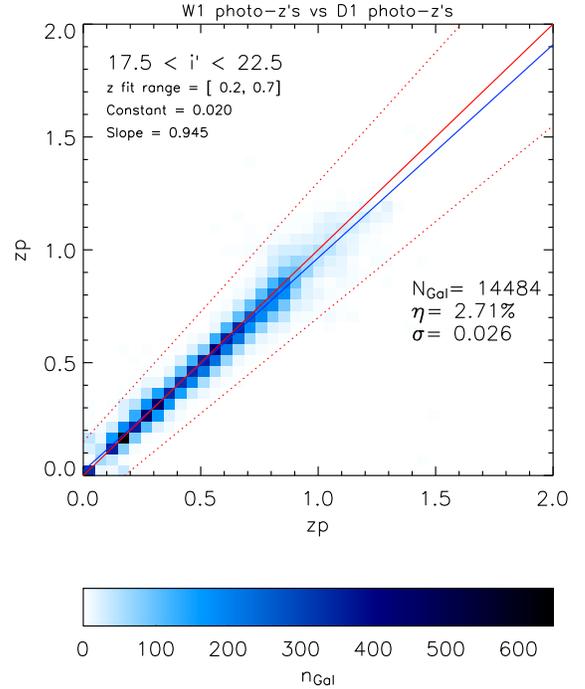}

   \caption{Photometric redshifts for the Wide field (W1, y-axis)
     compared with photometric redshifts for the Deep field (D1, x-axis).
     The sample is
     selected at $17.5 < i'_{AB} < 22.5$. We fit in the range [0.2,0.7]
     to better measure remaining systematic differences. These are
     small compared to the dispersion and we conclude there is a good
     consistency between the two samples.
 }  \label{W1T04}
 \end{figure}

 The figure \ref{W1T04} shows the W1 photometric redshift versus the D1 photometric redshift. 
  We find the dispersion between the wide and deep photometric redshift
 samples is $\sigma_{\Delta z/(1+z_s)}=0.026$, whilst the outlier
 rate is $\eta=2.71$ similar to values found for the D1 sample,
 demonstrating that the Wide and Deep photometric redshifts are well
 consistent with each other. 


 \section{Star/galaxy selection}
 \label{SecStargal}

 Bright samples ($i'_{AB} < 22.5$) can be highly contaminated by stars.
 For example in the CFHTLS W4 field stellar sources comprise up to
 50\% of sources and can be problematic for analyses sensitive to such
 mis-classifications. As in \cite {2006A&A...447..185S}, 
   we used the VVDS spectroscopic data to improve the 
   star/galaxy selection. 

 Thanks to purely flux limited selection of the VVDS spectroscopy 
   survey, we can test our ability to separate stars and galaxies at 
   the depth of the Wide and Deep CFHTLS surveys.
 We have derived an efficient technique to
 separate stars from galaxies based on both morphological and
 multi-colour information.

 \subsection{Size selection}

\begin{figure*}
 \begin{tabular}{c c}
 \centering\includegraphics[width=8cm]{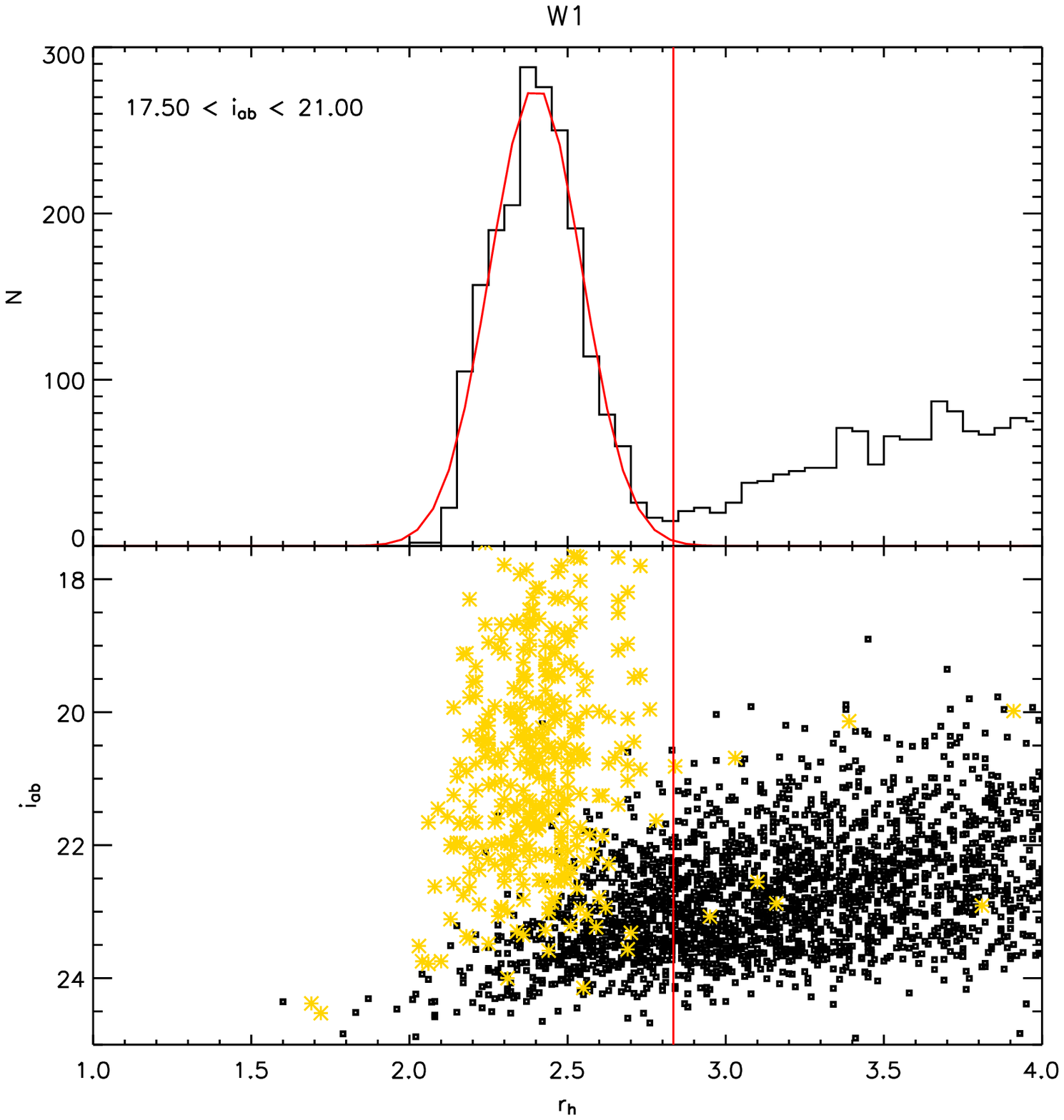} &
 \centering\includegraphics[width=8cm]{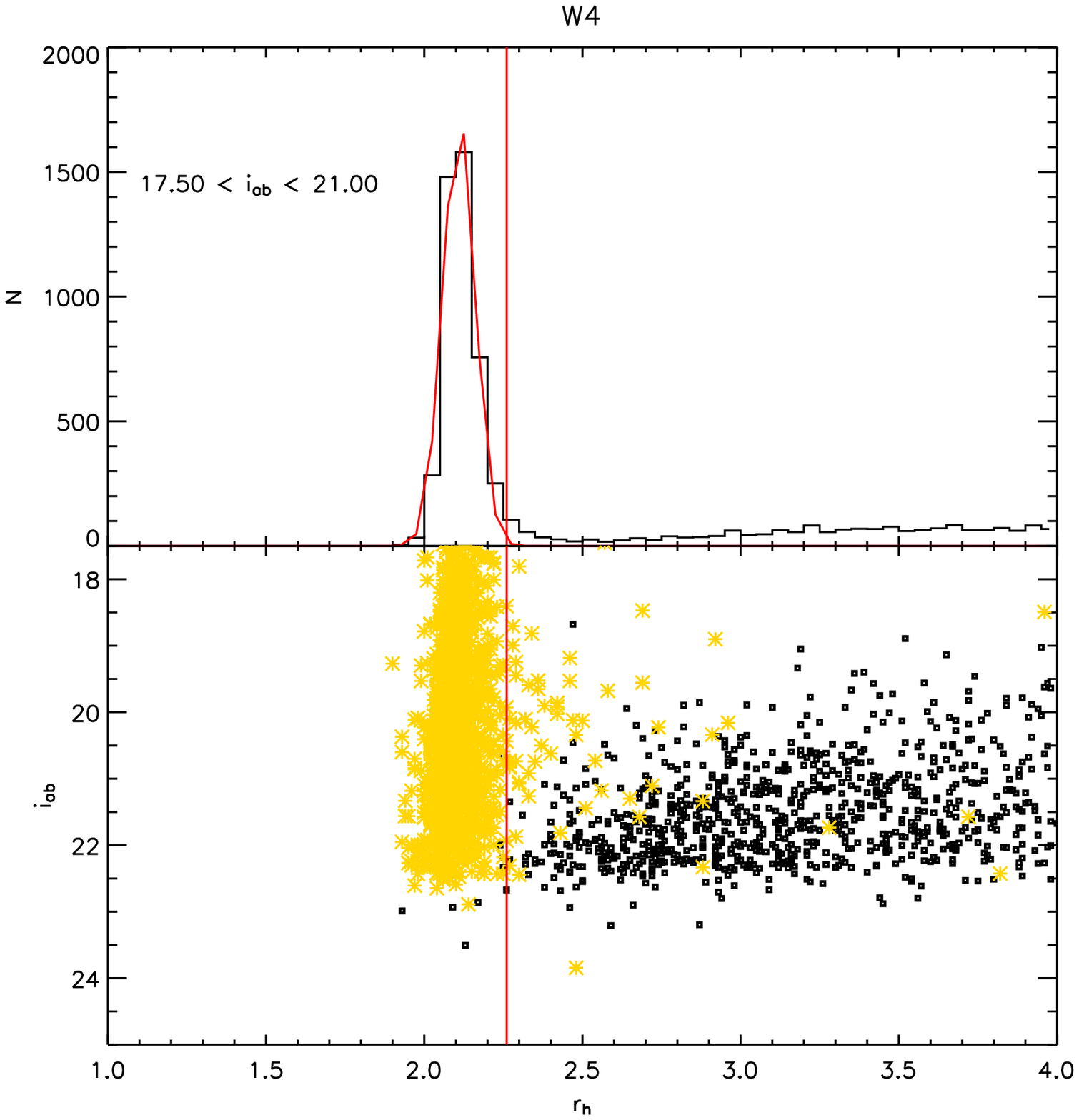}
 \end{tabular}
 \caption{Star/galaxy separation in one tile of the W1 field,  
 022539-041200 (left panel), 
 and one from the W4 field, 221706+002300 (right panel). 
 Top: $r_h$ distribution and the corresponding gaussian fitting. The
 red line is the $\mu_{r_h}+3\sigma_{r_h}$ cut.
 Bottom: $i'_{AB}$ vs $r_h$ in pixels. 
 Squares and asterisks are VVDS-Wide spectroscopically identified stars
 and galaxies respectively. 
 Both fields have
 different specifications; W1 overlaps with a deeper spectroscopic
 sample and has a low stellar density, whereas W4 has a larger
 proportion of stars but a brighter spectroscopic limit.}
 \label{sizemagwide}
 \end{figure*}

 A widespread technique to separate stars from galaxies involves
 comparing the size of sources with the local shape of the PSF.  Bright
 objects comparable in size to the local PSF are identified as stars
 while those with larger size are selected as galaxies. Figure
 \ref{sizemagwide} illustrates the difference in size between galaxies
 and stars for two tiles of the W1 and W4 fields.  The size of the
 sources is characterised by the half-light radius, $r_h$, which is
 defined as the radius which encloses $50\%$ of the object flux. The
 lower panels of figure \ref{sizemagwide} show the size
 ($r_h$)-magnitude ($i'_{AB}$) diagram \citep{1994ApJ...437...56F}.  
   Due to statistical measurement
 errors of $r_h$ and to the large PSF variation over the MegaCam field
 of view the dispersion of the $r_h$
 distribution is broad.

 Spectroscopic redshifts can be used to securely distinguish stars and
 galaxies and to clearly define star-galaxy selection criteria
 calibrated on bright sources.  We define a ``true'' star as an object
 with $z_s = 0$ and a ``true'' galaxy an object with a secure
 spectroscopic redshift $z_s > 0$.  Figure \ref{sizemagwide} shows the
 positions of spectroscopically confirmed stars (asterisks) and
 spectroscopically confirmed galaxies (black squares) in the $r_h$ vs
 $i'_{AB}$ diagram for two tiles in the wide fields. At bright
 magnitudes, a stringent cut in $r_h$ can reliably separate stars from
 galaxies. We then set this cut on each CFHTLS Deep field and Wide tile
 independently: at $i'_{AB} < 21$ the $r_h$ distribution is
 predominantly composed of stars.

 The histogram of $r_h$ values is close to a normal distribution,
 slightly skewed towards larger $r_h$ by the galaxies (top panel
 of Figure \ref{sizemagwide}). For this reason we fit the $r_h$
 distributions at $i'_{AB} < 21$ of all Deep and Wide tiles
 independently by a Gaussian. We denote $\mu_{r_h}$ and $\sigma_{r_h}$
 the best fit mean and standard deviation values respectively.  In the
 Gaussian approximation of the $r_h$ distribution, almost all stars should
 lie in the range $0 < r_h < \mu_{r_h}+3\sigma_{r_h}=r_{hlimit}$.

 %

 \subsection{Colour selection}

 A purely size-based selection as described in the previous section
 limits reliable star-galaxy classification to only the brightest
 sources.  As shown in Figure \ref{sizemagwide} the proportion of
 spectroscopically identified galaxies fainter than $i'_{AB}=21$
 which satisfies the size criterion $r_h \le r_{hlimit}$ increases
 towards fainter magnitudes, where galaxy and star populations become
 increasingly mixed. Adding an additional colour-based selection can
 improve the star-galaxy separation at fainter limits. 


 Following the galaxy spectral type classification based on the galaxy
 template fitting criteria (namely, $\chi^2_{galaxy}$), we also
 characterise each source by its stellar spectral type, based on a
 fitting of stellar templates (namely, $\chi^2_{star}$). We used a set
 of templates from \cite{1998PASP..110..863P} to estimate
 $\chi^2_{star}$.

 Ideally, a galaxy would satisfy the relation $\chi^2_{gal}
 <\chi^2_{star}$. However, $\chi^2_{galaxy}$ and $\chi^2_{star}$ were
 derived from independent SED libraries and with a different number of
 parameters.  The relevance and the statistical significance of the
 comparison between $\chi^2_{galaxy}$ and $\chi^2_{star}$ have been
 assessed by using the spectroscopic sample.  We used a free parameter
 to account for the differences between each $\chi^2$ estimate and
 found that the criterion $\chi^2_{gal} <\chi^2_{star}/2$ is robust and
 can be applied to all sources up to $i'_{AB}=23$.
  
\begin{figure}
 \centering\includegraphics[width=8cm]{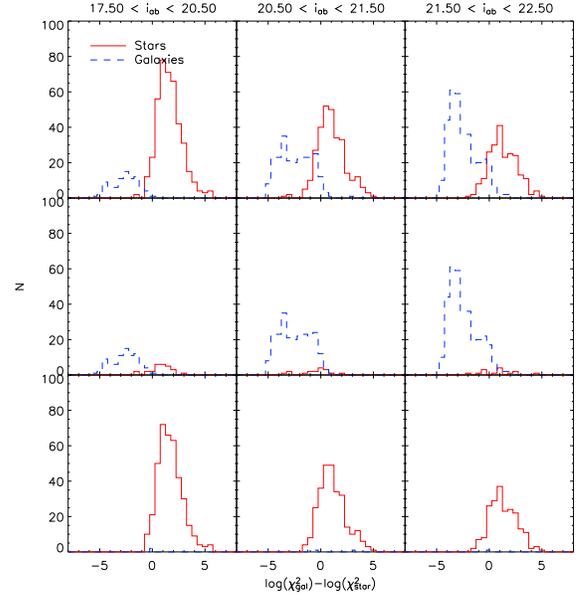}
 \caption{$\chi^2$ distributions for ``true'' stars and ``true'' galaxies
   (according to the VVDS Wide spectroscopy) in the field W4. Top: all
   objects. Middle: objects larger with $r_h > r_{hlimit}$. Bottom: objects
   with $r_h < r_{hlimit}$. if used alone, the $\chi^2$ selection or the
   $r_h$ selection lead to misidentification. On optimal way to
   separate stars from galaxies is to combine the two pieces of
   information.}
 \label{chi2dist}
 \end{figure}

 When only optical data ($u^*,g',r',i',z'$) are used, stellar and
 galaxy colors overlap (shown in figure \ref{chi2dist} is the $\chi^2$
 distributions for stars and galaxies in the spectroscopic sample).  As
 in the case of a purely size-based selection, a selection based solely
 on a multi-colour criterion cannot be used in isolation. The most
 robust star-galaxy separation consists in using both the $\chi^2$
 estimate and the size criterion. Each source is therefore defined by three
 parameters: its size, $r_h$, its galaxy index, $\chi^2_{galaxy}$, and
 its stellar index $\chi^2_{star}$.

 \subsection{Quality of the star/galaxy classification}

 \begin{table}
   \centering
   \caption{Star/galaxy selection results, the table shows
 the galaxy incompleteness (Inc.) and the star contamination 
 (Cont.) for the W1 field, with a large PSF size and the highly
 star crowded field, W4. ``Stars'' represents the intrinsic 
 proportion of stars in the field.}
  \label{starGalTable}
 \begin{center}
    \begin{tabular}{c c c c c}
      \multicolumn{4}{c}{}\\
      \multicolumn{4}{c}{Deep ($i'_{AB} < 24$)}\\
      \hline
      \hline
      Field & $r_{hlimit}$ & Stars & Inc. & Cont.\\
      \hline
      D1  & 2.7 & 11.0\% & 0.80\% &  1.1\%\\ 
      \hline 
    \end{tabular}
    \begin{tabular}{c c c c c}
      \multicolumn{4}{c}{}\\
      \multicolumn{4}{c}{W1 ($i'_{AB} < 22.5$)}\\
      \hline
      \hline
      Field & $r_{hlimit}$ & Stars & Inc. & Cont.\\
      \hline
      022539-050800 & 3.1 & 23.5\% & 3.1\% & 0.4\% \\
      022539-041200 & 2.8 & 18.8\% & 2.3\% & 0.9\% \\
      022929-041200 & 3.1 & 20.6\% & 2.9\% & 0.7\% \\
      All  & 3.0  & 19.0\% & 2.6\% &  1.0\%\\
      \hline
    \end{tabular}
     \begin{tabular}{c c c c c}
      \multicolumn{4}{c}{}\\
      \multicolumn{4}{c}{W4 ($i'_{AB} < 22.5$)}\\
      \hline
      \hline
      Field & $r_{hlimit}$ & Stars & Inc. & Cont.\\
      \hline
      221318-003100 & 2.2 & 57.2\% & 0.6\% & 7.7\% \\
      221318+002300 & 2.1 & 52.5\% & 0.0\% & 9.0\% \\
      221318+011900 & 2.6 & 44.7\% & 0.8\% & 2.5\% \\
      221706-003100 & 2.7 & 53.2\% & 1.0\% & 7.9\% \\
      221706+002300 & 2.3 & 51.9\% & 1.0\% & 6.4\% \\
      221706+011900 & 1.9 & 51.7\% & 0.4\% & 7.3\% \\
      222054-003100 & 2.5 & 52.6\% & 1.0\% & 6.1\% \\
      222054+002300 & 2.4 & 52.8\% & 1.2\% & 7.3\%\\
      222054+011900 & 2.1 & 45.8\% & 0.3\% & 6.3\%\\
       All & 2.3 & 51.8\% & 0.8\% & 6.9\% \\
       \hline
    \end{tabular}
 \end{center}
 \end{table}

 In order to assess the accuracy of the method we tested our
 selection against a strict spectroscopic selection. We defined the
 {\it incompleteness} of the parent galaxy sample as the number of
 galaxies lost after the selection compared to the number of galaxies
 and the {\it star contamination} of the final galaxy sample as the
 number of stars misidentified as galaxies compared to the number of
 estimated galaxies.

 At $i'_{AB} < 21$, stars are purely selected with their sizes. At
 fainter magnitude, we defined a galaxy as an object with $r_h \ge
 \mu_{r_h}+3\sigma_{r_h}$ \emph{or} $\chi^2_{gal} <\chi^2_{star}/2$
 (``\emph{or}'' is used to have a galaxy sample as complete as
 possible).  Stars are defined as  $r_h <
 \mu_{r_h}+3\sigma_{r_h}$ \emph{and} $\chi^2_{star} <2\chi^2_{gal}$.
 We stopped the selection at $i'_{AB} > 23$, where all objects are 
 flagged as galaxies.

 Table \ref{starGalTable} shows the results in the D1, W1 and W4 fields.
 The star contamination is strongly reduced even for the
 fields where the proportion of stars is more than 50\%.
 
 The methods works the best with a small ``PSF'', leading to a small
 $r_{hlimit}$ ($< 2.5~pixels$). 

 The colour-based selection is helpful to keep the 
   galaxy sample
 as complete as possible where the ``PSF'' is large, e.g. 
 $r_{hlimit}\sim3~pixels$
 (the incompleteness being 10\% instead of 3\%,otherwise).

 \subsection{Conclusion}

 The need for an accurate star/galaxy selection has arisen
 from the W4 field, where numerous stars contaminate the galaxy sample.
 We showed that the star/galaxy separation is simple for bright objects 
 ($i'_{AB} < 21$) but becomes less reliable when faint, small objects
 are mixed together. 
 We used the spectroscopic sample to build up and to assess a reliable 
 estimator for the nature of an object, using both the size and 
 color of objects.
 We succeeded in keeping an incompleteness always below 3\% 
 (in the worse ``PSF'' case) while we significantly reduced the 
 star contamination (from 50\% down to 7\% in average). 
 In the best cases (small ``PSF'' and a few stars in the field), 
 the method reaches 1.1\% of star contamination with less than 1\% 
 of incompleteness.
\section{Photometric redshift analysis}
\label{SecAnalysis}


From this Section onwards we consider only ``reliable'' photometric
redshifts. These are defined as redshift of a source estimated from
five photometric bands located in unmasked regions and which fulfilled
passed the size-SED star-galaxy classification described in the
previous Section.  The source must be in the following magnitude
range: $17.5< i'_{AB}<24$ for the Deep fields and $17.5<i'_{AB}<22.5$
for Wide fields.  Hereafter we focus on the reliable photometric redshift samples
to assess the quality redshift catalogues, to derive error estimates
and compute the the redshift distribution in both the Wide and Deep
fields.

\subsection{Error estimates}

``Le Phare'' computes an estimate of the photometric redshift
uncertainty from the redshift Probability Distribution Function (PDF)
of each object.  By determining the redshifts where $\Delta
\chi^2(z)=\chi^2(z)-\chi^2_{min}=1$, we determine a low ($z_{left}$)
and a high ($z_{right}$) redshift value corresponding to a 68\%
confidence interval.

We define the error estimated by ``Le Phare'' $\sigma_{z_p}$ as follows:

\begin{equation}
\ \sigma_{z_p}={|z_{left}(68\%)-z_{right}(68\%)|\over 2} \ .
\end{equation}

The reliability of this error estimate is examined as follows.  We
first compare $\sigma_{z_p}$ to the variance of the difference between
spectroscopic and photometric redshifts $\Delta z$\.  We then use
$\sigma_{z_p}$ to assess the accuracy of the photometric redshifts
over the entire range of magnitudes and redshifts. 

\begin{figure*}
  \centering
  \includegraphics[width=8cm]{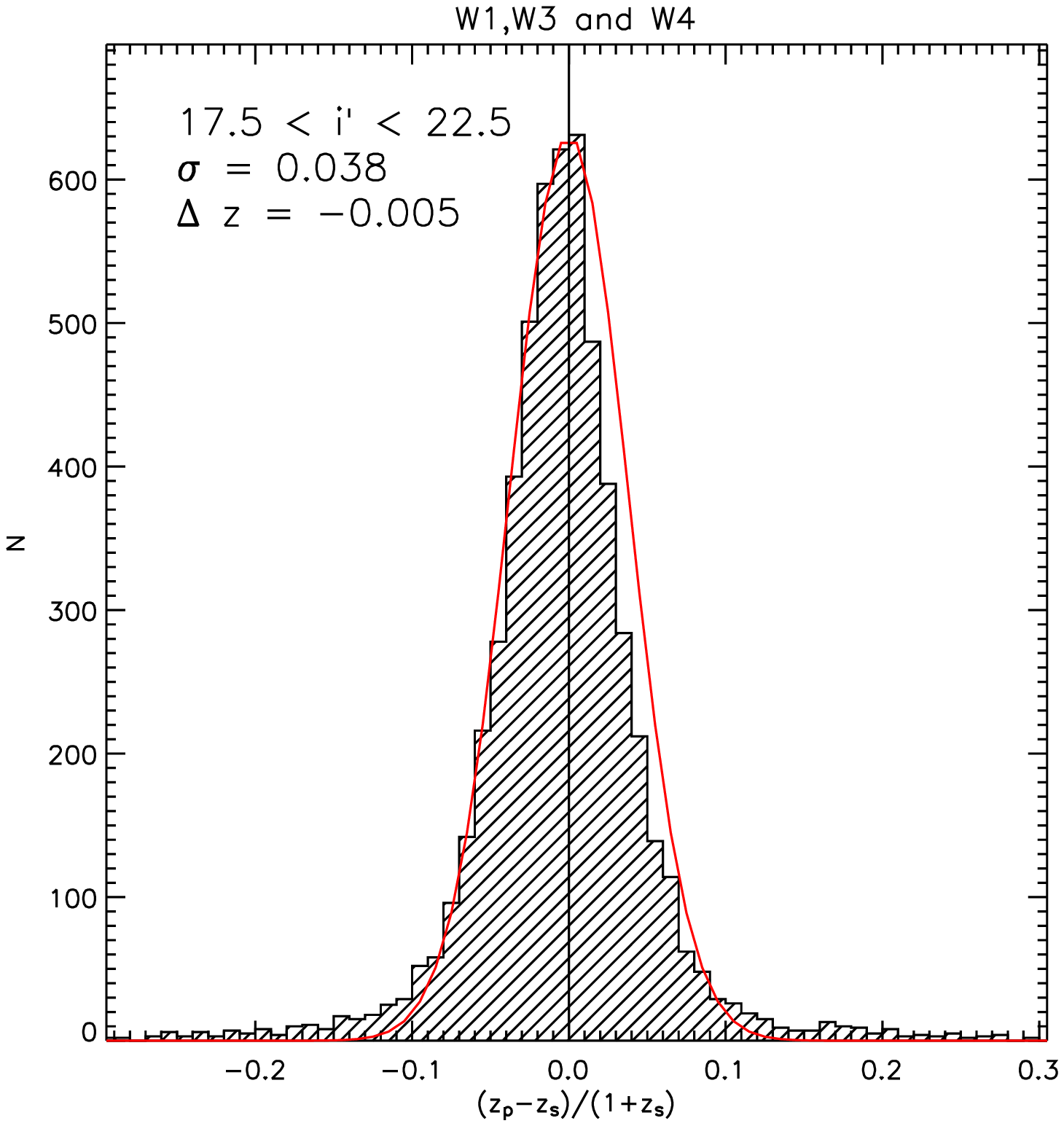}
  \includegraphics[width=8cm]{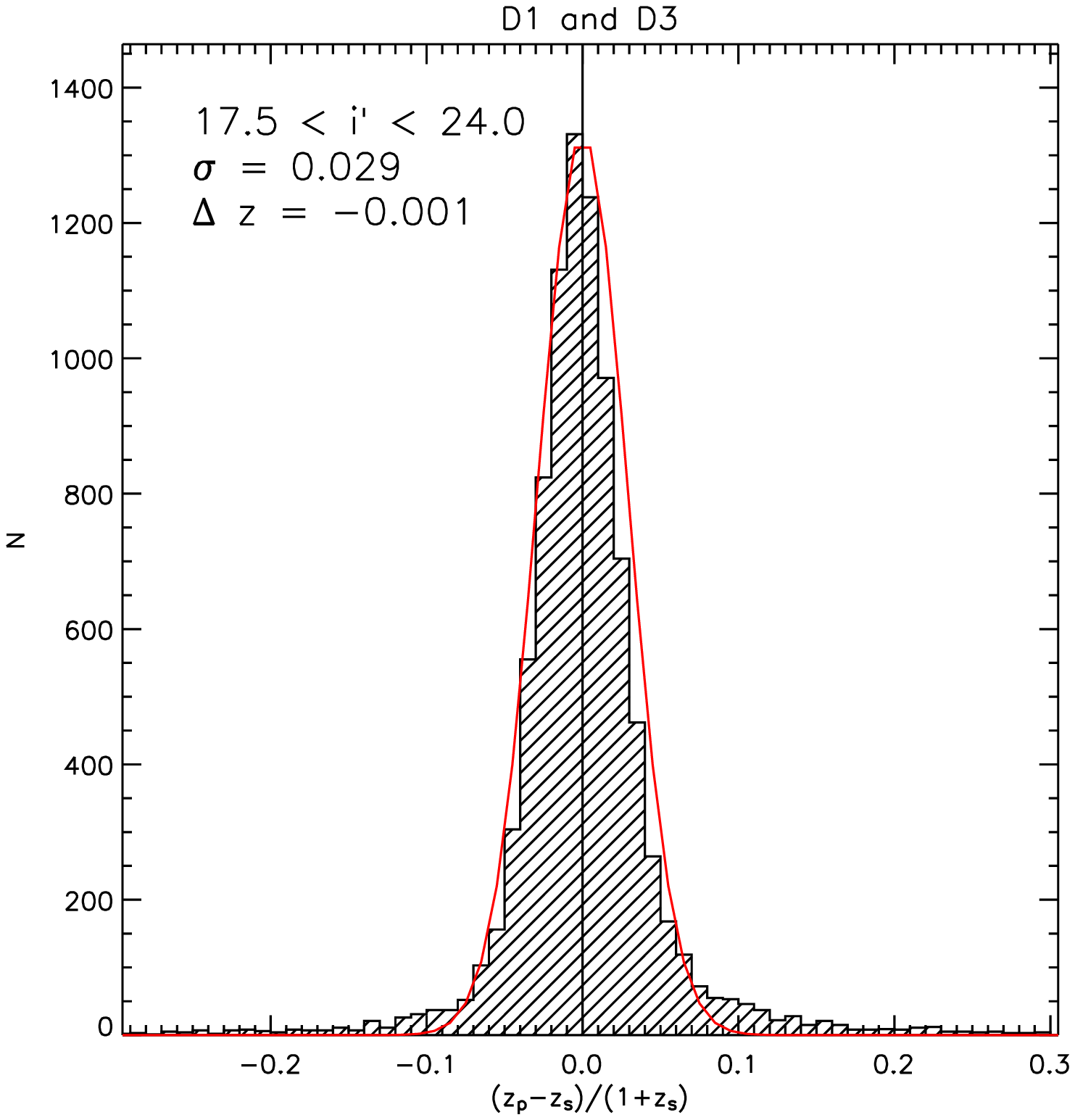}  
  \caption{Histogram of $\Delta z$ for the Wide (W1,W3 and W4, left)
    and the Deep (D1,D3, right) fields.  A Gaussian function with a
    mean of and a dispersion corresponding to the photometric redshift
    dispersion is superimposed. In both cases the error distribution
    is very well represented by this Gaussian function. }

\label{widedeepdisp}
\end{figure*}

Figure \ref{widedeepdisp} shows the distribution of differences
between the spectroscopic and the photometric redshifts ($\Delta z$),
for the wide W1, W3 and W4 fields (left) and for the deep D1 and D3
(right) fields.  On this Figure is overlaid a Gaussian with a mean of
zero and a standard deviation equal to the dispersion derived from the
data, that is $\sigma_{\Delta z-Wide/(1+z_s)}=0.038$ and
$\sigma_{\Delta z-Deep/(1+z_s)}=0.029$.  A very good agreement is
found in the Deep and Wide fields demonstrating that the errors are
nearly Gaussian. The distribution shows an extended tail and departs
from the Gaussian approximation for large errors which is caused by
the outliers. The distribution shows a slight bias in the Wide field
only of $\Delta z_{sys}=median((z_p-z_s)/(1+z_s))=-0.005$.  A closer
look at the bias as a function of the magnitude shows slightly
underestimated photometric redshifts for very bright objects ($i'_{AB} < 20.0$)
and overestimated photometric redshifts at fainter magnitude (see table
\ref{pzBias}), but this bias is always less than 1\%.



\begin{table}
\caption{Median photometric redshift bias in magnitude slices in the field W1.}
\begin{center}
   \begin{tabular}{c c c}
      \hline
      \hline
      $i'_{AB} $ & $\Delta z_{sys}$ \\
      \hline
      17.5-20.0 & 0.0070\\
      20.0-20.5 & -0.0016\\
      20.5-21.0 & -0.0055\\
      21.0-21.5 & -0.0080\\
      21.5-22.0 & -0.0078\\
      22.0-22.5 & -0.0073\\
      \hline
   \end{tabular}
\end{center}
\label{pzBias}
\end{table}

\begin{figure*}
  \centering
  \includegraphics[width=8cm]{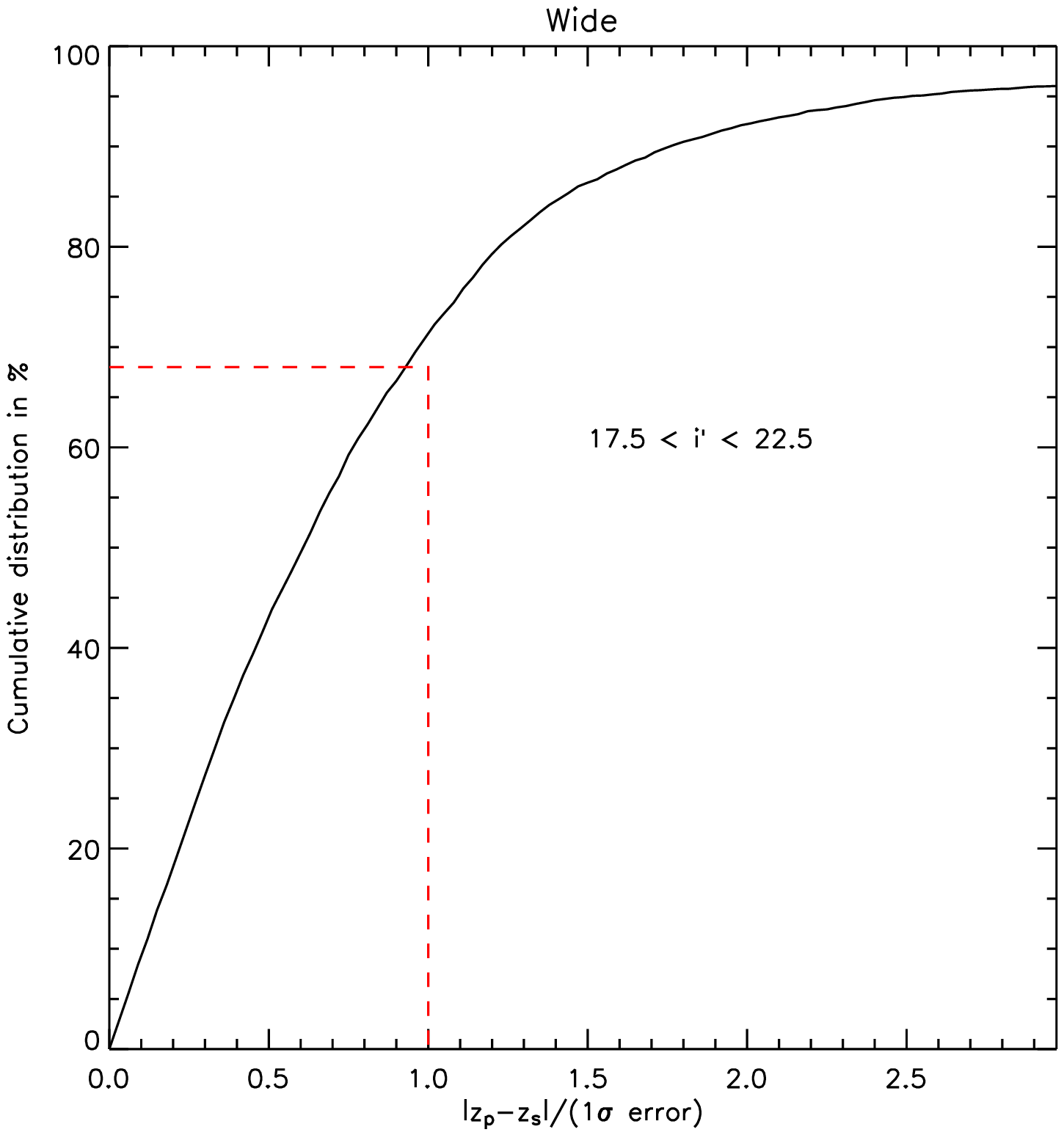}
  \includegraphics[width=8cm]{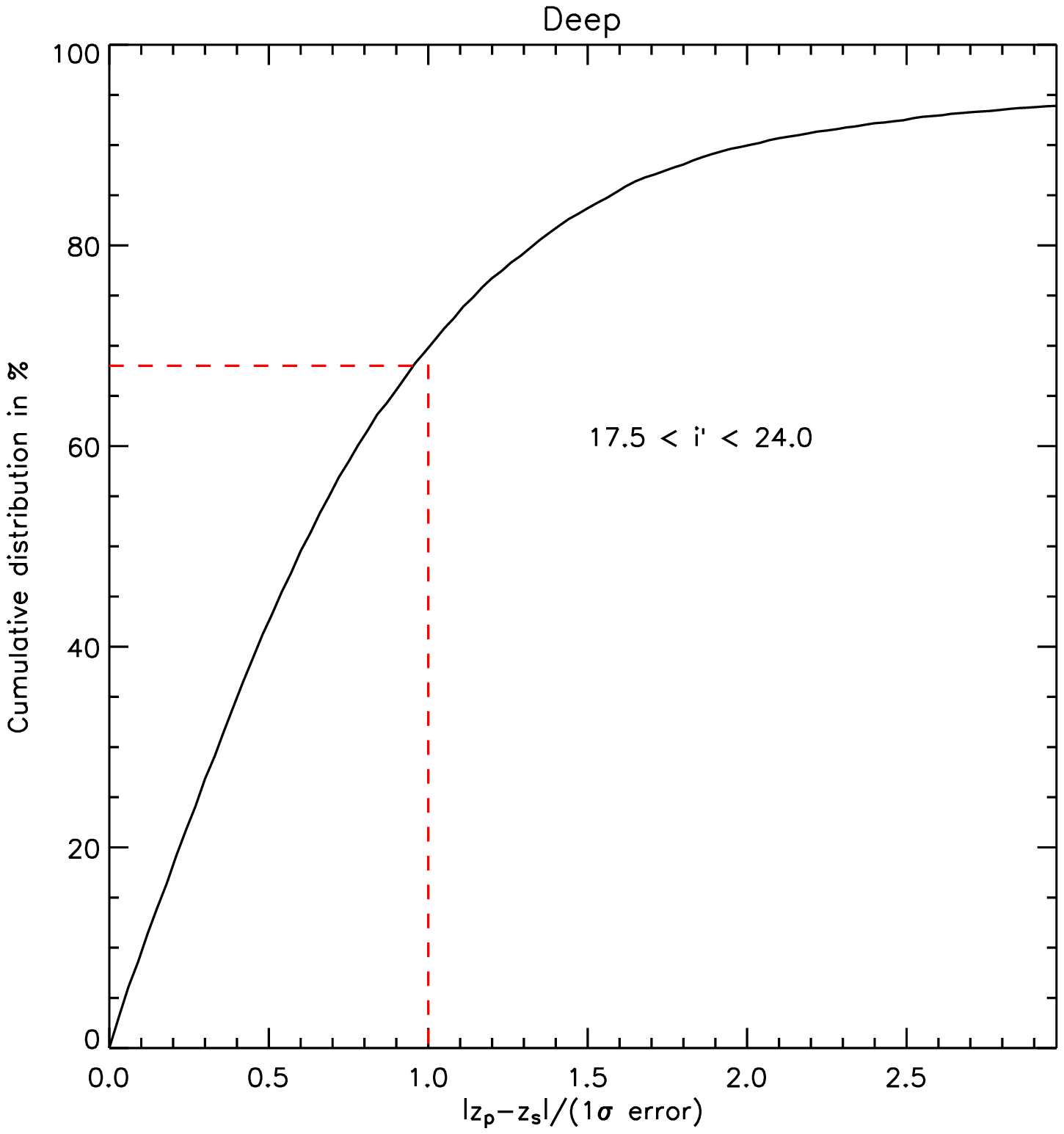}  
  \caption{Cumulative errors for the Wide (W1,W3 and W4) and the Deep
    (D1,D3) fields. As expected for a Gaussian error distribution, the
    photometric redshifts fall within $z_S\pm1\sigma_{z_p}$ 68\% of the time where
    $1\sigma_{z_p}$ is the estimated error.}
\label{WideDeepCumul}
\end{figure*}

The cumulative distributions of the estimated errors compared to the
photometric/spectroscopic error can also be computed.  Assuming a
Gaussian errors distribution we expect that 68\% of the objects have a
photometric redshift within the range $z_s \pm 1\sigma_{z_p}$,
recalling that $\sigma_{z_p}$ is the uncertainty estimated by {``Le
  Phare''}. Figure \ref{WideDeepCumul} shows the cumulative
distribution of $|z_p-z_s|/(1\sigma_{z_p}$.  From the two panels, we
find that 71.3\% of sources in the Wide and 70.8\% in the Deep
satisfy the relation $|z_p-z_s| < 1\sigma_{z_p}$, showing the
reliability of the Gaussian approximation of error estimates.  For
the remainder of this work photometric redshift errors will be
derived from $\sigma_{z_p}$ of ``Le Phare''.

\begin{figure}
\centering
  \includegraphics[width=8cm]{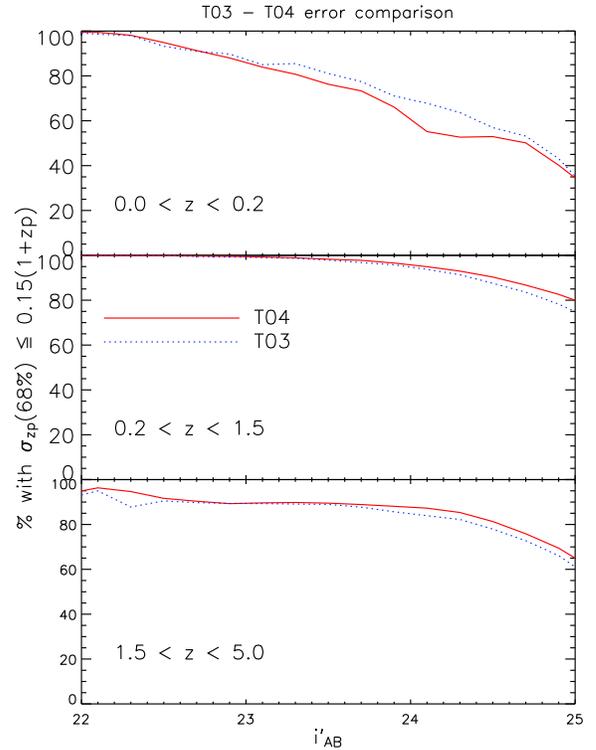}
  \caption{Photometric redshifts errors as function of magnitude and
    redshift for the Deep fields for this study (solid line) and for
    I06 (dotted line).}
\label{T04T03error}
  \end{figure}

We can use this error model to investigate the accuracy of photometric
redshifts over the complete magnitude and redshift ranges of our
catalogues. Figure \ref{T04T03error} shows the fraction of photometric
redshifts in the Deep survey with $\sigma_{z_p} < 0.15\times(1+z_p)$
as function of $i'_{AB}$ apparent magnitude. For photometric redshifts
without outliers one would expect to find almost all objects in this
range.  As seen from the comparisons with spectroscopic redshifts a
small number of objects have catastrophic errors at $z < 0.2$ and $z >
1.5$ where degeneracies in the colour-template space becomes important.

The solid and dotted lines show the photometric redshift errors for
the galaxies in common between T0003 (I06) and T0004 (this work).
Both studies show comparable results with a slight improvement in
T0004 for $z > 0.2$.

\begin{figure}
  \centering
  \includegraphics[width=8cm]{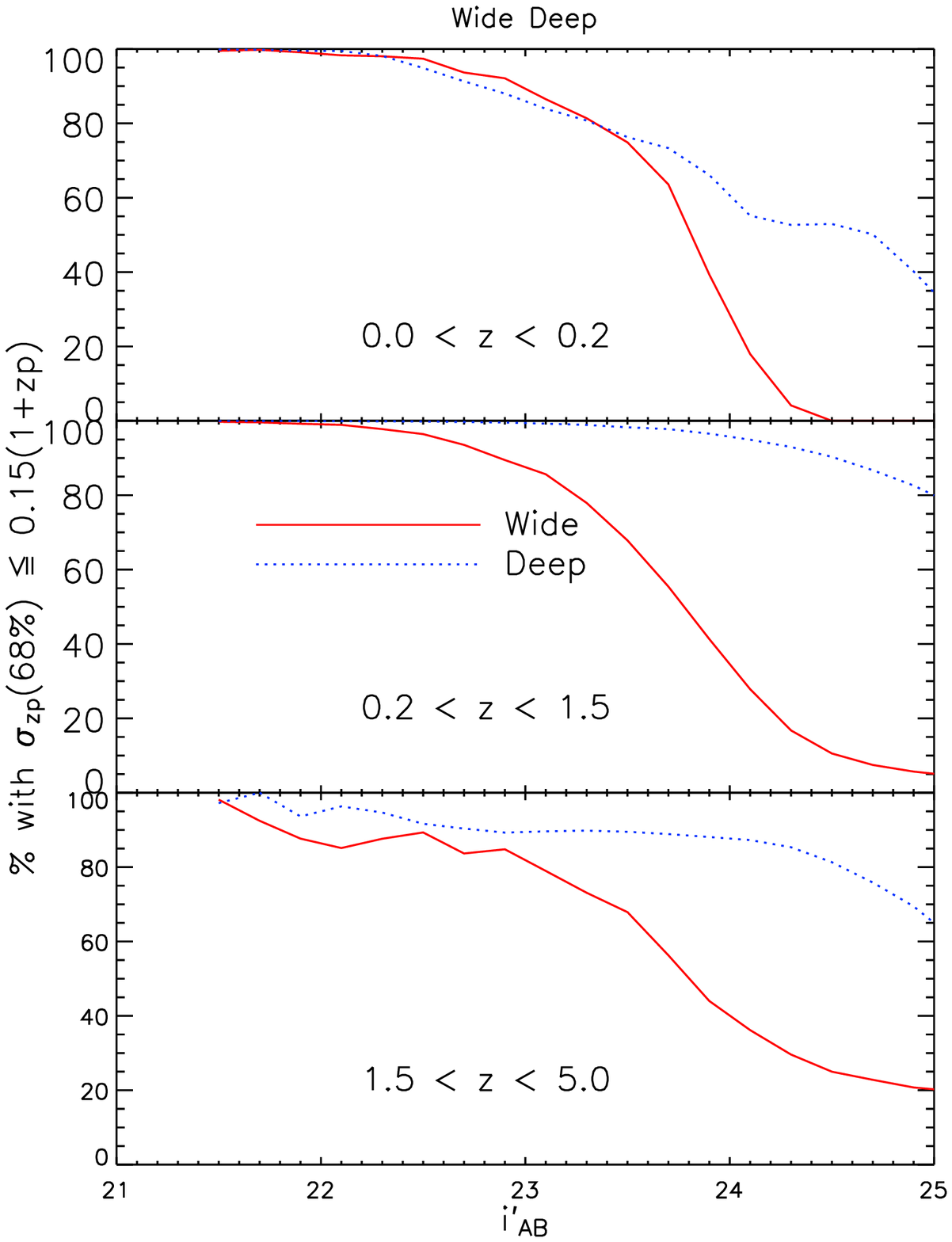}
  \includegraphics[width=8cm]{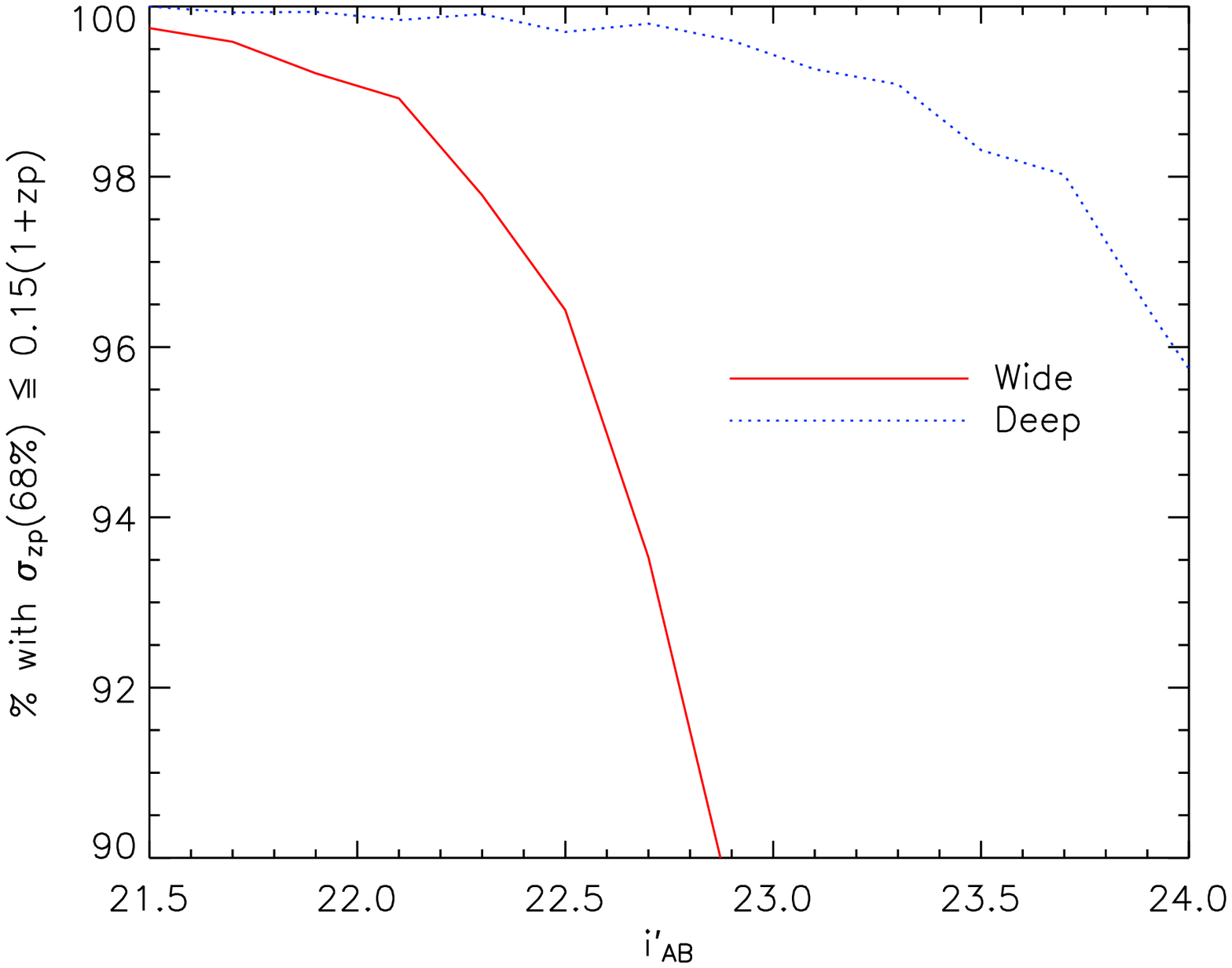}
  \caption{Photometric redshift errors as function of magnitude and
    redshift for both the Wide and the Deep fields (upper panel).
    Lower panel: A zoom of the redshift range $0.2 < z <1.5$.}
 \label{WideDeeperror}
\end{figure}

The comparison between photometric redshift errors in Wide and Deep
fields is shown in Figure\ref{WideDeeperror}.  95\% of sources have an
error within $\pm0.15\times(1+z_p)$, for $i'_{AB} < 24$ in the Deep
field and $i'_{AB} < 22.5$ in the Wide fields.  This result is
consistent with the comparisons with spectroscopic redshifts where the
outlier rate never exceeds 5\%.

\begin{figure}
 \centering
  \includegraphics[width=8cm]{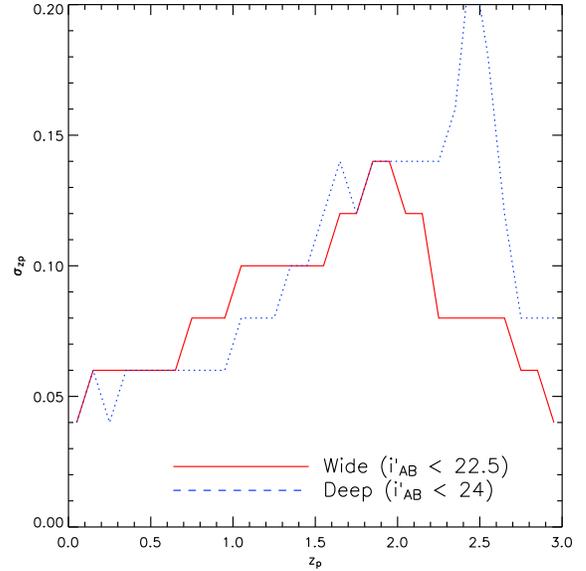}  
  \caption{Estimated photometric redshift error for the Wide (W1) and
    Deep (D1) fields as function of redshift. Errors are binned in
    redshift with an interval = 0.04, which explained the steps in the
    distribution. Few objects only are in the range $z>2$ in the Wide
    field so errors are less reliable in this range.}

 \label{errorvsz} 
 \end{figure}

The dependence of photometric redshift errors on redshift is
illustrated in figure \ref{errorvsz}. As already pointed out in I06,
the photometric redshift errors in the Deep survey is lowest for $z <
1.5$.

\subsection{Redshift distribution}


The redshift distributions of the T0004 Deep and Wide sources were
derived from the histogram of photometric redshifts.

We took into account three sources of uncertainties in the redshift
distribution: the uncertainties on the photo-$z$, the cosmic variance
and the Poisson errors.

We derived the photo-$z$ uncertainties on the redshift bins as follows.
Following \cite{2008A&A...479....9F}, we model the PDF by
 a normalised Gaussian distribution 
 with {\it rms} $\sigma_{z_p}=|z_{left}(68\%)-z_{right}(68\%)|/2$ and 
{\it mean} $z_p$ for each galaxy.
From each individual PDF we randomly drew a redshift and 
built the histogram of the redshift distribution 
for the whole sample. We repeated the process 
100 times and computed the dispersion for each redshift bin.

Cosmic variance and Poisson errors are combined into one single 
error value, where the field-to-field variance is estimated directly
 from the data.
In the Deep field, we simply computed the field-to-field scatter
 over the four independent fields and divided it by $\sqrt 4$.

In the Wide field, the effective area is different for W1, W3 and W4,
so we estimated the field-to-field variance in a different way. We
cut W1 into compact subareas and computed the field-to-field variance
as function of their angular size (see Fig. ]ref{cv}).
We then extrapolated the results
to derive the field-to-field variance corresponding to the full size
of each Wide field. 
Due to a correlation between adjacent subareas,
the field-to-field variance is likely understimated.  As for the Deep
field, we divided by $\sqrt 3$ the error, assuming the 3 Wide fields
are independent.

\begin{figure}
  \centering
  \includegraphics[width=8cm]{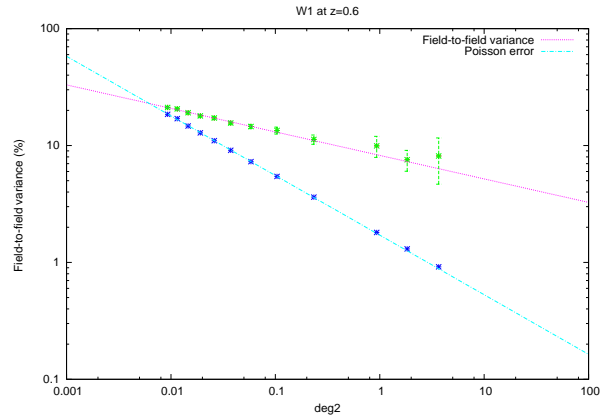}
  \caption{Relative field-to-field variance of subareas in W1 as function 
    of the area (overplotted is the Poisson error).
    Results are extrapolated to derive an estimate of the 
    total field-to-field variance in the Wide field. Uncertainties on
    the field-to-field variance were derived from the data using a Jackknife 
    estimator.}

  \label{cv}
\end{figure}

Finally we fit the redshift distribution by an analytic function.
 We use the parameterised form proposed by \cite{2001A&A...374..757V}:

 \begin{equation} 
n\left(z\right)=\frac{\beta}{z_0\Gamma\left(\frac{1+\alpha}{\beta}\right)}\left(\frac{z}{z_0}\right)^{\alpha}exp\left(-\left(\frac{z}{z_0}\right)^{\beta}\right)
 \end{equation}

and we fit over the range $0.0 < z < 1.5$ for the Wide redshift
distribution and over the range $0.0 < z < 2.5$ for the Deep one.



\begin{figure}
  \centering
  \includegraphics[width=8cm]{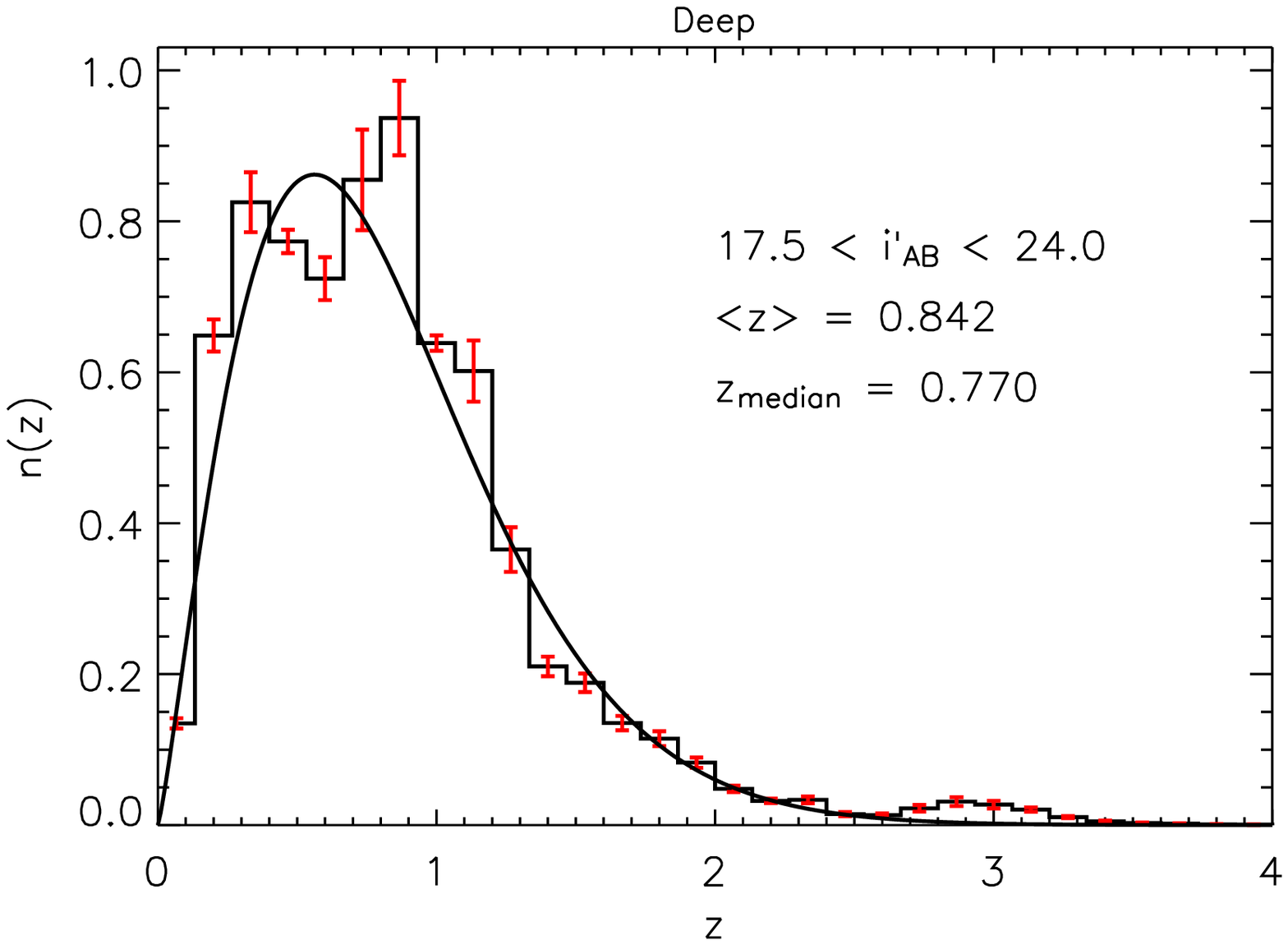}
    \includegraphics[width=8cm]{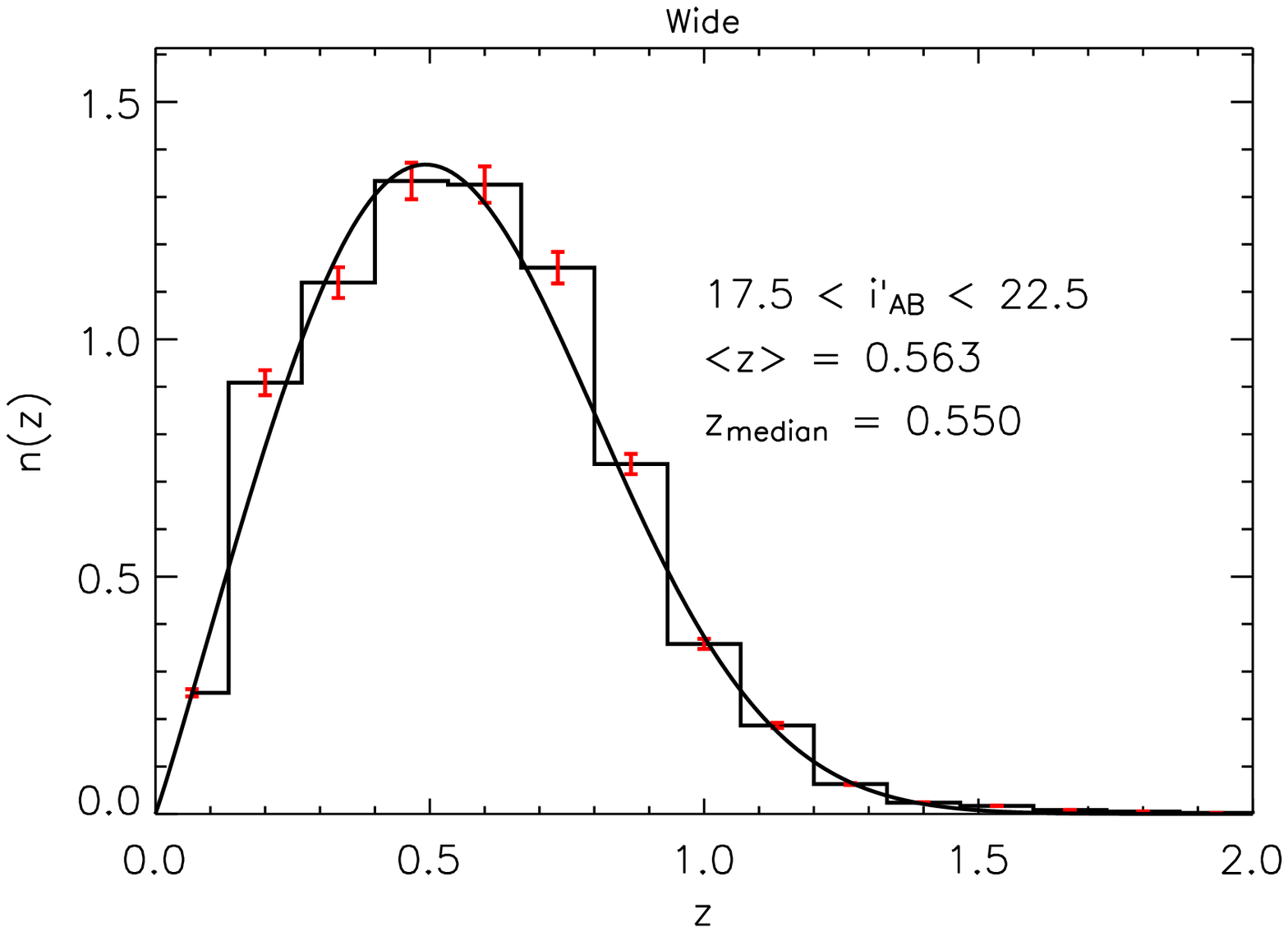}
  \caption{Redshift distribution in the Deep (top) and Wide (bottom) 
fields. Error bars comprise the photo-$z$ error and the field-to-field 
variance (cosmic variance and Poisson error). Error bars show
the uncertainty for the full Deep field (top),computed from the 
field-to-field variance between D1,D2,D3 and D4 and multiplied
 by a factor $1/\sqrt{4}$, since all 4 Deep field are independent.
 In the Wide field, the field-to-field variance 
 were computed for several subareas and extrapolated to
 the size of the Wide fields and finally divided by $\sqrt 3$,
as the three Wide fields are also independent.}
  \label{nz}
\end{figure}

Figure \ref{nz} shows the redshift distribution derived for the
fields W1, W3, W4 and D1, D2, D3, D4 grouped in a Wide and Deep
redshift distributions, respectively.  As expected, the cosmic
variance is lower in the Wide field which covers $35~\deg^2$.

We estimated the mean redshift of the Deep and Wide fields 
for several limiting magnitude.
 Table \ref{meanz} shows the mean and median redshift
computed directly from the data for several 
limiting magnitudes. As expected, the mean and
 median redshifts increase as the
limiting magnitude of the sample increases.

\begin{figure}
  \centering
  \includegraphics[width=8cm]{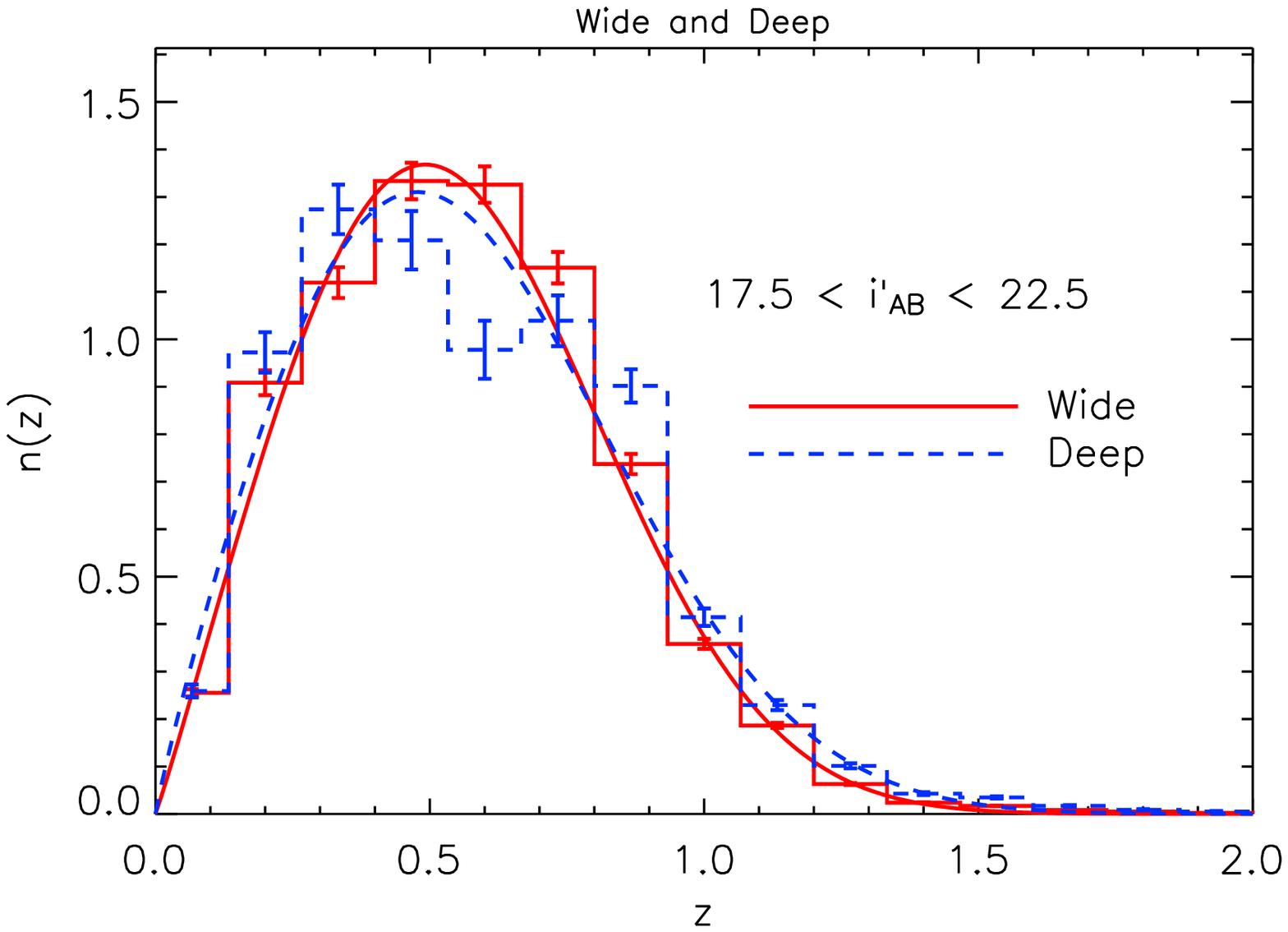}
  \includegraphics[width=8cm]{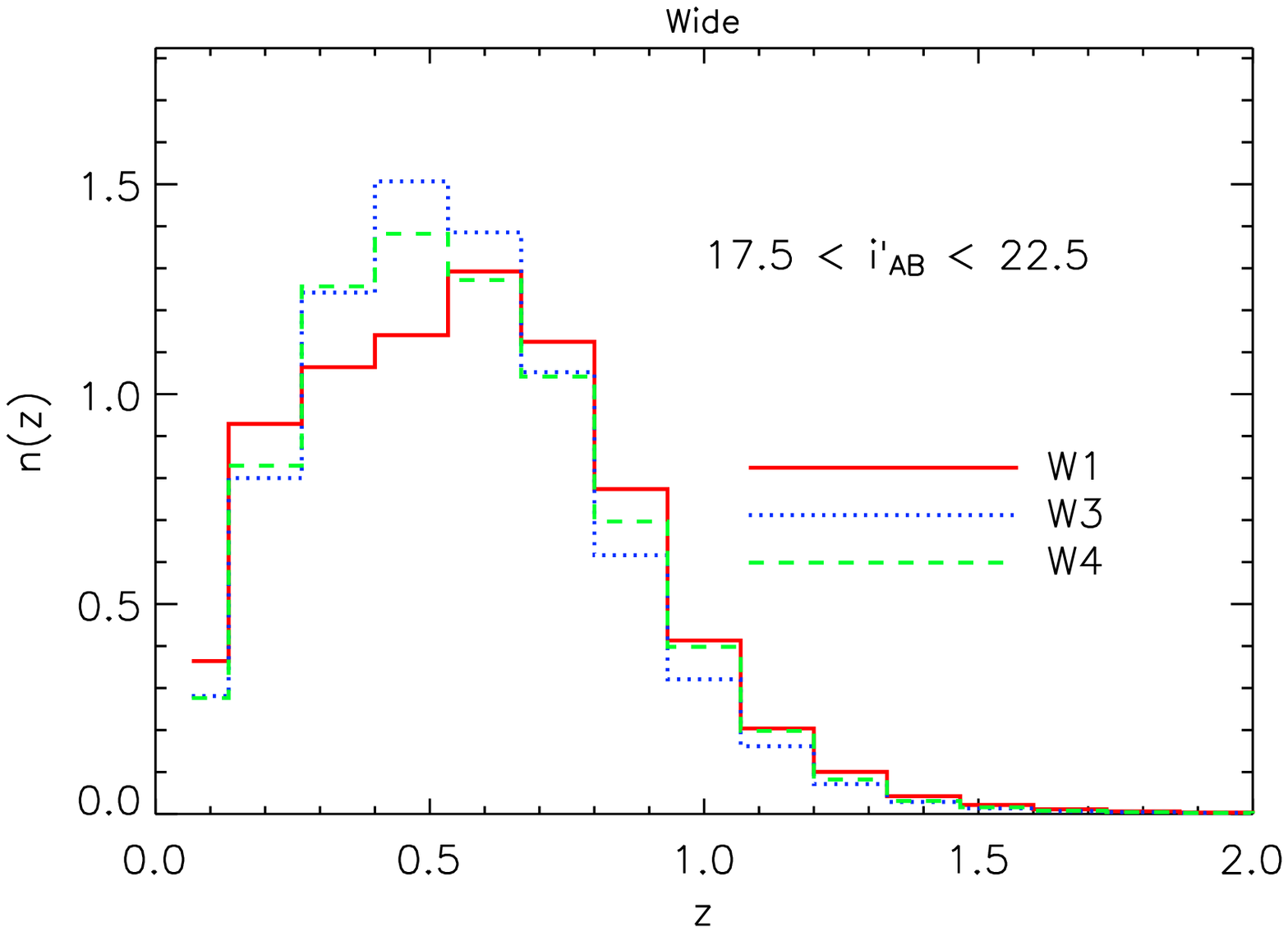}
   \caption{Distribution of the photometric redshifts ($i'_{AB} < 22.5$) computed 
    on the new data release (T0004) for the Wide fields, plotted separately
    on the lower panel and plotted combined on the upper panel.}
  \label{nzT04}
\end{figure}

The top panel of Figure \ref{nzT04} compares the Wide and Deep
redshift distributions and their fit for a same magnitude selection at
$17.5 < i'_{AB} < 22.5$. The redshift
distributions of the Deep and Wide fields agree well, when a same
magnitude limit is adopted.


\begin{table}
  \centering
  \caption{Mean and median redshifts in the Deep and Wide fields,
    for several limiting magnitudes. Both values are computed directly 
    from the data.}

\begin{center}
   \begin{tabular}{c c c}
     \multicolumn{3}{c}{Deep}\\
      \hline
      \hline
      $i'_{AB} < $ & $<z>$ & $z_{median}$\\
      \hline
     
      20.5 & 0.36  & 0.34 \\
      21.0 & 0.40 & 0.37 \\
      21.5 &  0.46 & 0.42 \\
      22.0 & 0.51 & 0.47\\
      22.5 & 0.58 & 0.54\\
      23.0 & 0.65 & 0.62 \\
      23.5 &  0.74 & 0.70\\
      24.0 &  0.84 & 0.77\\
      24.5 & 0.93 & 0.83 \\
      25.0 & 1.00 & 0.88\\
      \hline
   \end{tabular}
 \begin{tabular}{c c c}
   \multicolumn{3}{c}{Wide}\\

      \hline
      \hline
      $i'_{AB} < $ & $<z>$ & $z_{median}$\\
      \hline
      20.5 & 0.37 & 0.35 \\
      21.0 & 0.41 & 0.39 \\ 
      21.5 & 0.46 & 0.44 \\
      22.0 & 0.51 & 0.49 \\
      22.5 & 0.56 & 0.55 \\
      23.0 & 0.62 & 0.60 \\
      23.5 & 0.68 & 0.65 \\
      24.0 & 0.73 & 0.70 \\
      24.5 & 0.77 & 0.75 \\ 
      25.0 & 0.81 & 0.70 \\
      \hline
   \end{tabular}
\label{meanz}

\end{center}
\end{table}




\section*{Conclusions}

We computed photometric redshifts for the release T0004 of the Deep
and Wide CFHT Legacy Survey produced by {\sc Terapix}.  In this work,
only T0004 fields observed in all filters ($u^*$, $g'$, $r'$, $i'$,
and $z'$) relevant for photo-$z$ studies have been selected.  Our
photometric catalogues cover four independent Deep fields of
$1~\deg^2$ each (D1 to D4) and $35~\deg^2$ distributed over
three independent Wide fields (W1,W3 and W4).

We define ``reliable'' photometric redshifts  as redshifts derived for sources 
  classified as galaxies according to a joined size-SED star-galaxy classification criterion,
  located  in unmasked regions with
five-band photometric data and with $\chi^2_{gal} < 1000$. These
objects have a magnitude in the range   $17.5 <i'_{AB} < 22.5$ and 
  $17.5 <i'_{AB} < 24$  for the Wide and Deep samples, respectively.

 In total, the photometric redshift catalogues contain 244,701 reliable
redshifts in the Deep field within the magnitude range $17.5 < i'_{AB}
< 24$ and 592,891 reliable redshifts in the Wide fields within the
range $17.5 < i'_{AB} < 22.5$.

Our photometric redshift method is based on a SED template fitting
procedure using the code ``Le\_Phare''. Following Ilbert et al (2006)
spectroscopic redshifts were used to optimise our SED templates and to
correct the systematic offsets between the observations and SEDs. A
Bayesian prior is used to break the color-redshift degeneracies.

A total of 16,983 spectroscopic redshifts from the VVDS Deep, VVDS
Wide, DEEP2 and zCOSMOS spectroscopic surveys provide a spectroscopic
coverage for each field (except D4). The large quantity of
spectroscopic data enabled an independent systematic offset correction
for each field. A mean correction was applied to D4. These offsets
vary only a small amount between the fields, on the order of 0.01 mag,
demonstrating the excellent field-to-field stability of the CFHTLS
photometry.

We estimated the accuracy of our photometric redshifts by comparing
them to spectroscopic catalogues. The Deep field photometric redshifts
were compared to the VVDS and DEEP2 spectroscopic samples limited at
$17.5 < i'_{AB} < 24$. We found a stable dispersion of 0.028-0.030 and
an outlier rate of about 3-4\%. For the Wide fields, the dispersion is
0.036-0.039 and the outlier rate is about 3-4\% in the range $17.5 <
i'_{AB} < 22.5$ (comparing with VVDS Deep, VVDS Wide, zCOSMOS and
DEEP2). The systematic bias between the input spectroscopic 
   samples and the photometric redshifts derived with {\it Le Phare} keeps 
  below 1\% over $17.5 < i'_{AB} < 22.5$ for the Wide fields and 
  no significant bias is found in the Deep fields.
  We noticed that the dispersion is constant or slighlty
  increasing for $S/N > 40$, with a more significant increase at lower
  $S/N$ values.  The outlier rate increases dramatically in the Wide
  beyond $i'_{AB} = 23$ (5\% and 10\% at $i'_{AB} <23$ and $i'_{AB}
<24$).  Our results are comparable to similar analyses done in the
same CFHTLS Wide fields by \cite{2008arXiv0811.2239E} and
\cite{2008arXiv0811.3211v1B}.

Each CFHTLS Wide field is composed of several $1~\deg^2$
tiles. Unfortunately, the spectroscopic coverage is not sufficient to
perform a calibration of the systematic offsets for each tile.  We
investigated the effect of tile-to-tile photometric offsets on the
accuracy of our photometric redshifts. Taking an uncertainty of 0.03
magnitudes for the photometric calibration in each band, we found that
the photometric redshifts can be degraded up to 14\% in the Wide and
21\% in the Deep. Future wide-field spectroscopic surveys such as
  VIPERS will allow to reduce this problem.

Some catalogues in the Wide fields contain 50\% of stars at $i'_{AB}
<22.5$. A reliable galaxy/star separation is important to
scientifically exploit these catalogues. We combined both the size and
multi-colour data to perform a robust star/galaxy separation. The star
selection is purely based a size critea for the brightest sources
($i'_{AB} < 21$). Since some galaxies are unresolved at fainter
magnitudes, we added multi-colour information to avoid rejecting
galaxies. Using the spectroscopic data to assess the quality of our
classification, we found that our criteria reduces the stellar
contamination from 50\% (in W4, the field with the highest stellar
density) to 7\%, while keeping a galaxy sample more than 98\%
complete. Star contamination is as low as 1-2\% in some other fields,
like W1 or D1. 

Finally, we used the spectroscopic redshifts to evaluate the 68\%
error estimate computed by ``Le Phare'' for the photometric
redshifts. The error distribution, found to be very well approximated
by a Gaussian, was in excellent agreement with the real photo-$z$
dispersion.  Confident with the 68\% error estimate for each
photometric redshift, we determined the accuracy of our photometric
redshifts over a larger magnitude and redshift. Our photometric
redshifts are most accurate in the range $0.2 < z < 1.5$. 


The redshift distributions for the Deep and Wide fields have been
modeled using empirical formula of \cite{2001A&A...374..757V}. The
distributions and the $z_{median}$ and mean $<z>$ derived from the
Deep and Wide samples are consistent with I06,
\cite{2008A&A...479....9F} as well as the recent CFHTLS-Wide studies
of \cite{2008arXiv0811.2239E} and \cite{2008arXiv0811.3211v1B}.

The photometric redshifts calculated here will be an essential
  tool in realising the full potential of the CFHTLS survey. In
  particular, cosmological parameter estimation, galaxy-galaxy
  lensing, galaxy biasing studies and investigation of the halo
  occupation function for galaxies will greatly benefit from this
  homogeneous and well-calibrated set of photometric
  redshifts. Additionally, upcoming large spectroscopic surveys like
  VIPERS will benefit from the increased observing efficiency that
  pre-selection using photometric redshifts can offer.


Our CFHTLS T0004 Deep and Wide photometric redshifts are made
available to the community through the {\sc Terapix}\footnote{{\tt
    http://terapix.iap.fr}} and CENCOS\footnote{{\tt
    http://cencos.oamp.fr/CFHTLS}} databases.


%

\section*{Acknowledgements}
\label{sec:acknowledgement}
We thank the CFHT, {\sc Terapix} and CADC for their assistance and
considerable work in producing the CFHTLS data. We also acknowledge
the VVDS consortium for providing the spectroscopic redshift
catalogues. This work has made use of the VizieR catalogue access
tool, CDS, Strasbourg. We thank S. Colombi, T. Erben, H. Hildebrandt,
B. M\'enard, S. Seitz and L. van Waerbeke for useful discussions.  MK
is supported by the CNRS ANR ``ECOSSTAT'', contract number
ANR-BLAN-0283-04.  We acknowledge the CNRS-Institut National des
Sciences de l'Univers (INSU) and the French Progamme National de
Cosmologie (PNC) for their support for the CFHTLS. This research was
supported by ANR grant ``ANR-07-BLAN-0228''.

\bibliographystyle{aa}
\bibliography{aamnemonic,references}

\begin{thebibliography}{42}
\expandafter\ifx\csname natexlab\endcsname\relax\def\natexlab#1{#1}\fi

\bibitem[{{Arnouts} {et~al.}(1999){Arnouts}, {Cristiani}, {Moscardini},
  {Matarrese}, {Lucchin}, {Fontana}, \& {Giallongo}}]{1999MNRAS.310..540A}
{Arnouts}, S., {Cristiani}, S., {Moscardini}, L., {et~al.} 1999, \mnras, 310,
  540

\bibitem[{{Arnouts} {et~al.}(2002){Arnouts}, {Moscardini}, {Vanzella},
  {Colombi}, {Cristiani}, {Fontana}, {Giallongo}, {Matarrese}, \&
  {Saracco}}]{2002MNRAS.329..355A}
{Arnouts}, S., {Moscardini}, L., {Vanzella}, E., {et~al.} 2002, \mnras, 329,
  355

\bibitem[{{Ball} {et~al.}(2008){Ball}, {Brunner}, {Myers}, {Strand}, {Alberts},
  \& {Tcheng}}]{2008ApJ...683...12B}
{Ball}, N.~M., {Brunner}, R.~J., {Myers}, A.~D., {et~al.} 2008, \apj, 683, 12

\bibitem[{{Ben{\'{\i}}tez}(2000)}]{2000ApJ...536..571B}
{Ben{\'{\i}}tez}, N. 2000, \apj, 536, 571

\bibitem[{{Benjamin} {et~al.}(2007){Benjamin}, {Heymans}, {Semboloni}, {van
  Waerbeke}, {Hoekstra}, {Erben}, {Gladders}, {Hetterscheidt}, {Mellier}, \&
  {Yee}}]{2007MNRAS.381..702B}
{Benjamin}, J., {Heymans}, C., {Semboloni}, E., {et~al.} 2007, \mnras, 381, 702

\bibitem[{{Berg{\'e}} {et~al.}(2008){Berg{\'e}}, {Pacaud}, {R{\'e}fr{\'e}gier},
  {Massey}, {Pierre}, {Amara}, {Birkinshaw}, {Paulin-Henriksson}, {Smith}, \&
  {Willis}}]{2008MNRAS.385..695B}
{Berg{\'e}}, J., {Pacaud}, F., {R{\'e}fr{\'e}gier}, A., {et~al.} 2008, \mnras,
  385, 695

\bibitem[{{Bertin}(2006)}]{2006ASPC..351..112B}
{Bertin}, E. 2006, in Astronomical Society of the Pacific Conference Series,
  Vol. 351, Astronomical Data Analysis Software and Systems XV, ed.
  C.~{Gabriel}, C.~{Arviset}, D.~{Ponz}, \& S.~{Enrique}, 112--+

\bibitem[{{Bertin} \& {Arnouts}(1996)}]{1996A&AS..117..393B}
{Bertin}, E. \& {Arnouts}, S. 1996, \aaps, 117, 393

\bibitem[{{Bertin} {et~al.}(2002){Bertin}, {Mellier}, {Radovich}, {Missonnier},
  {Didelon}, \& {Morin}}]{2002ASPC..281..228B}
{Bertin}, E., {Mellier}, Y., {Radovich}, M., {et~al.} 2002, in Astronomical
  Society of the Pacific Conference Series, Vol. 281, Astronomical Data
  Analysis Software and Systems XI, ed. D.~A. {Bohlender}, D.~{Durand}, \&
  T.~H. {Handley}, 228--+

\bibitem[{{Boulade} {et~al.}(2000){Boulade}, {Charlot}, {Abbon}, {Aune},
  {Borgeaud}, {Carton}, {Carty}, {Desforge}, {Eppele}, {Gallais}, {Gosset},
  {Granelli}, {Gros}, {de Kat}, {Loiseau}, {Mellier}, {Ritou}, {Rousse},
  {Starzynski}, {Vignal}, \& {Vigroux}}]{2000SPIE.4008..657B}
{Boulade}, O., {Charlot}, X., {Abbon}, P., {et~al.} 2000, in Society of
  Photo-Optical Instrumentation Engineers (SPIE) Conference Series, Vol. 4008,
  Society of Photo-Optical Instrumentation Engineers (SPIE) Conference Series,
  ed. M.~{Iye} \& A.~F. {Moorwood}, 657--668

\bibitem[{{Brimioulle} {et~al.}(2008){Brimioulle}, {Lerchster}, {Seitz},
  {Bender}, \& {Snigula}}]{2008arXiv0811.3211v1B}
{Brimioulle}, F., {Lerchster}, M., {Seitz}, S., {Bender}, R., \& {Snigula}, J.
  2008, ArXiv 0811.3211v1

\bibitem[{{Brodwin} {et~al.}(2006){Brodwin}, {Brown}, {Ashby}, {Bian}, {Brand},
  {Dey}, {Eisenhardt}, {Eisenstein}, {Gonzalez}, {Huang}, {Jannuzi},
  {Kochanek}, {McKenzie}, {Murray}, {Pahre}, {Smith}, {Soifer}, {Stanford},
  {Stern}, \& {Elston}}]{2006ApJ...651..791B}
{Brodwin}, M., {Brown}, M.~J.~I., {Ashby}, M.~L.~N., {et~al.} 2006, \apj, 651,
  791

\bibitem[{{Bruzual} \& {Charlot}(2003)}]{2003MNRAS.344.1000B}
{Bruzual}, G. \& {Charlot}, S. 2003, \mnras, 344, 1000

\bibitem[{{Coleman} {et~al.}(1980){Coleman}, {Wu}, \&
  {Weedman}}]{1980ApJS...43..393C}
{Coleman}, G.~D., {Wu}, C.-C., \& {Weedman}, D.~W. 1980, \apjs, 43, 393

\bibitem[{{Davis} {et~al.}(2003){Davis}, {Faber}, {Newman}, {Phillips},
  {Ellis}, {Steidel}, {Conselice}, {Coil}, {Finkbeiner}, {Koo}, {Guhathakurta},
  {Weiner}, {Schiavon}, {Willmer}, {Kaiser}, {Luppino}, {Wirth}, {Connolly},
  {Eisenhardt}, {Cooper}, \& {Gerke}}]{2003SPIE.4834..161D}
{Davis}, M., {Faber}, S.~M., {Newman}, J., {et~al.} 2003, in Presented at the
  Society of Photo-Optical Instrumentation Engineers (SPIE) Conference, Vol.
  4834, Discoveries and Research Prospects from 6- to 10-Meter-Class Telescopes
  II. Edited by Guhathakurta, Puragra. Proceedings of the SPIE, Volume 4834,
  pp. 161-172 (2003)., ed. P.~{Guhathakurta}, 161--172

\bibitem[{{Davis} {et~al.}(2007){Davis}, {Guhathakurta}, {Konidaris}, {Newman},
  {Ashby}, {Biggs}, {Barmby}, {Bundy}, {Chapman}, {Coil}, {Conselice},
  {Cooper}, {Croton}, {Eisenhardt}, {Ellis}, {Faber}, {Fang}, {Fazio},
  {Georgakakis}, {Gerke}, {Goss}, {Gwyn}, {Harker}, {Hopkins}, {Huang},
  {Ivison}, {Kassin}, {Kirby}, {Koekemoer}, {Koo}, {Laird}, {Le Floc'h}, {Lin},
  {Lotz}, {Marshall}, {Martin}, {Metevier}, {Moustakas}, {Nandra}, {Noeske},
  {Papovich}, {Phillips}, {Rich}, {Rieke}, {Rigopoulou}, {Salim},
  {Schiminovich}, {Simard}, {Smail}, {Small}, {Weiner}, {Willmer}, {Willner},
  {Wilson}, {Wright}, \& {Yan}}]{2007ApJ...660L...1D}
{Davis}, M., {Guhathakurta}, P., {Konidaris}, N.~P., {et~al.} 2007, \apjl, 660,
  L1

\bibitem[{{Erben} {et~al.}(2008){Erben}, {Hildebrandt}, {Lerchster}, {Hudelot},
  {Benjamin}, {van Waerbeke}, {Schrabback}, {Brimioulle}, {Cordes}, {Dietrich},
  {Holhjem}, {Schirmer}, \& {Schneider}}]{2008arXiv0811.2239E}
{Erben}, T., {Hildebrandt}, H., {Lerchster}, M., {et~al.} 2008, ArXiv 0811.2239

\bibitem[{{Fahlman} {et~al.}(1994){Fahlman}, {Kaiser}, {Squires}, \&
  {Woods}}]{1994ApJ...437...56F}
{Fahlman}, G., {Kaiser}, N., {Squires}, G., \& {Woods}, D. 1994, \apj, 437, 56

\bibitem[{{Fu} {et~al.}(2008){Fu}, {Semboloni}, {Hoekstra}, {Kilbinger}, {van
  Waerbeke}, {Tereno}, {Mellier}, {Heymans}, {Coupon}, {Benabed}, {Benjamin},
  {Bertin}, {Dor{\'e}}, {Hudson}, {Ilbert}, {Maoli}, {Marmo}, {McCracken}, \&
  {M{\'e}nard}}]{2008A&A...479....9F}
{Fu}, L., {Semboloni}, E., {Hoekstra}, H., {et~al.} 2008, \aap, 479, 9

\bibitem[{{Garilli} {et~al.}(2008){Garilli}, {Le F{\`e}vre}, {Guzzo},
  {Maccagni}, {Le Brun}, {de La Torre}, {Meneux}, {Tresse}, {Franzetti},
  {Zamorani}, {Zanichelli}, {Gregorini}, {Vergani}, {Bottini}, {Scaramella},
  {Scodeggio}, {Vettolani}, {Adami}, {Arnouts}, {Bardelli}, {Bolzonella},
  {Cappi}, {Charlot}, {Ciliegi}, {Contini}, {Foucaud}, {Gavignaud}, {Ilbert},
  {Iovino}, {Lamareille}, {McCracken}, {Marano}, {Marinoni}, {Mazure},
  {Merighi}, {Paltani}, {Pell{\`o}}, {Pollo}, {Pozzetti}, {Radovich}, {Zucca},
  {Blaizot}, {Bongiorno}, {Cucciati}, {Mellier}, {Moreau}, \&
  {Paioro}}]{2008A&A...486..683G}
{Garilli}, B., {Le F{\`e}vre}, O., {Guzzo}, L., {et~al.} 2008, \aap, 486, 683

\bibitem[{{Hildebrandt} {et~al.}(2008){Hildebrandt}, {Wolf}, \&
  {Ben{\'{\i}}tez}}]{2008A&A...480..703H}
{Hildebrandt}, H., {Wolf}, C., \& {Ben{\'{\i}}tez}, N. 2008, \aap, 480, 703

\bibitem[{{Ilbert} {et~al.}(2006){Ilbert}, {Arnouts}, {McCracken},
  {Bolzonella}, {Bertin}, {Le F{\`e}vre}, {Mellier}, {Zamorani}, {Pell{\`o}},
  {Iovino}, {Tresse}, {Le Brun}, {Bottini}, {Garilli}, {Maccagni}, {Picat},
  {Scaramella}, {Scodeggio}, {Vettolani}, {Zanichelli}, {Adami}, {Bardelli},
  {Cappi}, {Charlot}, {Ciliegi}, {Contini}, {Cucciati}, {Foucaud}, {Franzetti},
  {Gavignaud}, {Guzzo}, {Marano}, {Marinoni}, {Mazure}, {Meneux}, {Merighi},
  {Paltani}, {Pollo}, {Pozzetti}, {Radovich}, {Zucca}, {Bondi}, {Bongiorno},
  {Busarello}, {de La Torre}, {Gregorini}, {Lamareille}, {Mathez}, {Merluzzi},
  {Ripepi}, {Rizzo}, \& {Vergani}}]{2006A&A...457..841I}
{Ilbert}, O., {Arnouts}, S., {McCracken}, H.~J., {et~al.} 2006, \aap, 457, 841

\bibitem[{{Ilbert} {et~al.}(2008){Ilbert}, {Capak}, {Salvato}, {Aussel},
  {McCracken}, {Sanders}, {Scoville}, {Kartaltepe}, {Arnouts}, {Le Floc'h},
  {Mobasher}, {Taniguchi}, {Lamareille}, {Leauthaud}, {Sasaki}, {Thompson},
  {Zamojski}, {Zamorani}, {Bardelli}, {Bolzonella}, {Bongiorno}, {Brusa},
  {Caputi}, {Carollo}, {Contini}, {Cook}, {Coppa}, {Cucciati}, {de la Torre},
  {de Ravel}, {Franzetti}, {Garilli}, {Hasinger}, {Iovino}, {Kampczyk},
  {Kneib}, {Knobel}, {Kovac}, {Le Borgne}, {Le Brun}, {Le Fevre}, {Lilly},
  {Looper}, {Maier}, {Mainieri}, {Mellier}, {Mignoli}, {Murayama}, {Pello},
  {Peng}, {Perez-Montero}, {Renzini}, {Ricciardelli}, {Schiminovich},
  {Scodeggio}, {Shioya}, {Silverman}, {Surace}, {Tanaka}, {Tasca}, {Tresse},
  {Vergani}, \& {Zucca}}]{2008arXiv0809.2101I}
{Ilbert}, O., {Capak}, P., {Salvato}, M., {et~al.} 2008, ArXiv 0809.2101

\bibitem[{{Ilbert} {et~al.}(2005){Ilbert}, {Tresse}, {Zucca}, {Bardelli},
  {Arnouts}, {Zamorani}, {Pozzetti}, {Bottini}, {Garilli}, {Le Brun}, {Le
  F{\`e}vre}, {Maccagni}, {Picat}, {Scaramella}, {Scodeggio}, {Vettolani},
  {Zanichelli}, {Adami}, {Arnaboldi}, {Bolzonella}, {Cappi}, {Charlot},
  {Contini}, {Foucaud}, {Franzetti}, {Gavignaud}, {Guzzo}, {Iovino},
  {McCracken}, {Marano}, {Marinoni}, {Mathez}, {Mazure}, {Meneux}, {Merighi},
  {Paltani}, {Pello}, {Pollo}, {Radovich}, {Bondi}, {Bongiorno}, {Busarello},
  {Ciliegi}, {Lamareille}, {Mellier}, {Merluzzi}, {Ripepi}, \&
  {Rizzo}}]{2005A&A...439..863I}
{Ilbert}, O., {Tresse}, L., {Zucca}, E., {et~al.} 2005, \aap, 439, 863

\bibitem[{{Kilbinger} {et~al.}(2008){Kilbinger}, {Benabed}, {Guy}, {Astier},
  {Tereno}, {Fu}, {Wraith}, {Coupon}, {Mellier}, {Balland}, {Bouchet},
  {Hamana}, {Hardin}, {McCracken}, {Pain}, {Regnault}, {Schultheiss}, \&
  {Yahagi}}]{2008arXiv0810.5129K}
{Kilbinger}, M., {Benabed}, K., {Guy}, J., {et~al.} 2008, ArXiv 0810.5129

\bibitem[{{Kinney} {et~al.}(1996){Kinney}, {Calzetti}, {Bohlin}, {McQuade},
  {Storchi-Bergmann}, \& {Schmitt}}]{1996ApJ...467...38K}
{Kinney}, A.~L., {Calzetti}, D., {Bohlin}, R.~C., {et~al.} 1996, \apj, 467, 38

\bibitem[{{Le F{\`e}vre} {et~al.}(2005{\natexlab{a}}){Le F{\`e}vre}, {Guzzo},
  {Meneux}, {Pollo}, {Cappi}, {Colombi}, {Iovino}, {Marinoni}, {McCracken},
  {Scaramella}, {Bottini}, {Garilli}, {Le Brun}, {Maccagni}, {Picat},
  {Scodeggio}, {Tresse}, {Vettolani}, {Zanichelli}, {Adami}, {Arnaboldi},
  {Arnouts}, {Bardelli}, {Blaizot}, {Bolzonella}, {Charlot}, {Ciliegi},
  {Contini}, {Foucaud}, {Franzetti}, {Gavignaud}, {Ilbert}, {Marano}, {Mathez},
  {Mazure}, {Merighi}, {Paltani}, {Pell{\`o}}, {Pozzetti}, {Radovich},
  {Zamorani}, {Zucca}, {Bondi}, {Bongiorno}, {Busarello}, {Lamareille},
  {Mellier}, {Merluzzi}, {Ripepi}, \& {Rizzo}}]{2005A&A...439..877L}
{Le F{\`e}vre}, O., {Guzzo}, L., {Meneux}, B., {et~al.} 2005{\natexlab{a}},
  \aap, 439, 877

\bibitem[{{Le F{\`e}vre} {et~al.}(2005{\natexlab{b}}){Le F{\`e}vre},
  {Vettolani}, {Garilli}, {Tresse}, {Bottini}, {Le Brun}, {Maccagni}, {Picat},
  {Scaramella}, {Scodeggio}, {Zanichelli}, {Adami}, {Arnaboldi}, {Arnouts},
  {Bardelli}, {Bolzonella}, {Cappi}, {Charlot}, {Ciliegi}, {Contini},
  {Foucaud}, {Franzetti}, {Gavignaud}, {Guzzo}, {Ilbert}, {Iovino},
  {McCracken}, {Marano}, {Marinoni}, {Mathez}, {Mazure}, {Meneux}, {Merighi},
  {Paltani}, {Pell{\`o}}, {Pollo}, {Pozzetti}, {Radovich}, {Zamorani}, {Zucca},
  {Bondi}, {Bongiorno}, {Busarello}, {Lamareille}, {Mellier}, {Merluzzi},
  {Ripepi}, \& {Rizzo}}]{2005A&A...439..845L}
{Le F{\`e}vre}, O., {Vettolani}, G., {Garilli}, B., {et~al.}
  2005{\natexlab{b}}, \aap, 439, 845

\bibitem[{{Lilly} {et~al.}(2007){Lilly}, {Le F{\`e}vre}, {Renzini}, {Zamorani},
  {Scodeggio}, {Contini}, {Carollo}, {Hasinger}, {Kneib}, {Iovino}, {Le Brun},
  {Maier}, {Mainieri}, {Mignoli}, {Silverman}, {Tasca}, {Bolzonella},
  {Bongiorno}, {Bottini}, {Capak}, {Caputi}, {Cimatti}, {Cucciati}, {Daddi},
  {Feldmann}, {Franzetti}, {Garilli}, {Guzzo}, {Ilbert}, {Kampczyk}, {Kovac},
  {Lamareille}, {Leauthaud}, {Borgne}, {McCracken}, {Marinoni}, {Pello},
  {Ricciardelli}, {Scarlata}, {Vergani}, {Sanders}, {Schinnerer}, {Scoville},
  {Taniguchi}, {Arnouts}, {Aussel}, {Bardelli}, {Brusa}, {Cappi}, {Ciliegi},
  {Finoguenov}, {Foucaud}, {Franceschini}, {Halliday}, {Impey}, {Knobel},
  {Koekemoer}, {Kurk}, {Maccagni}, {Maddox}, {Marano}, {Marconi}, {Meneux},
  {Mobasher}, {Moreau}, {Peacock}, {Porciani}, {Pozzetti}, {Scaramella},
  {Schiminovich}, {Shopbell}, {Smail}, {Thompson}, {Tresse}, {Vettolani},
  {Zanichelli}, \& {Zucca}}]{2007ApJS..172...70L}
{Lilly}, S.~J., {Le F{\`e}vre}, O., {Renzini}, A., {et~al.} 2007, \apjs, 172,
  70

\bibitem[{{Magnier} \& {Cuillandre}(2004)}]{2004PASP..116..449M}
{Magnier}, E.~A. \& {Cuillandre}, J.-C. 2004, \pasp, 116, 449

\bibitem[{{McCracken} {et~al.}(2008){McCracken}, {Ilbert}, {Mellier}, {Bertin},
  {Guzzo}, {Arnouts}, {Le F{\`e}vre}, \& {Zamorani}}]{2008A&A...479..321M}
{McCracken}, H.~J., {Ilbert}, O., {Mellier}, Y., {et~al.} 2008, \aap, 479, 321

\bibitem[{{Mobasher} {et~al.}(2007){Mobasher}, {Capak}, {Scoville}, {Dahlen},
  {Salvato}, {Aussel}, {Thompson}, {Feldmann}, {Tasca}, {Lefevre}, {Lilly},
  {Carollo}, {Kartaltepe}, {McCracken}, {Mould}, {Renzini}, {Sanders},
  {Shopbell}, {Taniguchi}, {Ajiki}, {Shioya}, {Contini}, {Giavalisco},
  {Ilbert}, {Iovino}, {Le Brun}, {Mainieri}, {Mignoli}, \&
  {Scodeggio}}]{2007ApJS..172..117M}
{Mobasher}, B., {Capak}, P., {Scoville}, N.~Z., {et~al.} 2007, \apjs, 172, 117

\bibitem[{{Parker} {et~al.}(2007){Parker}, {Hoekstra}, {Hudson}, {van
  Waerbeke}, \& {Mellier}}]{2007ApJ...669...21P}
{Parker}, L.~C., {Hoekstra}, H., {Hudson}, M.~J., {van Waerbeke}, L., \&
  {Mellier}, Y. 2007, \apj, 669, 21

\bibitem[{{Pickles}(1998)}]{1998PASP..110..863P}
{Pickles}, A.~J. 1998, \pasp, 110, 863

\bibitem[{{Prevot} {et~al.}(1984){Prevot}, {Lequeux}, {Prevot}, {Maurice}, \&
  {Rocca-Volmerange}}]{1984A&A...132..389P}
{Prevot}, M.~L., {Lequeux}, J., {Prevot}, L., {Maurice}, E., \&
  {Rocca-Volmerange}, B. 1984, \aap, 132, 389

\bibitem[{{Rowan-Robinson} {et~al.}(2008){Rowan-Robinson}, {Babbedge},
  {Oliver}, {Trichas}, {Berta}, {Lonsdale}, {Smith}, {Shupe}, {Surace},
  {Arnouts}, {Ilbert}, {Le F{\'e}vre}, {Afonso-Luis}, {Perez-Fournon},
  {Hatziminaoglou}, {Polletta}, {Farrah}, \& {Vaccari}}]{2008MNRAS.386..697R}
{Rowan-Robinson}, M., {Babbedge}, T., {Oliver}, S., {et~al.} 2008, \mnras, 386,
  697

\bibitem[{{Schlegel} {et~al.}(1998){Schlegel}, {Finkbeiner}, \&
  {Davis}}]{1998ApJ...500..525S}
{Schlegel}, D.~J., {Finkbeiner}, D.~P., \& {Davis}, M. 1998, \apj, 500, 525

\bibitem[{{Schultheis} {et~al.}(2006){Schultheis}, {Robin}, {Reyl{\'e}},
  {McCracken}, {Bertin}, {Mellier}, \& {Le F{\`e}vre}}]{2006A&A...447..185S}
{Schultheis}, M., {Robin}, A.~C., {Reyl{\'e}}, C., {et~al.} 2006, \aap, 447,
  185

\bibitem[{{Szalay} {et~al.}(1999){Szalay}, {Connolly}, \&
  {Szokoly}}]{1999AJ....117...68S}
{Szalay}, A.~S., {Connolly}, A.~J., \& {Szokoly}, G.~P. 1999, \aj, 117, 68

\bibitem[{{Tereno} {et~al.}(2008){Tereno}, {Schimd}, {Uzan}, {Kilbinger},
  {Vincent}, \& {Fu}}]{2008arXiv0810.0555T}
{Tereno}, I., {Schimd}, C., {Uzan}, J.-P., {et~al.} 2008, ArXiv 0810.0555

\bibitem[{{Van Waerbeke} {et~al.}(2001){Van Waerbeke}, {Mellier}, {Radovich},
  {Bertin}, {Dantel-Fort}, {McCracken}, {Le F{\`e}vre}, {Foucaud},
  {Cuillandre}, {Erben}, {Jain}, {Schneider}, {Bernardeau}, \&
  {Fort}}]{2001A&A...374..757V}
{Van Waerbeke}, L., {Mellier}, Y., {Radovich}, M., {et~al.} 2001, \aap, 374,
  757

\bibitem[{{Wolf} {et~al.}(2003){Wolf}, {Meisenheimer}, {Rix}, {Borch}, {Dye},
  \& {Kleinheinrich}}]{2003A&A...401...73W}
{Wolf}, C., {Meisenheimer}, K., {Rix}, H.-W., {et~al.} 2003, \aap, 401, 73

\end{thebibliography}
\end{document}